\documentclass[12pt]{article}
\pdfoutput=1
\usepackage{float}
\usepackage{amssymb}
\usepackage{amsmath}
\usepackage{latexsym}
\usepackage[usenames]{color}
\usepackage[mathscr]{eucal}
\usepackage{fancybox}
\usepackage{simplewick}
\usepackage{comment}
\usepackage{cite}
\usepackage{framed}
\definecolor{shadecolor}{rgb}{0.9,0.9,0.95}
\usepackage{setspace}
\usepackage[normalem]{ulem}
\definecolor{darkgreen}{rgb}{0,0.5,0}
\definecolor{darkblue}{cmyk}{0.9,0.9,0,0}
\definecolor{darkred}{rgb}{0.6,0,0.3}

\usepackage{graphicx}
\usepackage[curve,matrix,arrow,color]{xy}
\usepackage{tikz}
\usetikzlibrary{intersections}
\usetikzlibrary{er,positioning}
\usetikzlibrary{arrows,shapes}
\usetikzlibrary{decorations.markings}
\usepackage[setpagesize=false,pagebackref=false, linktocpage, bookmarksopen=true, colorlinks=true, linkcolor=darkblue,citecolor=darkblue,urlcolor=black]{hyperref}
\usepackage{hyperref}

\renewcommand{\Im}{{\rm Im}}
\renewcommand{\Re}{{\rm Re}}
\newcommand{\mm}{\mathfrak{m}}
\renewcommand{\thefootnote}{\arabic{footnote}}

\def\eqref#1{(\ref{#1})}

\def\beq{\begin{equation}}
\def\eeq{\end{equation}}

\newcommand{\rQ}{\mathrm{Q}}

\numberwithin{equation}{section}
\newcommand{\aSt}{\mathcal{W}}

\newcommand{\AStTL}[2]{\aSt_{#1,#2}}

\newcommand{\StJTL}[2]{\mathcal{W}_{#1,#2}}
\newcommand{\bAStTL}[2]{\overline{\mathcal{W}}_{#1,#2}}
\newcommand{\ATL}[1]{\mathsf{T}^{\rm a}_{#1}}

\newcommand{\HN}{\mathcal{H}}

\newcommand{\m}{\mathfrak{m}}
\newtheorem{theorem}{Theorem}[section]

\newtheorem{corollary}[theorem]{Corollary}

\newtheorem{definition}[theorem]{Definition}
\newtheorem{example}[theorem]{Example}
\newtheorem{lemma}[theorem]{Lemma}

\newtheorem{remark}[theorem]{Remark}

\begin{document}
\thispagestyle{empty}

\renewcommand{\thefootnote}{\fnsymbol{footnote}}
\setcounter{page}{1}
\setcounter{footnote}{0}
\setcounter{figure}{0}
%%%%%%%%%%%%%%%%%%%%%%%%%%%%%%%%%%%%%%%%%%%%%%%%%%%%%%%%%%%%%%%%%%%%%%%%%%%%%%%%%%%%%%%%%%%%%%%%%%%
\begin{flushright}
\tt USTC-ICTS/PCFT-22-03
\end{flushright}
\vspace{0.7cm}
\begin{center}
\Large{\textbf{Geometric Algebra and Algebraic Geometry\\ of Loop and Potts Models}}

\vspace{1.3cm}

\normalsize{\textrm{Janko B\"ohm$^1$, Jesper Lykke
Jacobsen$^{2,3,4}$, Yunfeng Jiang$^{{5,6}}\footnote{Corresponding author}$, Yang Zhang$^{7,8}$}}
\\ \vspace{1cm}
\footnotesize{\textit{
$^{1}$Department of Mathematics, Technische Universit\"at Kaiserslautern, 67663 Kaiserslautern, Germany
\\
$^{2}$Laboratoire de Physique de l'\'Ecole Normale Sup\'erieure, ENS, Universit\'e PSL, \\
CNRS, Sorbonne Universit\'e, Universit\'e de Paris, F-75005 Paris, France \\
$^{3}$Sorbonne Universit\'e, \'Ecole Normale Sup\'erieure, CNRS, \\
Laboratoire de Physique (LPENS), F-75005 Paris, France\\
$^{4}$Institut de Physique Th\'eorique, Paris Saclay, CEA, CNRS, 91191 Gif-sur-Yvette, France
\\
$^{5}$School of Physics, Southeast University, Nanjing 211189, China\\
$^{6}$Shing-Tung Yau Center, Southeast University, Nanjing 210096, China
\\
$^{7}$Interdisciplinary Center for Theoretical Study, University of Science and Technology of China,
Hefei, Anhui 230026, China\\
$^{8}$Peng Huanwu Center for Fundamental Theory, Hefei, Anhui 230026, China
}
\vspace{1cm}
}

\par\vspace{1.0cm}

\textbf{Abstract}\vspace{2mm}
\end{center}
\noindent
We uncover a connection between two seemingly separate subjects in integrable models: the representation theory of the affine Temperley-Lieb algebra, and the algebraic structure of solutions to the Bethe equations of the XXZ spin chain. We study the solution of Bethe equations analytically by computational algebraic geometry, and find that the solution space encodes rich information about the representation theory of Temperley-Lieb algebra. Using these connections, we compute the partition function of the completely-packed loop model and of the closely related random-cluster Potts model, on medium-size lattices with toroidal boundary conditions, by two quite different methods. We consider the partial thermodynamic limit of infinitely long tori and analyze the corresponding condensation curves of the zeros of the partition functions. Two components of these curves are obtained analytically in the full thermodynamic limit.
%We study the exact partition function of the Temperley-Lieb algebra loop model and Potts model by two very different approaches. The first method is \emph{algebraic} in nature which combines representation theory of Temperley-Lieb algebra and explicit transfer matrix computation; The second method is more \emph{geometric} which exploits Bethe ansatz and computational algebraic geometry.
\setcounter{page}{1}
\renewcommand{\thefootnote}{\arabic{footnote}}
\setcounter{footnote}{0}
\setcounter{tocdepth}{2}
\newpage
\tableofcontents

%%%%%%%%%%%%%%%%%%%%%%%%%%%%%%%%%%%%%%%%%%%%%%%%%%%%%%%%%%%%%%
\section{Introduction}
\label{sec:intro}
%%%%%%%%%%%%%%%%%%%%%%%%%%%%%%%%%%%%%%%%%%%%%%%%%%%%%%%%%%%%%%
Integrable lattice model and computational algebraic geometry seem to be two distant subjects \emph{a priori}. However, a bit of thought shows that this should not be the case. Computational algebraic geometry stems from studying solutions of algebraic equations. At the same time, such algebraic equations are ubiquitous in integrable lattice models, where they arise as the famous Bethe ansatz equations. Therefore, it is natural to expect that computational algebraic geometry should play a useful role in the study of integrable models.

Surprisingly, this possibility has not been pursued until very recently \cite{Jiang:2017phk}.%
\footnote{See \cite{langlands1995algebro,langlands1997aspects} for the application of algebraic geometry to the inhomogeneous XXZ spin chain. Here we are making a connection to \emph{computational} algebraic geometry, whose focus and the techniques involved (such as Gr\"obner basis) are different from \cite{langlands1995algebro,langlands1997aspects}.}
It has been shown, as expected, that standard methods in computational
algebraic geometry, such as Gr\"obner bases and companion matrices,
provide powerful tools to study lattice integrable models. Important
applications involve testing the completeness of the Bethe ansatz, and
computing exact partition functions of the six-vertex
model at its isotropic point, both for torus \cite{Jacobsen:2018pjt}
and cylinder \cite{Bajnok:2020xoz} geometries. In this paper, we emphasize the relevance of these exact results for intermediate sizes of the lattice, which are not small enough to be tackled by hand or by brute force computational approaches, and yet not large enough to be approximated by the thermodynamic limit. More recently, these techniques have also been applied to a much wider class of observables, such as quantities arising in quench dynamics \cite{Jiang:2021krx}.

So far the development of the algebro-geometric method has focussed on the Bethe equations of XXX type, which can be written as polynomial equations of the Bethe roots. The current work is devoted to the generalization to Bethe equations of XXZ type. Written in terms of Bethe roots, the Bethe equations involve hyperbolic functions. By a change of variable, we can work with the exponential of Bethe roots, in terms of which the Bethe equations still take a polynomial form.

The XXZ-type Bethe equations show up in many integrable models. In this work, we are interested in two such models
that have connections with the affine Temperley-Lieb (aTL) algebra \cite{GrahamLehrer,MartinSaleur,MartinSaleur1},
an extension of the well-known Temperley-Lieb algebra \cite{TL71} to periodic boundary conditions. These models are the completely-packed O($\mathfrak{m}$) loop model \cite{BKW76} and the $Q$-state Potts model in the random-cluster formulation due to Fortuin and Kasteley (FK) \cite{FK72}, henceforth referred to simply as the {\em loop model} and the {\em Potts model}. Both of these models are fundamental models of statistical mechanics, and many specific values of $\mathfrak{m}$ and $Q$ harbor cases of particular physical interest. We shall consider both models on a square lattice with toroidal boundary conditions. A configuration in the loop model is a set of loops living on the lattice edges, subject to the constraints that each edge is covered by exactly one loop, and that loops do not cross at the vertices. The loop model partition function is then defined by attributing a weight $\mathfrak{m}$ to each loop and summing over the configurations. The configurations in the Potts model are exactly the same, but the weights are slightly different. Color each face of the graph defined by the loops in black or white, subject to the constraints that a fixed reference point (the origin) lives on a black face, and that adjacent faces have different colors. In this formulation, each black face coincides with an FK cluster. The Potts model partition function is then defined by attributing a weight $Q$ to each black face and summing over the configurations. Setting $Q = \mathfrak{m}^2$, the two models would be strictly equivalent if they were
defined on a planar lattice (use the Euler identity), but on the torus there are subtle differences (see below).

The loop model and the Potts model can be studied by means of representation theory of the affine Temperley-Lieb algebra \cite{GrahamLehrer,MartinSaleur,MartinSaleur1}, and more precisely a quotient thereof, known as the Jones-Temperley-Lieb algebra \cite{Jones,ReadSaleur07-1}, that we shall define precisely below.
Our goal is to study the torus partition function for these models. It is computed by taking the Markov traces \cite{Jones-index} of the transfer matrix in the so-called standard modules of the aTL algebra, denoted $\mathcal{W}_{j,\rho^2}$, which can be constructed in the basis of link patterns. The parameters $j$ and $\rho^2$ are defined by choosing a quantization scheme, in which one principal direction of the torus is taken as `space' and the other as (imaginary) `time'. Then $2j$ is the number of non-contractible loop strands, henceforth called {\em through-lines}, running along the time direction, while $\rho^2$ is the pseudo-momentum corresponding to the movement of through-lines across the periodic space direction.
The transfer matrices in the aTL standard modules are related to the transfer matrices of the XXZ spin chain with a diagonal twist, which can be solved by Bethe ansatz. The correct twist is related to $\rho^2$, so this parameter enters the Bethe equations.
In the framework of the Bethe ansatz, the eigenvalues of the transfer matrix are functions of Bethe roots which are physical solutions of the Bethe equations. Therefore one can expect physical solutions of the Bethe equations to contain information about the representation theory of the aTL algebra.

Confirming such a connection and offering a method to see it explicitly are part of the main results of the current work. To this end, we shall rely on analytical tools from computational algebraic geometry (CAG).
It is customary to parametrize $\mathfrak{m} = -q-q^{-1}$, with $q=-e^{i \gamma}$. %\textcolor{red}{[JJ: Make sure that the $-q$ convention is used throughout]}
We focus in this paper on the so-called {\em generic} situation where $q$ is not a root of unity (\emph{i.e.}, $\gamma \in \mathbb{R} \setminus \pi \mathbb{Q}$); the root of unity case will be discussed elsewhere.
In this context we have identified the following links between aTL algebra, AG and the Bethe ansatz:
\begin{enumerate}
\item The dimension of a standard module equals the number of physical solutions of the corresponding XXZ Bethe equations.
\item When the twist variable $\rho^2$ takes specific values that
  satisfy a certain resonance condition
  \cite{MartinSaleur1,GrahamLehrer}, the standard module
  $\mathcal{W}_{j,\rho^2}$ becomes reducible. This resonance condition
  also shows up in the Bethe equation in an interesting way. When
  $\rho^2$ satisfies the resonance condition, the corresponding
  \emph{Gr\"obner basis} of the Bethe equations has vanishing leading coefficients.
\item The transfer matrices constructed in the standard module can be identified with the \emph{companion matrix} of the transfer matrix of the twisted XXZ spin chain. In general they take different forms, but have the same eigenvalues.
\end{enumerate}
Based on these identifications, the torus partition function, or more precisely the transfer matrices that generate it, can be computed in two rather different ways. One way is based on the link-pattern basis construction of the standard module. Since it gives nice diagrammatical representations of the aTL algebra, we call this approach the \emph{geometric algebra} approach; the other approach is by a combination of Bethe ansatz and computational algebraic geometry, which we call the \emph{algebraic geometry} approach. We consider the torus partition function for both models on an $L\times N$ lattice with periodic boundary conditions in both directions (space and time). We give closed-form analytical results for $L=4,6$ and generic $N$. For $L=8,10$ we compute the partition function analytically for large $N\sim 1000$.\par

The partition functions that we obtain are polynomials with integer
coefficients. All information about such polynomials are contained in
their zeros over $\mathbb C$. We therefore investigate the distribution of the zeros of the partition functions. We find that for fixed $L$ and in the limit $N \to \infty$, the zeros condense on certain curves. We analyze these condensation curves for $L=6,8,10,12,14$ and compare them with the finite-$N$ results for the zeros for $L=6,8,10$.

The rest of the paper is structured as follows. We introduce the aTL algebra in section~\ref{sec:aTLalgebra}. Its connection with the algebraic Bethe Ansatz for the twisted XXZ chain is discussed in section~\ref{sec:XXZ}. The loop model and the Potts model will be introduced and discussed in sections~\ref{sec:Loopmodel} and \ref{sec:Pottsmodel}, respectively. We provide their connection with the XXZ chain, and also show how to produce their torus partition functions from aTL representation theory. To study the {\em physical} solutions of the Bethe equations, by which we mean solutions that also solve the original eigenvalue problem of the transfer matrix, it is more convenient to apply the rational $Q$-system approach. The generalization of this approach to the twisted case is a byproduct of our analysis and is given in section~\ref{sec:Qsystem}. We then apply algebraic geometry to analyze the rational $Q$-system and give the resonance condition from a different perspective in section~\ref{sec:resonance}. The closed form results for $L=4,6$ and generic $N$ are presented in section~\ref{sec:Closedform}. The distribution of zeros and the partition functions for $L=6,8,10$ are studied in section~\ref{sec:zeros}. For large $N$ they are in accord with the condensation curves, which we provide as well for the larger sizes $L=12,14$.
Furthermore we obtain two components of these curves analytically in the full thermodynamic limit.
We conclude and discuss future directions in section~\ref{sec:conclude}. The appendices contain more background and technical details for the main text.

%More precisely, we focus on two important models, which are the Temperley-Lieb loop model [...] and Potts model [...]. These models can be solved by Bethe ansatz whose corresponding BAE are of XXZ type. We shall compute the torus partition function of such models for intermediate size lattices. We exploit Bethe ansatz solution of the model, together with Gr\"obner basis and companion matrix approach from computational algebraic geometry.\par

%Finite size torus partition functions for these two models can also be computed by a quite different method without invoking Bethe ansatz. This method exploits the underlying algebraic structure of the model. The transfer matrices are constructed via representation theory of the affine Temperley-Lieb algebra. Remarkably, the same transfer matrices can be constructed by combining Bethe ansatz and algebraic geometry. As we will see, the physical solutions of BAE knows a lot about representation theory of the affine TL-algebra. This new connection is quite fascinating and will shed lights on both.

%%%%%%%%%%%%%%%%%%%%%%%%%%%%%%%%%%%%%%%%%%%%%%%%%%%%%%%%%%%%%%
\section{The affine Temperley-Lieb algebra}
\label{sec:aTLalgebra}
%%%%%%%%%%%%%%%%%%%%%%%%%%%%%%%%%%%%%%%%%%%%%%%%%%%%%%%%%%%%%%
In this section, we introduce the affine Temperley-Lieb algebra. We first give the definition and then study its finite-dimensional representations called standard modules.
\paragraph{Definitions.} We first define the usual Temperley-Lieb (TL) algebra, which is generated by $L-1$ generators $e_j$ $(j=1,\ldots,L-1)$, together with the identity $\mathbf{1}$, subject to the following relations \cite{TL71}
\begin{eqnarray}
\label{TL}
e_j^2&=&\mm\, e_j \,, \\\nonumber
e_je_{j\pm 1}e_j&=&e_j \,, \\\nonumber
e_je_k&=&e_ke_j\qquad(\mbox{for } j\neq k,~k\pm 1) \,.
\end{eqnarray}
%\end{subequations}
%
We shall here consider the case of periodic boundary conditions, for which
the indices $j$ are interpreted modulo $L$ (so in particular we also have an element $e_0$), and moreover
there exist additional elements $U$ and $U^{-1}$ which generate cyclic translations
by one site to the right and to the left, respectively. They satisfy the following relations
\begin{eqnarray}
\label{TL-u}
Ue_jU^{-1}&=&e_{j+1} \,, \\\nonumber
U^2e_{L-1}&=&e_1 \cdots e_{L-1} \,.
\end{eqnarray}
It is clear that $U^{\pm L}$ is a central element. The algebra generated by $e_1$ and $U^{\pm1}$ together with the relations \eqref{TL} and \eqref{TL-u} is called the \textit{affine} Temperley--Lieb algebra \cite{GrahamLehrer,MartinSaleur,MartinSaleur1} and will be denoted by $\ATL{L}$.
In the usual quantization scheme of two-dimensional Euclidean models, the site index $j$ is a discretization of the space direction, while the algebra action is a discretization of the imaginary time evolution.

\medskip

The remainder of this section reviews some well-known facts about the aTL algebra, along the lines of
\cite{looppaper}. We focus on the material that is essential for our purposes,
in order to keep the presentation self-contained.

\subsection{Diagrammatic representation}
The aTL algebra can be represented by particular diagrams
on an annulus. This is called the \emph{loop representation}. In the geometry where time propagates upwards (and which is natural from the point of view of the algebra)
the identity operator $\mathbf{1}$ and the TL generator $e_j$ are represented as
% (\textcolor{red}{YF: I'm a bit confused by this graph representation and how to get the graphs in Fig.2.1.})
%
\begin{equation} \label{splittings0}
 \mathbf{1} =
\raisebox{-5mm}{\begin{tikzpicture}[scale=0.6]
 \draw[thick] (0,-1)--(0,1);
 \draw[thick] (1,-1)--(1,1);
\end{tikzpicture}} \,,
\qquad\qquad
 e_j =
\raisebox{-5mm}{\begin{tikzpicture}[scale=0.6]
 \draw[thick] (0,-1)--(0,-0.5) arc(180:0:5mm and 4mm)--(1,-1);
 \draw[thick] (0,1)--(0,0.5) arc(180:360:5 mm and 4mm)--(1,1);
\end{tikzpicture}} \,,
\end{equation}
where we have only depicted the sites of index $j$ and $j+1$. The remaining sites evolve trivially as in $\mathbf{1}$.
One of the advantages of the loop representation is that we can attribute a non-local weight
\begin{equation}
 \mm = 2 \cos \gamma
\end{equation}
to each loop in the diagrammatic formulation. This is very useful for describing geometrical problems, such as percolation hulls or dense polymers. It is also closely related to the cluster representation of the $Q$-state Potts model with $Q=\mm^2$ \cite{JacobsenSaleur}, as we will discuss below.\par

\begin{figure}
\begin{center}
 \begin{tikzpicture}
 	\draw[thick, dotted] (-0.05,0.5) arc (0:10:0 and -7.5);
 	\draw[thick, dotted] (-0.05,0.55) -- (2.65,0.55);
 	\draw[thick, dotted] (2.65,0.5) arc (0:10:0 and -7.5);
	\draw[thick, dotted] (-0.05,-0.85) -- (2.65,-0.85);
	\draw[thick] (0,0) arc (-90:0:0.5 and 0.5);
	\draw[thick] (0.9,0.5) arc (0:10:0 and -7.6);
	\draw[thick] (1.65,0.5) arc (0:10:0 and -7.6);
	\draw[thick] (2.6,0) arc (-90:0:-0.5 and 0.5);
	\draw[thick] (0.5,-0.8) arc (0:90:0.5 and 0.5);
	\draw[thick] (2.1,-0.8) arc (0:90:-0.5 and 0.5);
	\end{tikzpicture}\;\;,
	\qquad\qquad
 \begin{tikzpicture}
 	\draw[thick, dotted] (-0.05,0.5) arc (0:10:0 and -7.5);
 	\draw[thick, dotted] (-0.05,0.55) -- (2.65,0.55);
 	\draw[thick, dotted] (2.65,0.5) arc (0:10:0 and -7.5);
	\draw[thick, dotted] (-0.05,-0.85) -- (2.65,-0.85);
	\draw[thick] (0,0) arc (-90:0:0.5 and 0.5);
	\draw[thick] (0.8,0.5) arc (-180:0:0.5 and 0.5);
	\draw[thick] (2.6,0) arc (-90:0:-0.5 and 0.5);
	\draw[thick] (0.5,-0.8) arc (0:90:0.5 and 0.5);
	\draw[thick] (1.8,-0.8) arc (0:180:0.5 and 0.5);
	\draw[thick] (2.1,-0.8) arc (0:90:-0.5 and 0.5);
	\end{tikzpicture}\;\;,
	\qquad\qquad
 \begin{tikzpicture}
 	\draw[thick, dotted] (-0.05,0.5) arc (0:10:0 and -7.5);
 	\draw[thick, dotted] (-0.05,0.55) -- (2.65,0.55);
 	\draw[thick, dotted] (2.65,0.5) arc (0:10:0 and -7.5);
	\draw[thick, dotted] (-0.05,-0.85) -- (2.65,-0.85);
	\draw[thick] (0,0.1) arc (-90:0:0.5 and 0.4);
	\draw[thick] (0,-0.1) arc (-90:0:0.9 and 0.6);
	\draw[thick] (2.6,-0.1) arc (-90:0:-0.9 and 0.6);
	\draw[thick] (2.6,0.1) arc (-90:0:-0.5 and 0.4);
	\draw[thick] (0.9,-0.8) arc (0:180:0.3 and 0.3);
	\draw[thick] (2.2,-0.8) arc (0:180:0.3 and 0.3);
	\end{tikzpicture}
\end{center}
\caption{Examples of affine diagrams for $L=4$, with the left and right sides of the framing rectangle identified, so as to form an annulus. The first diagram represents the generator $e_4$, the second is $e_2 e_4$, and the third is $e_4 e_3 e_1$.
\label{fig:aff-diag}}
\end{figure}
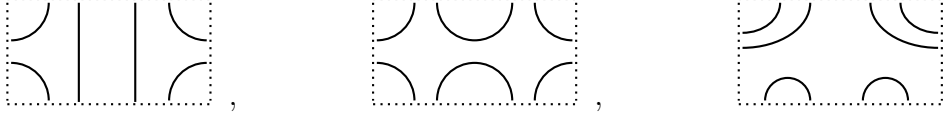

A general basis element in the algebra of diagrams corresponds to a diagram of $L$ sites on the inner and $L$ on
the outer boundary of the annulus. We will always restrict to even $L$ for simplicity. The sites are connected in pairs,
and only configurations that can be represented using simple curves inside the annulus that do not cross are allowed. Such
diagrams  are called \textit{affine}.  Examples of affine diagrams are shown in Fig.~\ref{fig:aff-diag}, where we draw them in a slightly different geometry: we cut the annulus and transform it to a rectangle, which we call \textit{framing}, with the sites labeled from left to right.\par

An important parameter is the number of {\em through-lines}, which we denote by $2j$; each through-line is a simple curve connecting
a site on the inner and a site on the outer boundary of the
annulus; the $2j$ sites on the inner boundary attached to a through-line we call {\em free} or
{\em non-contractible}. The inner (resp.\ outer) boundary of the annulus corresponds to the bottom (resp.\ top) side of the framing rectangle.

Multiplication of two affine diagrams, $a$ and $b$, is defined in a natural
way, by joining the inner boundary of the annulus containing $a$ to the outer boundary of the annulus containing $b$, and
removing the interior sites. Accordingly, $ab$ is obtained by joining the bottom side of $a$'s framing rectangle to the top side of $b$'s framing
rectangle, and removing the corresponding joined sites. Whenever a closed contractible loop is
produced in the process of multiplying diagrams, this loop must be
replaced by a numerical factor~$\mm$.

\subsection{Standard modules}
\label{sec:stdmod}

With the defining relations \eqref{TL}--\eqref{TL-u}, the algebra $\ATL{L}$ is infinite-dimensional. There are two mechanisms at play \cite{Jones,ReadSaleur07-1} that prevent the algebra from being finite-dimensional:
\begin{enumerate}
 \item any loop which is non-homotopic to a point (with respect to the periodic space direction) cannot be reduced to a number (such as $\mathfrak{m}$) by the mere application of the relations (\ref{TL})--(\ref{TL-u}) and such loops may hence accumulate in arbitrary numbers, and
 \item through-lines may wind around the periodic space direction, in one direction or the other, an arbitrary number of times.
\end{enumerate}

To describe lattice models with a finite number of degrees of freedom per site we shall need some finite-dimensional {\em representations} of $\ATL{L}$. These are the so-called {\em standard modules} $\AStTL{j}{e^{i\phi}}$, which depend on two parameters.  In terms of diagrams, the first parameter defines the number of
through-lines $2j$, with $j=0,1,\ldots, \frac{L}{2}$. We also stipulate that if an action contracts two or more free sites, a situation which reduces the number of through-lines, the result is zero. This implies that the action of $\ATL{L}$ does not connect standard modules with different $j$. Notice that for a given
non-zero value of $j$, it is possible, using the  action of the algebra,  to cyclically
permute the free sites. This motivates the introduction of a
{\em pseudomomentum}, which we parametrize by the second parameter $\phi$:
whenever a through-line winds around the annulus once, we attribute to it a phase $e^{\pm i\phi/2}$, where the sign plus (resp.\ minus) is for a clockwise (resp.\ counterclockwise) winding.
%By definition,
%whenever $2j$ through-lines wind counterclockwise around
%the annulus $l$ times, we can unwind them at the price of a factor
%$e^{ijl\phi}$; similarly, for clockwise winding, the phase is $e^{-i jl\phi}$ \cite{MartinSaleur,MartinSaleur1}. Stated more simply, there is a phase $e^{\pm i\phi/2}$ per winding through-line.

A more convenient formulation of $\AStTL{j}{e^{i\phi}}$ can be obtained as follows. As the free sites are not allowed to be contracted, the pairwise connections among non-free sites on the inner boundary cannot be changed by the action of the algebra. This part of the diagrammatic information is thus inessential and can be omitted. Even so, the representation remains infinite-dimensional, in particular because the $2j$ through-lines connecting the inner and outer boundaries may still spiral around the periodic boundary condition. As a first step to obtain the desired finite-dimensional representation we therefore decree that a full turn of the inner boundary with respect to the outer one is irrelevant. Such a turn is equivalent to a full turn of all the through-lines, so by definition of the pseudomomentum we should set
\begin{equation}
 (e^{\pm i\phi/2})^{2j} = 1 \,.
\end{equation}
Having made this choice, it is now enough to concentrate on the upper halves of the affine diagrams, obtained by cutting the affine diagrams across its $2j$ through-lines. Each upper half is then called a {\em link state}, and for simplicity the ``half'' through-lines attached to the free sites on the outer boundary of the annulus (or top side of the framing rectangle) are still called through-lines. The phase
\begin{equation} \label{def_rho}
 \rho \equiv e^{i\phi/2}
\end{equation}
(resp.\ $\rho^{-1} = e^{-i\phi/2}$) is now attributed each time one of these through-lines moves through the periodic boundary condition of the framing rectangle in the rightward (resp.\ leftward) direction. The quantisation condition on the pseudomomentum then reads
\begin{equation} \label{twist-quant}
 (\rho^2)^j = 1 \,.
\end{equation}
With these conventions, it is readily seen that the Temperley-Lieb algebra action obtained by stacking the affine diagrams on top of the link states gives rise to exactly the same representations $\AStTL{j}{\rho^2}$ as defined above.

The representations $\AStTL{j}{e^{i\phi}}$ over $\ATL{L}(\mm)$  having through-lines (\emph{i.e.}, with $j>0$) are now finite-dimensional. Their dimensions are easily found by counting the link states, giving
 \begin{equation}\label{eq:dj}
 \hat{d}_{j}=
 \binom{L}{\frac{L}{2}+j} \,,\qquad \mbox{for } j>0 \,.
 \end{equation}
Note that these dimensions do not depend on $\phi$ (but representations with
different $e^{i\phi}$ are not isomorphic).
These standard modules $\AStTL{j}{e^{i\phi}}$ are also called
{\em cell modules} in the seminal work~\cite{GrahamLehrer}.

The situation without through-lines (\emph{i.e.}, when $j=0$) is a bit different.
There is no pseudomomentum, but representations are still characterized by a parameter other than $j$, which now specifies the weight given to loops which are non-contractible, \emph{i.e.}, which wrap the periodic space direction of the lattice.
Notice that since the curves cannot cross, non-contractible loops cannot coexist with through-lines, and hence are not possible for $j > 0$. Parametrizing the weight of a non-contractible loop as $\mathfrak{n} = z+z^{-1}$, the corresponding standard module of  $\ATL{N}(\mm)$ is denoted  $\AStTL{0}{z^2}$. This module is isomorphic to $\AStTL{0}{z^{-2}}$.

It is natural to require that $\mathfrak{n}=\m$, \emph{i.e.}, to set $z = q$, so that contractible and non-contractible loops get the same weight. Imposing this is however not without consequences for the nature of the corresponding aTL representation, as is most easily seen by focussing on the dimension.
If we do not distinguish between contractible and non-contractible loops, the corresponding link state needs not distinguish whether two sites are connected strictly inside the framing rectangle, or by going through the periodic boundary condition.%
\footnote{To be precise, the distinction is whether the link connecting the two sites crosses the periodic boundary condition an even or an odd number of times. By deforming the curves, these crossing numbers can always be reduced to 0 or 1.}
Taking therefore all connections strictly inside the framing rectangle, we arrive at the same dimension as in the usual (non-periodic) TL algebra, which is
\begin{equation}
 \bar{d}_0=\binom{L}{\frac{L}{2}}-\binom{L}{\frac{L}{2}+1} \,,\qquad \mbox{for } j=0 \,.
 \label{dim-bard0}
\end{equation}
We denote the corresponding representation by $\bAStTL{\! 0}{q^2}$, and below we shall make clear how it is related to $\AStTL{0}{z^2}$ from an algebraic point of view.

To avoid too much drawing, it is convenient to adopt a no-frill notation for specifying the link states. We denote the (half) through-lines by a bar and an arc
connecting two sites by a pair of matching parentheses. An arc represented by a matching $)($ pair then straddles the periodic boundary condition, whereas a $()$ pair does not.
For instance, the link states describing the upper halves of the diagrams in Figure~\ref{fig:aff-diag} are
written as $)||($, $)()($ and $))(($ respectively.

\subsection{Resonances and quotient modules}
\label{sec:resonance}
We now parametrize % \textcolor{red}{[JJ: Check carefully that the $-q$ convention is used systematically.]}
\begin{equation}
\label{m_param}
 \mm= -q-q^{-1} = 2 \cos \gamma \,.
\end{equation}
The standard modules $\AStTL{j}{e^{i\phi}}$ are irreducible for generic values of $q$ and $\phi$.
However, degeneracies appear whenever the following {\em resonance criterion} is satisfied \cite{MartinSaleur1,GrahamLehrer}:%
\footnote{In \cite{GrahamLehrer} this criterion appears with some extra liberty in the form of certain $\pm$ signs, but we shall not need these signs here.}
 \begin{eqnarray}\label{deg-st-mod}
 \rho^2 \equiv e^{i\phi}&=&q^{2j+2k},\qquad
 \mbox{for } k\in\mathbb{Z}_+ \,.
 \end{eqnarray}
 The representation $\AStTL{j}{q^{2j+2k}}$ then becomes reducible, and contains a submodule isomorphic to
 $\AStTL{j+k}{q^{2j}}$. The quotient $\AStTL{j}{q^{2j+2k}} / \AStTL{j+k}{q^{2j}}$ is generically irreducible, with
 dimension
 \begin{equation} \label{bard}
  \bar{d}_j := \hat{d}_j-\hat{d}_{j+k} \,, \qquad \mbox{for } j > 0 \,.
 \end{equation}
When $q$ is a root of unity, there are infinitely many solutions to \eqref{deg-st-mod}, leading to a complex pattern of degeneracies.

We now return to the case $j=0$ without through-lines and with non-contractible loops having the weight $\mathfrak{n} = z + z^{-1}$. If we make the identification
\begin{equation}
\label{def_z}
 z = e^{i\phi/2}
\end{equation}
the resonance criterion \eqref{deg-st-mod} still applies. Imposing this therefore
leads to the module $\AStTL{0}{q^2}$, which is reducible
even for generic $q$. Indeed, \eqref{deg-st-mod} is satisfied with $j=0$, $k=1$, and hence $\AStTL{0}{q^2}$ contains a submodule isomorphic to
$\AStTL{1}{1}$ (and more precisely $\rho = -1$, so that $e^{i \phi} = \rho^2 = 1$). Taking the quotient $\AStTL{0}{q^2}/\AStTL{1}{1}$ leads to a simple module  for generic $q$ which we denote by
\begin{equation} \label{quotientmodule}
 \bAStTL{\! 0}{q^2} = \AStTL{0}{q^2}/\AStTL{1}{1} \,.
\end{equation}
This module is isomorphic to $\bAStTL{\! 0}{q^{-2}}$. As already anticipated in (\ref{dim-bard0}), it has the dimension
\begin{equation}
 \bar{d}_0=\binom{L}{\frac{L}{2}}-\binom{L}{\frac{L}{2}+1} \,,
\end{equation}
in agreement with the general formula \eqref{bard} for $k=1$.

We now discuss in some more detail the geometrical meaning of the difference between $\AStTL{0}{q^2}$ and $\bAStTL{0}{q^2}$, which was already sketched above. In the latter case, one only keeps track of which sites are connected to which in the diagrams, while in the former, one also keeps information of how the connectivities wind around the periodic direction of the annulus (the ambiguity does not arise when there are through-lines propagating). Formally, this corresponds to the existence of a surjection $\psi$ between different quotients of the $\ATL{L}$ algebra:
\begin{equation}\label{psi-ex}
  \xymatrix@C=8pt@R=1pt@M=-5pt@W=-2pt{
  &&	\mbox{}\quad\xrightarrow{{\mbox{}\quad\psi\quad} }\quad &\\
  & {
 \begin{tikzpicture}
  %%%%%%%%% Frame %%%%%%%%
 	\draw[thick, dotted] (-0.05,0.5) arc (0:10:0 and -7.5);
 	\draw[thick, dotted] (-0.05,0.55) -- (2.65,0.55);
 	\draw[thick, dotted] (2.65,0.5) arc (0:10:0 and -7.5);
	\draw[thick, dotted] (-0.05,-0.85) -- (2.65,-0.85);
%%%%%%%%%%%%	
	\draw[thick] (0,0.1) arc (-90:0:0.5 and 0.4);
	\draw[thick] (0,-0.1) arc (-90:0:0.9 and 0.6);
	\draw[thick] (2.6,-0.1) arc (-90:0:-0.9 and 0.6);
	\draw[thick] (2.6,0.1) arc (-90:0:-0.5 and 0.4);
%%%%%%%%%%%%%	
	\draw[thick] (0.5,-0.8) arc (0:90:0.5 and 0.5);
	\draw[thick] (1.8,-0.8) arc (0:180:0.5 and 0.5);
	\draw[thick] (2.1,-0.8) arc (0:90:-0.5 and 0.5);
	\end{tikzpicture}
\quad}&
	%\ar[r]
%	\xrightarrow{{\mbox{}\quad\psi\quad} }
%	\ar@{->}[r]^(0.45){\mbox{}\qquad\psi\qquad}
	& {	\quad
  \begin{tikzpicture}
 %%%%%%%%% Frame %%%%%%%%
 	\draw[thick, dotted] (-0.05,0.5) arc (0:10:0 and -7.5);
 	\draw[thick, dotted] (-0.05,0.55) -- (2.65,0.55);
 	\draw[thick, dotted] (2.65,0.5) arc (0:10:0 and -7.5);
	\draw[thick, dotted] (-0.05,-0.85) -- (2.65,-0.85);
%%%%%%%%%%%%	
	\draw[thick] (0.5,0.5) arc (-180:0:0.8 and 0.56);
%	\draw[thick] (1.0,0.5) arc (0:10:-44.5 and -7.6);
%	\draw[thick] (1.65,0.5) arc (0:10:38.5 and -7.6);
	\draw[thick] (0.8,0.5) arc (-180:0:0.5 and 0.4);
%	\draw[thick] (2.6,0.2) arc (-90:0:-0.5 and 0.3);
%%%%%%%%
	\draw[thick] (2.1,-0.8) arc (0:180:0.8 and 0.56);
	\draw[thick] (1.8,-0.8) arc (0:180:0.5 and 0.4);
	\end{tikzpicture}}
%  &&&
  }
\end{equation}
The definition of link states as the upper halves of the affine diagrams also makes sense for $j=0$. The representation $\AStTL{0}{q^2}$ requires keeping
track of whether each pairwise connection between the sites on the outer boundary goes through the periodic boundary condition, whereas in the quotient module $\bAStTL{0}{q^2}$ this information is omitted. In both cases, it is easy to see that the number of link states coincides with the dimension $d_0$ or $\bar{d}_0$, respectively.

\medskip

The quotient $\bAStTL{0}{q^2}$ is just one example of representations that appear more generally in $\ATL{L}$ when $q$
is still generic, but $z$ takes particular values \cite{GrahamLehrer,PerioFusion}.
Indeed, the standard module $\AStTL{j'}{z'}$ has a non-zero homomorphism to $\AStTL{j}{z}$,
\begin{equation}\label{cell-emb}
\StJTL{j'}{z'}\hookrightarrow\StJTL{j}{z} \,,
\end{equation}
if and only if  $j'-j \in \mathbb{N}_0$ and the pairs $(j',z')$ and $(j,z)$ satisfy
%\textcolor{red}{[JJ: Do we finally ever use the signs? There was a footnote above saying that no, and the above formula omitted them squarely.]}
\begin{equation}\label{eq:emb-cond}
(z')^2 = (-q)^{2\epsilon j} \qquad \text{and}\qquad z^2 = (-q)^{2\epsilon j'},\qquad \text{for} \quad  \epsilon=\pm1.
\end{equation}
When $q$ is not a root of unity, there is at most one solution to (\ref{cell-emb}). When there is one, the module $\StJTL{j}{z}$ is not
irreducible, but has a unique proper irreducible submodule isomorphic to $\StJTL{j'}{z'}$. One can then obtain a simple module by taking the quotient
\begin{equation}
\bAStTL{\! j}{z}\equiv\StJTL{j}{z}/\StJTL{j'}{z'}\label{Wbar-def}
\end{equation}
of dimension
\begin{equation}
\label{eq:dbarj}
\overline{d}_j = \dim \bAStTL{\! j}{z} = \binom{L}{\frac{L}{2}+j} - \binom{L}{\frac{L}{2}+j'} \,.
\end{equation}

The quotient $\bAStTL{\! 0}{q^2}$ appearing above is the simplest example (with $j=0$, $z=q^2$ and $j'=1$, $z'=1$) of this situation, and it is the {\em only}
such quotient that will turn out to be relevant for the Potts model at generic $q$.

%%%%%%%%%%%%%%%%%%%%%%%%%%%%%%%%%%%%%%%%%%%%%%%%%%%%%%%%%%%%%%
\section{The XXZ spin chain}
\label{sec:XXZ}
%%%%%%%%%%%%%%%%%%%%%%%%%%%%%%%%%%%%%%%%%%%%%%%%%%%%%%%%%%%%%%
There exists another representation of the aTL algebra which is called the {\em XXZ representation}. It makes manifest a connection to an integrable model,
the twisted XXZ spin chain. This connection is of paramount importance to us, since the application of the AG approach goes via the Bethe equations
of an integrable model.

To make the link, we first recall the definition of the XXZ spin chain in the framework of algebraic Bethe ansatz. The transfer matrix of the XXZ spin chain can be defined by
\begin{equation} \label{transfer}
 t(u) = {\rm tr}_{a} \, R_{{a},1}(u) R_{{a},2}(u) \cdots R_{{a},L}(u) \,.
\end{equation}
where $R_{a,n}(u)$ is the $R$-matrix acting between the auxiliary space and the $n$-th site of the chain. The corresponding $\check{R}$-matrix, defined by $\check{R}_{a,n}(u)=P_{a,n}R_{a,n}(u)$
where $P$ denotes the permutation operator,
can be written in terms of the Temperley-Lieb generators as
\begin{equation}
 \label{checkR-aTL}
 \check{R}_{{a},j} (u) = \sin(\gamma-u) \mathbf{1} + \sin(u) e_j \,,
\end{equation}
where $u$ denotes the spectral parameter, and the crossing parameter is $\gamma\in [0,\pi]$.
The auxiliary space represents a horizontal row of lattice edges, and $t(u)$ is the transfer matrix going from one row of vertical edges to the next.

This presence of the auxiliary space implies a change of geometry with respect to the algebraic consideration made in section~\ref{sec:aTLalgebra},
namely that imaginary time does no longer flow upwards but rather towards the North-East. For instance, in the present setup the diagrammatic
rendering of the aTL generators, formerly drawn as (\ref{splittings0}), now looks like
\begin{equation} \label{splittings}
 \mathbf{1} =
\raisebox{-5mm}{\begin{tikzpicture}[scale=0.6]
 \draw[thick] (0,0)--(0.8,0) arc(270:360:2mm)--(1,1);
 \draw[thick] (1,-1)--(1,-0.2) arc(180:90:2mm)--(2,0);
\end{tikzpicture}} \,,
\qquad\qquad
 e_j =
\raisebox{-5mm}{\begin{tikzpicture}[scale=0.6]
 \draw[thick] (0,0)--(0.8,0) arc(90:0:2mm)--(1,-1);
 \draw[thick] (1,1)--(1,0.2) arc(180:270:2mm)--(2,0);
\end{tikzpicture}} \,.
\end{equation}

If we take the {\em isotropic} value $u = \gamma/2$ of the spectral parameter in (\ref{checkR-aTL}), we have $\check{R}_{{a},j} (u) \propto \mathbf{1} + e_j$ and the transfer matrix
covers the lattice by the {\em locally} equal-weighted linear combination of those two operators.
This isotropic value is particularly important for the applications to the loop and Potts model, to be presented in the following sections.
Recall that in the loop representation of section~\ref{sec:aTLalgebra},
the commutation relations satisfied by the $e_j$ gave rise to a non-local weight $\m$ per loop. The amount of non-locality is controlled by $\gamma$ and goes away
at the special value $\gamma = \frac{\pi}{3}$, since then $\m = 1$ and the non-local loops are simply not counted.

The corresponding XXZ spin chain ensues by expanding $t(u)$ around another special value, the {\em fully anisotropic} value $u=0$. To first order in this expansion, the
logarithmic derivative of $t(u)$ becomes the XXZ Hamiltonian
\begin{equation}\label{H_phi}
\HN = -\frac{\gamma}{\pi \sin \gamma} \sum^{N}_{j=1} (e_j-e_\infty) \,.
\end{equation}
Here the prefactor is chosen to ensure relativistic invariance at low energy, and  $e_\infty$ is a constant energy density that
cancels out extensive contributions to the ground state. Its value is
\begin{equation}\label{e_inf}
e_\infty = \sin \gamma \: I_0 \,,
\end{equation}
with $I_0$ given by the integral \cite{KooSaleur}
\begin{equation} \label{I0-def}
I_0=\int^\infty_{-\infty} \frac{\sinh(\pi-\gamma)t }{\sinh(\pi t)\cosh(\gamma t)} \, \mathrm{d}t \,.
\end{equation}

In \eqref{transfer} and \eqref{H_phi}, the generators $e_j$ can be taken to act in different representations of the aTL algebra $\ATL{L}(\m)$.
The loop representation has already been shown in (\ref{splittings}). But from the usual formulation of the XXZ chain in terms of spins $| \! \uparrow \rangle$ and $| \! \downarrow \rangle$,
it is more natural to consider the {\em XXZ representation} in which the $e_j$ act on $(\mathbb{C}^{2})^{\otimes L}$ with
\begin{equation}
e_j = -\sigma_j^{-}\sigma_{j+1}^{+}-\sigma_j^{+}\sigma_{j+1}^{-}
-\frac{\cos\gamma}{2}\sigma_j^{z}\sigma_{j+1}^{z} -\frac{i\sin\gamma}{2}(\sigma_j^{z}-\sigma_{j+1}^{z})+\frac{\cos\gamma}{2} \,,
\label{TLspin}
\end{equation}
where the $\sigma_j$ are the Pauli matrices, so the Hamiltonian is the familiar XXZ spin chain
\begin{equation}\label{Pauliham}
\HN = \frac{\gamma}{2 \pi \sin \gamma}\sum^{N}_{j=1}   \left[ \sigma^x_j \sigma^x_{j+1} + \sigma^y_j \sigma^y_{j+1} + \Delta (\sigma^z_j \sigma^z_{j+1} - 1 ) + 2 e_\infty \right]
\end{equation}
with anisotropy parameter
\begin{equation}
 \Delta = \cos(\gamma) \,.
\end{equation}

In the spin basis, the generator $e_j$ then acts on sites $j, j+1$ (with periodic boundary conditions) as
\begin{equation} e_j = \cdots \otimes\mathbf{1}\otimes
\left(\begin{array}{cccc}
0 & 0 & 0 & 0 \\
0 & -q^{-1} & -1 & 0\\
0& -1 & -q & 0\\
0 & 0 & 0 & 0
\end{array}\right) \otimes \mathbf{1}\otimes \cdots \,,
\end{equation}
where the quantum-group related parameter $q$ is defined as
\begin{equation}
 q = -e^{i \gamma} \,.
\end{equation}

One may introduce a twist in the spin chain without changing the expression (\ref{H_phi}),
by modifying the expression of the Temperley-Lieb generator acting between the first and last spin with a twist parametrized by $\phi$.
In terms of the Pauli matrices, this twist imposes the boundary conditions $\sigma^z_{L+1}=\sigma^z_1$ and $\sigma^{\pm}_{L+1}=e^{\mp i \phi} \sigma^{\pm}_1$.
% \textcolor{red}{Verify whether this needs a change due to the $-q$ convention.}
The value of the energy density $e_\infty$ is independent of $\phi$ and is still given by (\ref{e_inf}).

%%%%%%%%%%%%%%%%%%%%%%%%%%%%%%%%%%%%%%%%%%%%%%%%%%%%%%%%%%%%%%
\section{The loop model}
\label{sec:Loopmodel}
%%%%%%%%%%%%%%%%%%%%%%%%%%%%%%%%%%%%%%%%%%%%%%%%%%%%%%%%%%%%%%
We are now ready to define the loop model precisely and place it in the context of lattice algebras
(see section~\ref{sec:aTLalgebra}) and integrable models (see section~\ref{sec:XXZ}).

Consider a square lattice of width $L$ and height $N$, with the usual (`axial') orientation, corresponding to auxiliary spaces
along the periodic horizontal (or `space') direction. On this lattice, draw a set of loops by splitting each vertex in one of the two
ways shown in (\ref{splittings}). We wish to give the same weight (viz., one) to each splitting, and up to an unimportant proportionality
constant this is accomplished by the $\check{R}$-matrix (\ref{checkR-aTL}) at the isotropic point $u = \frac{\gamma}{2}$. The corresponding
row-to-row transfer matrix (\ref{transfer}) then adds one row to the lattice and, at the same time, generates all
possible configurations of splittings.

We recall from the introduction that the loop model is defined by giving a weight $\m^\ell$ to each configuration, where $\ell$ is the
number of loops. We further wish to impose toroidal boundary conditions on the $L \times N$ lattice. The periodic boundary
conditions in the horizontal direction is accounted for by the trace in (\ref{transfer}). The aTL algebra can handle this boundary
condition, and at the same time it ensures that each closed loop gets the correct weight $\m$.

It follows that $t \left( \gamma/2 \right)^N$ generates the partition function of the loop model, up to one last subtlety.
Namely, we need also to impose periodic boundary conditions in the vertical (or `time') direction, giving still a weight $\m$
per loop. In fact, $t \left( \gamma/2 \right)^N$ is a weighted sum of words in the algebra, such as those shown in
figure~\ref{fig:aff-diag}, so what remains is to ``glue'' the top and the bottom of the framing rectangle of each diagram generated by the algebra, and replace each loop in
the glued diagram by the weight $\mathfrak{m}$. For instance, the gluing of the three diagrams shown in Figure~\ref{fig:aff-diag} will result in the respective weights $\mathfrak{m}^3$, $\mathfrak{m}^2$ and $\mathfrak{m}$.
This may sound like a difficult task, but fortunately there is an algebraic construction that does just what we need.

\subsection{Markov trace}

The algebraic construction enabling this gluing is called the Markov trace \cite{Jones-index},
denoted ${\rm Tr}$ to distinguish it from the usual matrix trace ${\rm tr}$.
For the partition function $Z(L,N)$ on the $L \times N$ torus we then have
\begin{equation}
 Z(L,N) = {\rm Tr} \, \left[ t \left( \frac{\gamma}{2} \right)^N \right] \,.
\end{equation}
Below we shall always assume that both $L$ and $N$ are even.

We shall need the following fundamental result, which is well-known for the case of open transverse boundary conditions (cylinder geometry), but
which carries over to periodic boundary conditions (toroidal geometry) as well: The Markov trace ${\rm Tr}$ can be decomposed as a linear
combination of usual matrix traces ${\rm tr}$ over the standard modules.
The coefficients in this decomposition, which we shall call {\em eigenvalue multiplicities}, are denoted $\Lambda_{j,m}$.

For $j \in \mathbb{N}$ and $m$ a positive divisor of $j$, the multiplicity $\Lambda_{j,m}$ turns out to be
\begin{equation}
 \Lambda_{j,m} = 2 \sum_{d>0 \, : \, d|j} \frac{\mu\left( \frac{m}{m\wedge d}\right)\varphi\left( \frac{j}{d} \right)}
{j \, \varphi\left( \frac{m}{m\wedge d} \right)} \: T_{2d} \left( \frac{\m}{2} \right) \,.
\label{defLambda}
\end{equation}
Here, the sum is over all positive divisors of $j$.
Also, $m \wedge d$ denotes the greatest common divisor of $m$ and $d$, and
$\mu$ and $\varphi$ are respectively the M\"obius function and Euler's totient function~\cite{hardy}. The
M\"obius function $\mu$ is defined by $\mu(n)=(-1)^r$, if $n$ is an integer that is a product
$n=\prod_{i=1}^r p_i$ of $r$ {\em distinct} primes, $\mu(1)=1$, and $\mu(x)=0$ otherwise or if $x$
is not an integer. Euler's totient function $\varphi$ is defined for positive integers $n$
as the number of integers $n'$ such that $1\leq n'\leq n$ and $n\wedge n'=1$.
Finally, $T_n(x)$ is the $n$'th order Chebyshev polynomial of the first kind.
If we parameterise the loop weight $\m$ as in \eqref{m_param} then
\begin{equation} \label{Chebyshev}
 T_{2d} \left( \frac{\m}{2} \right) = \cos(2 d \gamma) \,.
\end{equation}

The crucial formula \eqref{defLambda} originated in sketchy form in \cite{DFSZ}, was first written as above
in the appendix of \cite{RS01} and justified by a field-theoretic reasoning, and finally a rigorous combinatorial derivation
was given in \cite{Richard}.

\subsection{Torus partition function}
We can now state the general result for the decomposition of the torus partition function over the standard modules.
Recall first that the twist variable $\rho^2$ has to be a $j$'th root of unity by the quantisation condition \eqref{twist-quant}.
We can write the possible solutions as
\begin{equation}
\label{eq:quantizerho2}
 \rho^2 = \omega_k \equiv \exp\left( \frac{2 \pi i k}{j} \right) \,, \quad \mbox{for } k=1,2,\ldots,j \,.
\end{equation}

Let $m$ be a positive divisor of $j$. Define $\kappa(m)$ to be the subset of $k \in \{1,2,\ldots,j\}$, such that
$k \in \kappa(m)$ if and only if $m$ is the {\em smallest} integer satisfying $k m \in \mathbb{N} j$.
For example, with $j=12$ we have
\begin{eqnarray}
 \kappa(1) &=& \{ 12 \} \,, \nonumber \\
 \kappa(2) &=& \{ 6 \} \,, \nonumber \\
 \kappa(3) &=& \{ 4,8 \} \,, \nonumber \\
 \kappa(4) &=& \{ 3,9 \} \,, \nonumber \\
 \kappa(6) &=& \{ 2,10 \} \,, \nonumber \\
 \kappa(12) &=& \{ 1,5,7,11 \} \,. \nonumber
\end{eqnarray}
Note that $\{1,2,\ldots,j\}$ is the disjoint union $\cup_{m|j} \kappa(m)$.

We then claim that within $\ATL{L}$ we can express the torus partition function as follows:
\begin{equation}
\label{Z-loop}
 Z(L,N) = {\rm tr}_{\AStTL{0}{q^2}} \left[ t \left( \frac{\gamma}{2} \right)^N \right]
 + \sum_{j=1}^{L/2} \sum_{m | j \atop m \ge 1} \Lambda_{j,m}
 \sum_{k \in \kappa(m)} {\rm tr}_{\AStTL{j}{\rho^2 = \omega_k}} \left[ t \left( \frac{\gamma}{2} \right)^N \right] \,.
\end{equation}
Note that the first term has multiplicity one, while the multiplicities of the remaining terms are the $\Lambda_{j,m}$ defined in \eqref{defLambda}.
We also notice that the total multiplicity for a fixed $j$ is simply
\begin{equation}
 \Lambda_j \equiv \sum_{m | j \atop m \ge 1} |\kappa(m)| \; \Lambda_{j,m} = 2 \; T_{2j} \left( \frac{\mathfrak{m}}{2} \right) \,,
\end{equation}
with the Chebyshev polynomial being defined in \eqref{Chebyshev}. Thus, in the classical (\emph{i.e.}, not ${q}$-deformed) limit $\mathfrak{m} = -2$
we have $\Lambda_j = 2$ for any $j \ge 1$. Since $T_{2j} \left( \frac{\mathfrak{m}}{2} \right)$ is a polynomial in $\m^2$, the same is true for $\m=2$.

\subsection{Explicit results}
\label{sec:loopExplicit}
To illustrate our general discussions on the torus partition function of the loop model, we give some explicit examples in this subsection. These include the closed-form expressions of the torus partition functions $Z_{\text{loop}}(2,N)$ and $Z_{\text{loop}}(4,N)$.

When assembling the ingredients corresponding to (\ref{Z-loop}), it is convenient to let $t_0$ denote the matrix representation of the transfer matrix $t$ acting in the representation $\AStTL{0}{q^2}$,
while $t_{j,\rho}$ denotes $t$ acting in $\AStTL{j}{\rho^2}$.

\subsubsection{Results for $L=2$}
For $L=2$, we have $j=0,1$. Therefore, we can define the standard modules $\AStTL{0}{q^2}$ and $\AStTL{1}{\rho^2}$.

Within $\AStTL{0}{q^2}$ we pick the link-state basis $\{ (),)( \}$. The (isotropic) transfer matrix reads
% (\textcolor{red}{YF: Explain a little bit how the transfer matrix is constructed.})
%
\begin{equation}
 t_0 = \left(\begin{array}{cc} \mathfrak{m} & 2+\mathfrak{m} \\ 2+\mathfrak{m} & \mathfrak{m}\end{array}\right)
\end{equation}
with eigenvalues
\begin{equation}
 \label{eq:evt0L2}
 \left \lbrace 2+2\mathfrak{m},-2 \right \rbrace \,.
\end{equation}
To obtain this result (and those given below) it is important to realise that the construction (\ref{transfer}) with an auxiliary space is
compatible with the aTL formalism. The crucial part is to perform the trace over the auxiliary space in the loop representation. This is done by decomposing $t$ as a product of $\check{R}$-matrices.
Before acting with the first $\check{R}$-matrix of a row, we need to insert a pair of mutually connected extra sites, \emph{i.e.}, to append $()$ to the left of each link state. The second of those sites plays the
role of the auxiliary space $a$ on which the first factor $\check{R}_{a,1}$ acts. The space corresponding to $a$ then moves to the right after each $\check{R}$ multiplication, and
after acting with the product of all $\check{R}$-matrices, the `open' (non-traced) auxiliary space is represented by the first and last sites.
Identifying those two sites by acting with the corresponding aTL operator $e_{0}$ attributes the correct weight, corresponding to the trace operation ${\rm tr}_a$ in (\ref{transfer}). We may then finally delete the two extra sites, and
relabel the remaining $L$ sites, so as to obtain a completed row.

Within $\AStTL{1}{\rho^2}$ the only link state is $||$, and the transfer matrix is
\begin{equation}
 \label{eq:t1L2}
 t_{1,\rho} =  \rho+\rho^{-1}\,.
\end{equation}

The partition function (\ref{Z-loop}) is then
\begin{eqnarray}
\label{Z2M}
 Z_{\text{loop}}(2,N) &=& {\rm tr}\, \big(t_0\big)^N + \Lambda_{1,1} \, {\rm tr} \, \big(t_{1,1} \big)^N \nonumber \\
 &=& (2+2\mathfrak{m})^N + (-2)^N + (\mathfrak{m}^2-2) \times 2^N \,,
\end{eqnarray}
where we have set $\rho = 1$ in $t_{1,\rho}$.

\subsubsection{Results for $L=4$}

For $L=4$, we $j=0,1,2$. The corresponding standard modules are $\AStTL{0}{q^2}$, $\AStTL{1}{\rho^2}$ and $\AStTL{2}{\rho^2}$.

Within $\AStTL{0}{q^2}$ we pick the link-state basis
\begin{equation}
 \{ ()() , (()) , ())(, )((), )()(, ))(( \}
\end{equation}
and the transfer matrix reads
\begin{equation}
\label{t0_L4}
 t_0 = \left(\begin{array}{cccccc}
 \mm & 2\mm+\mm^2 & 4+2\mm & 4+2\mm & 4+4\mm+\mm^2 & 2\mm+\mm^2 \\
 0 & \mm & 2+\mm & 2+\mm & 2+\mm & 0 \\
 2+\mm & 2+\mm & \mm & 0 & 0 & 2+\mm \\
 2+\mm & 2+\mm & 0 & \mm & 0 & 2+\mm \\
 4+4\mm+\mm^2 & 4+2\mm & 2\mm+\mm^2 & 2\mm+\mm^2 & \mm & 4+2\mm \\
 0 & 0 & 2+\mm & 2+\mm & 2+\mm & \mm
 \end{array}\right).
\end{equation}
The eigenvalues can be found analytically, and they read
\begin{equation}
\label{t0evals}
 \{ 8+7\mm+\mm^2,-\mm-\mm^2 \} \cup \{ -8-3\mm, \mm,\mm,\mm \} \,.
\end{equation}

Within $\AStTL{1}{\rho^2}$ we pick the link-state basis
\begin{equation}
 \{ ()|| , |()|, ||() , )||( \}
\end{equation}
and the transfer matrix reads
\begin{equation}
\label{t1_L4}
 t_{1,\rho} = \left(\begin{array}{cccc}
 \rho+\rho^{-1} & 2 \rho^{-1}+\mm \rho^{-1} & 2\rho^{-1}+\mm \rho^{-1} & 2+\mm \\
 2\rho+\mm \rho & \rho+\rho^{-1} & 2\rho^{-1}+\mm \rho^{-1} & 2+\mm \\
 2\rho+\mm \rho & 2\rho+\mm \rho & \rho+\rho^{-1} & 2+\mm \\
 2+\mm & 2+\mm & 2+\mm & \rho+\rho^{-1}
 \end{array}\right) \,.
\end{equation}
The eigenvalues for $\rho=1$ read
\begin{equation}
 \{ 8+3\mm,-\mm,-\mm,-\mm \} \,,
\end{equation}
while those for $\rho=-1$ all have the opposite sign. Note that the latter eigenvalues correspond to the second of the sets in \eqref{t0evals}.
This is consistent with the general result (\ref{cell-emb}) that $\AStTL{0}{q^2}$ contains a submodule isomorphic to $\AStTL{1}{\rho^2}$ with $\rho=-1$
[set $j=0$, $j'=1$, $z=q^2$ and $z'=\rho^2$ in (\ref{eq:emb-cond})].
Although this permits us to define a quotient module $\bAStTL{0}{q^2}$, of dimension $6-4=2$ in this case, it is the full module
$\AStTL{0}{q^2}$ that we need to construct the partition function.

Finally, within $\AStTL{2}{\rho^2}$ the only link state is $||||$, and the transfer matrix is
\begin{equation}
 t_{2,\rho} = \rho+\rho^{-1}\,.
\end{equation}

The partition function (\ref{Z-loop}) is then
\begin{eqnarray}
\label{Z4M}
 Z_{\text{loop}}(4,N) &=& {\rm tr}\, \big(t_0\big)^N + \Lambda_{1,1} \, {\rm tr} \, \big(t_{1,1} \big)^N + \Lambda_{2,1} \, {\rm tr} \, \big(t_{2,1} \big)^N + \Lambda_{2,2} \, {\rm tr} \, \big(t_{2,i} \big)^N
 \nonumber \\
 &=& (8+7\m+\m^2)^N + 3 \m^N + (-\m-\m^2)^N + (-8-3\m)^N \nonumber \\
 &+& (\m^2-2) \times \left[ (8+3\m)^N + 3 (-\m)^N \right] + \frac12 (\m^4-3\m^2) \times 2^N \,.
\end{eqnarray}
Note that $t_{2,i}$ is zero in this case; this will obviously not be so for higher values of $L$.

From \eqref{Z2M} and \eqref{Z4M} we can make the crucial check of modular invariance
\begin{equation}
 Z_{\text{loop}}(2,4) = Z_{\text{loop}}(4,2) = 16 \m^4 + 64 \m^3 + 112 \m^2 + 64 \m \,.
\end{equation}

\subsubsection{Results for $L \ge 6$}
For $L=6$, the analysis is similar, which leads to the following partition function
\begin{eqnarray}
\label{Z6M}
 Z_{\text{loop}}(6,N) &=& {\rm tr}\, \big(t_0\big)^N + \Lambda_{1,1} \, {\rm tr} \, \big(t_{1,1} \big)^N + \Lambda_{2,1} \, {\rm tr} \, \big(t_{2,1} \big)^N + \Lambda_{2,2} \, {\rm tr} \, \big(t_{2,i} \big)^N \nonumber \\
 & & + \Lambda_{3,1} \big(t_{3,1} \big)^N + \Lambda_{3,3} \left[ \big(t_{3,e^{i \pi / 3}} \big)^N + \big(t_{3,e^{2 i \pi / 3}} \big)^N \right] \,.
\end{eqnarray}
We see here the first manifestation of the regrouping of several traces with a common multiplicity, mirroring the fact that $\kappa(3) = \{1,2\}$.

Using the computer algebra system {\sc Mathematica} and {\sc Singular}
\cite{DGPS}, we have generated the matrix representations of the transfer matrix within each standard module for sizes $L=6,8,10$.
From this we have constructed the corresponding $Z(L,N)$ for $L \le 10$. % (\textcolor{red}{YF: Shall we give these results in some ancillary file?})
To give one example, we find that
\begin{eqnarray}
 Z_{\text{loop}}(6,4) &=& 786432 \mathfrak{m} + 2916256 \mathfrak{m}^2 + 4685824 \mathfrak{m}^3 + 4346912 \mathfrak{m}^4 + 2603712 \mathfrak{m}^5 +
 1063398 \mathfrak{m}^6 \nonumber \\
 & & + 304032 \mathfrak{m}^7 + 61242 \mathfrak{m}^8 + 8560 \mathfrak{m}^9 + 798 \mathfrak{m}^{10} +
 48 \mathfrak{m}^{11} + 2 \mathfrak{m}^{12} \,,
 \label{Z64-loop}
\end{eqnarray}
and we can check that $Z_{\text{loop}}(6,4) = Z_{\text{loop}}(4,6)$. More generally, we have checked the modular invariance, $Z_{\text{loop}}(L,N) = Z_{\text{loop}}(N,L)$,
for all values $L,N \le 10$. Some of the results for $Z_{\text{loop}}(L,L)$ are given in Appendix~\ref{app:resloop}.

\subsection{Further comments}
Before ending the section, let us make several comments.
\paragraph{Special values of $\mathfrak{m}$.}
Although a closed-form expression for $Z(L,N)$ as a function of
$\mathfrak{m}$ for arbitrary $L$ and $N$ is beyond reach, at special values of $\mathfrak{m}$ the partition function can be written down. Two important cases are $\mathfrak{m}=0,1$.

For $\mathfrak{m} = 0$, we have obviously $Z_{\text{loop}}(L,N) = 0$ for any values of $L,M$, because each configuration consists of at least one loop. A non-trivial quantity is however provided by
\begin{equation}
 E(L,N) = \left. \frac{{\rm d}}{{\rm d}\m} Z(L,N) \right|_{\m=0} \,,
\end{equation}
which gives the number of configurations with precisely one loop. These are known in physics as dense polymers and in graph theory as Eulerian circuits, here on the $L \times N$ torus.
Unlike in the planar case, we do not have the bijection between dense polymers and spanning trees---note that the latter can be counted by the matrix-tree theorem and exploiting properties of the
discrete Laplacian \cite{Kirchhoff1847} (see also \cite{CJSSS} and references therein).
We shall however have more to say about the numbers $E(L,N)$ in section~\ref{sec:Potts-results} below.
For $L=N$ they can be read off from the first coefficient of the results in Appendix~\ref{app:resloop}.

For $\mathfrak{m} = 1$, we have the trivial result $Z(L,N) = 2^{LN}$, since in this case the number of loops is not counted and we have
independently one of the two configurations \eqref{splittings} at each vertex.

Apart from this we have made two non-trivial observations, based on all $Z(L,N)$ with $L,N \le 10$. They read
\begin{subequations}
\begin{eqnarray}
 Z(L,N) &=& 0 \,, \quad \quad \quad \mbox{for } \mathfrak{m} = -1 \,, \label{m-minus1-loop} \\
 Z(L,N) &=& 2^{L+N} \,, \quad \ \mbox{for } \mathfrak{m} = -2 \,. \label{m-minus2-loop}
\end{eqnarray}
\end{subequations}
We shall be able to prove (\ref{m-minus1-loop}) in section~\ref{sec:Pottsmodel} by a reasoning that involves the Potts model.
However (\ref{m-minus2-loop}) remains elusive, and it is particularly intriguing that the power of two is reminiscent of a surface effect, despite the torus having of course no boundary.

\paragraph{Momentum sectors.}
We have further block-diagonalized the three $L=4$ transfer matrices with respect to lattice momentum. This
leads to the same results, but using at most matrices of dimension 2.
Note that there are two momenta in this problem: the pseudomomentum of the through-lines (controlled by the parameter $\rho$)
and the lattice momentum. In the corresponding XXZ spin chain the pseudomomentum is controlled by the twist  of the chain,
while the lattice momentum has its usual significance.

\paragraph{Quotients and further resonances.}
We wish to illustrate as well the more general construction of quotient modules in \eqref{Wbar-def}. The reason is that the corresponding singular values of
$z$ in $\StJTL{j}{z^2}$ also appear from the point of view of the
$QQ$-relations, to be discussed in section~\ref{sec:Qsystem}. As we
will see, such special values lead to vanishing leading coefficients in the Gr\"obner basis.

To provide this illustration, we first generalize the construction of the module $\StJTL{0}{q^2}$ so that contractible loops still have a weight
$\mathfrak{m} =q +q^{-1}$, while non-contractible loops now have a different weight $\mathfrak{n} = z + z^{-1}$. In this way \eqref{t0_L4} is replaced by
\begin{equation}
\label{t0L4bis}
 t_0 = \left(\begin{array}{cccccc}
 \mathfrak{n} & 2\mathfrak{n}+\mm\mathfrak{n} & 4+2\mm & 4+2\mm & 4+4\mm+\mm^2 & 2\mathfrak{n}+\mm\mathfrak{n} \\
 0 & \mathfrak{n} & 2+\mm & 2+\mm & 2+\mm & 0 \\
 2+\mm & 2+\mm & \mathfrak{n} & 0 & 0 & 2+\mm \\
 2+\mm & 2+\mm & 0 & \mathfrak{n} & 0 & 2+\mm \\
 4+4\mm+\mm^2 & 4+2\mm & 2\mathfrak{n}+\mm\mathfrak{n} & 2\mathfrak{n}+\mm\mathfrak{n} & \mathfrak{n} & 4+2\mm \\
 0 & 0 & 2+\mm & 2+\mm & 2+\mm & \mathfrak{n}
 \end{array}\right).
\end{equation}

We have already seen the resonance criterion \eqref{eq:emb-cond} at work in the case $(j,j') = (0,1)$, corresponding to $z' = 1$ and $z = -q$. We saw then
that 4 of the eigenvalues in $t_0$ coincided with those of $t_1$, so that the corresponding quotient module $\bAStTL{\! \! j}{(-q)^2}$ of dimension 2 could be defined.

Another case allowed by \eqref{eq:emb-cond} is $(j,j') = (0,2)$, corresponding to $z' = 1$ and $z = q^2$. We have then
\begin{equation}
 \mathfrak{n} = q^2 + q^{-2} = \mm^2 - 2 \,.
\end{equation}
Inserting this into \eqref{t0L4bis} we find the eigenvalues
\begin{equation}
 \left\{ 2 , \mm^2-2,\mm^2-2,2+6\mm+3\mm^2,\frac12 \left(-12-6\mm+\mm^2 \pm(\mm+2)\sqrt{32-7\mm^2}\right) \right\}
\end{equation}
The first of these eigenvalues (namely $2$) is also the (unique) eigenvalue of $t_{2,\rho=1}$, and another quotient module \eqref{Wbar-def} of dimension 5 can be defined.

Finally, we consider \eqref{eq:emb-cond} with $(j,j') = (1,2)$, corresponding to $z'=-q$ and $z =q^2$. The eigenvalues of $t_{1,q^2}$ can be found
from \eqref{t1_L4} as
\begin{equation}
 \left\{ -\mm,\mm(2\mm+3), q^{-2} \pm i q^{-2}(q-1)(q+1)^3 + q^{-1}(-1+q^3-q^2-2q) \right\} \,.
\end{equation}
The first of these eigenvalues (namely $-\mm$) coincides with the (unique) eigenvalue of $t_{2,\rho=-q}$, and so we can define a quotient module \eqref{Wbar-def} of dimension 3.

%%%%%%%%%%%%%%%%%%%%%%%%%%%%%%%%%%%%%%%%%%%%%%%%%%%%%%%%%%%
\section{The Potts model}
\label{sec:Pottsmodel}
%%%%%%%%%%%%%%%%%%%%%%%%%%%%%%%%%%%%%%%%%%%%%%%%%%%%%%%%%%%
Having discussed the torus partition function of the loop model, we turn to the Potts model in this section.
It will turn out possible to construct the torus partition function of the Potts model from the same ingredients
that have been employed to study the loop model \cite{DFSZ}.

\subsection{Potts model as a loop model}
\label{sec:Potts-as-loop}

The $Q$-state Potts model on the graph $G=(V,E)$ with vertex set $V$ and edge set $E$ is originally defined for $Q \in \mathbb{N}^*$ by \cite{Potts52}
\begin{equation}
 Z_{\rm Potts} = \sum_{\{\sigma\}} \exp \left( \sum_{(ij) \in E} K \delta_{\sigma_i,\sigma_j} \right) \,,
\end{equation}
where the $Q$-component spins $\sigma_i = 1,\ldots,Q$ interact along edges with coupling constant $K$ through the Kronecker symbol ($\delta_{x,y} = 1$ if $x = y$, and $\delta_{x,y} = 0$ otherwise).
Using the identity ${\rm e}^{K \delta_{\sigma_i,\sigma_j}} = 1 + v \delta_{\sigma_i,\sigma_j}$ with $v = {\rm e}^K - 1$ (valid because $\delta_{\sigma_i,\sigma_j} = 0$ or $1$), expanding the product over edges of
this two-term expression, and performing the sum over spins $\{\sigma\}$, this can be rewritten in the random-cluster form due to Fortuin and Kasteleyn (FK) \cite{FK72}
\begin{equation}
 \label{eq:Z-FK}
 Z_{\rm Potts} = \sum_{A \subseteq E} v^{|A|} Q^{k(A)} \,,
\end{equation}
where $|A|$ and $k(A)$ denote respectively the number of elements in the edge subset $A$ and the corresponding number
of connected components (often called FK clusters) in the induced graph $G_A = (V,A)$. It is important to notice that the FK form makes sense for any $Q \in \mathbb{R}$ (or even $Q \in \mathbb{C}$).

This can be further transformed into a loop model on the medial lattice $G_{\rm m}=(V_{\rm m},E_{\rm m})$ with vertices $V_{\rm m}$ located on the edges of $G$, and edges $E_{\rm m}$ between vertex pairs
in $V_{\rm m}$ whose corresponding edges in $E$ are indicident on a common vertex in $V$ \cite{BKW76}. Take now $G_{\rm m}$ to be the same $L \times N$ square lattice on which we defined the loop
model in section~\ref{sec:Loopmodel}, which, as the reader will recall, was axially oriented (\emph{i.e.}, with horizontal and vertical edges). Supposing again $L$ and $N$ to be even, it is easy to see that there exists a
graph $G$ of which $G_{\rm m}$ is the medial: indeed, $G$ is another square lattice, but diagonally oriented (\emph{i.e.}, rotated through $\frac{\pi}{4}$ with respect to $G_{\rm m}$ and scaled up by a factor $\sqrt{2}$).
There is a bijection between FK clusters on $G$ and loops on $G_{\rm m}$. Indeed, split the vertices $V_{\rm m}$ in one of two ways, as in (\ref{splittings}), so that the loops intersect none of the edges in $A$
and all of the edges in $E \setminus A$. Stated less precisely, but more intuitively, the loops wrap tightly around the FK clusters.

Now, if $G$ were a {\em planar} graph, application of the Euler relation would give \cite{BKW76}
\begin{equation} \label{Z-Potts-planar}
 Z_{\rm Potts} = \mm^{|V|} \sum_{A \subseteq E} \left(\frac{v}{\mm}\right)^{|A|} \mm^{\ell(A)} \,,
\end{equation}
where $\ell(A)$ is the number of loops, and $Q = \mm^2$. Remarkably, at the critical point on the square lattice, which is $v_{\rm c} = \mm$ \cite{Baxter73},
the weight conjugate to $|A|$ becomes one, so, up to the unimportant overall factor, $Z_{\rm Potts}$ is just the partition function of a loop model with weight $\m$
per loop. This reasoning however breaks down (as does the Euler relation) on the torus, so although the Potts model can still be formulated in terms of loops
the precise weighting is slightly different.

\subsection{Torus partition function}
On the torus, the decomposition of the Potts-model partition function with respect to the aTL algebra is subject to a number of subtleties,
which have their root in the non-planarity. This can however be solved by a subtle reweighting of the FK configurations in which there exists a
cluster (necessarily unique) that wraps around both periodic directions. In the original paper \cite{DFSZ} such a doubly-wrapping FK cluster was
called a `cluster with cross topology'.

%(\textcolor{red}{YF: This part is slightly dense and hard to understand for readers without backgrounds in Potts model.})
The end result is very similar to \eqref{Z-loop} and reads \cite{DFSZ}
\begin{eqnarray} \label{Z-Potts}
 Z_{\rm Potts}(L,N) &=& \mm^{L N / 2} \left\{ {\rm tr}_{\AStTL{0}{q^2}} \left[ t \left( \frac{\gamma}{2} \right)^N \right]
 + \frac{Q-1}{2} {\rm tr}_{\AStTL{0}{-1}} \left[ t \left( \frac{\gamma}{2} \right)^N \right] \right. \nonumber \\
 &+& \left. \sum_{j=1}^{L/2} \sum_{m | j \atop m \ge 1} \widetilde{\Lambda}_{j,m}
 \sum_{k \in \kappa(m)} {\rm tr}_{\AStTL{j}{\rho^2 = \omega_k}} \left[ t \left( \frac{\gamma}{2} \right)^N \right] \right\} \,.
\end{eqnarray}
The overall factor $\mm^{L N / 2}$ obviously has the same origin as the overall factor in (\ref{Z-Potts-planar}).
Much more interestingly, the second term in the bracket is new with respect to \eqref{Z-loop} and is related to the doubly-wrapping FK clusters.
Let us sketch the derivation of the form (\ref{Z-Potts}), giving an argument a bit different from that presented in the original paper \cite{DFSZ}.
The Euler relation used to derive (\ref{Z-Potts-planar}) on a planar graph {\em almost} holds on the torus, the only difference being
that two loops are `lost' in configurations with a doubly-wrapping FK cluster, as is readily seen by a bit of drawing. These wrapping clusters correspond to a situation in which
non-contractible loops are disallowed, whence the particular role of $\AStTL{0}{-1}$ (in which $z^2 = -1$ corresponds to
a vanishing weight $\mathfrak{n} = z + z^{-1} = 0$ of the non-contractible loops). This sector must acquire an extra weight of $Q = \m^2$
to make up for the two `lost' loops in the Euler-type argument, explaining the factor $Q-1$. The further factor of $\frac12$ is explaining by the fact (well known from aTL representation theory,
but which can also be derived by a duality argument \cite{Jacobsen14}) that exactly
when $\mathfrak{n} = 0$ the representation $\AStTL{0}{z^2}$ breaks up as the direct sum of two isomorphic representations, of which we
need only one copy. The same kind of argument shows that the eigenvalue multiplicities are affected by the reweighting
and become \cite{DFSZ,RS01,Richard}
\begin{equation}
\label{eq:shiftedLambda}
 \widetilde{\Lambda}_{j,m} = \Lambda_{j,m}(\mm) + \frac{Q-1}{2} \Lambda_{j,m}(0) \,,
\end{equation}
where $\Lambda_{j,m}(\mm)$ now denotes the same quantity as \eqref{defLambda}.

It is interesting to notice that $\widetilde{\Lambda}_{1,1} = -1$. This implies that in the expression \eqref{Z-Potts} the $j=1$ term in the sum (for which $m=1$ is the only divisor) can be subtracted from the first, $\AStTL{0}{q^2}$ term \cite{looppaper}.
This subtraction is precisely what defines the quotient module \eqref{quotientmodule}. An expression
equivalent to \eqref{Z-Potts} is then obtained by replacing $\AStTL{0}{q^2}$ by the quotient $\bAStTL{0}{q^2}$ and starting the sum at $j=2$. This implies,
somewhat surprisingly, that $\AStTL{1}{1}$ does not contribute to the partition function of the Potts model, although it contributes to the loop model.
Conversely, as we have seen, $\AStTL{0}{-1}$ contributes to the Potts model but not to the loop model.

\subsection{Explicit results}
\label{sec:Potts-results}

As for the loop model, we have used the computer-generated matrix representations of the transfer matrix within each standard module
to compute the corresponding $Z_{\rm Potts}(L,N)$ for $L \le 10$. As an example, we have
\begin{eqnarray}
\label{Z64-Potts}
 Z_{\rm Potts}(6,4) &=& 393216 \mm^{13} + 1810336 \mm^{14} + 3679232 \mm^{15}  + 4388576 \mm^{16}\nonumber \\ &+&
 3453776 \mm^{17} + 1920258 \mm^{18}
 + 794064 \mm^{19} + 255294 \mm^{20} + 65984 \mm^{21}\nonumber \\ &+& 13854 \mm^{22} + 2312 \mm^{23} + 289 \mm^{24} + 24 \mm^{25} + \mm^{26} \,,
\end{eqnarray}
and we can check that $Z_{\rm Potts}(6,4) = Z_{\rm Potts}(4,6)$. More generally, we have checked the modular invariance, $Z_{\rm Potts}(L,N) = Z_{\rm Potts}(N,L)$,
for all values $L,N \le 10$. Some of the results for $Z(L,L)$ are given in Appendix~\ref{app:respotts}.

Recall that the graph $G$ on which the original Potts model (\ref{eq:Z-FK}) is defined is a tilted square lattice with $|V| = \frac{LN}{2}$ spins and $|E| = LN$ edges.
Since we have $v = \m$ and $Q = \m^2$, the lowest power of $\m$ is obtained from configurations with a single cluster and the least possible number of edges,
of weight $v^{|V|-1} Q = \m^{\frac{LN}{2}+1}$, corresponding to spanning trees. It is not hard to see that no other types of configurations can produce such a small power of $\m$, so
the coefficient of $\m^{\frac{LN}{2}+1}$ counts the number of spanning trees on $G$. The highest power of $\m$ is meanwhile obtained from configurations with the
largest possible number of edges, of weight $v^{|E|} Q = \m^{LN+2}$. Since there is only one such configuration, the coefficient of $ \m^{LN+2}$ is one.
All configurations with strictly more than one cluster correspond to terms with non-extremal powers of $\m$.

\paragraph{Special values of $\mathfrak{m}$.}
For $\m=0$ the number of spanning trees can hence be found from the derivative
\begin{equation}
 S(L,N) = s! \left. \frac{{\rm d}^{s}}{{\rm d}\m^s} Z(L,N) \right|_{\m=0} \,, \qquad \mbox{with } s=\frac{LN}{2}+1 \,.
\end{equation}
Let ${\cal L}_G$ denote the discrete Laplacian of the corresponding graph $G$, and let ${\cal L}'_G$ be any one of its minors,
obtained by deleting some row (\emph{e.g.}, the last one) and its corresponding column.
We have verified that $S(L,N) = \det {\cal L}'_G$, in agreement with the matrix-tree theorem
\cite{Kirchhoff1847,CJSSS}.
By comparing \eqref{Z64-loop} with \eqref{Z64-Potts}---or the results of
Appendix~\ref{app:resloop} with those of Appendix~\ref{app:respotts}---we see that, in fact, the number of
Eulerian circuits is precisely twice the number of spanning trees:
\begin{equation}
 \label{eq:E=2S}
 E(L,N) = 2 S(L,N) \,.
\end{equation}
This is of course no coincidence. Indeed, an Eulerian circuit either wraps tightly around a spanning tree on the
graph $G$, {\em or} around a spanning tree of the dual $G^*$.
On a planar graph the dual of a spanning tree is another spanning tree, so the {\em same} Eulerian circuit sees
one tree on its inside and another on its outside. However, on a torus a spanning tree is homotopic to a point,
so its dual contains a double-wrapping FK cluster, hence cannot be a spanning tree (and vice versa). So we
have just proven that, in general, $E(L,N)$ is the number of spanning trees on $G$ plus the number of spanning
trees on the dual $G^*$. For the case of a selfdual lattice ($G = G^*$), such as the square lattice considered here,
we infer \eqref{eq:E=2S}.

For $\m=1$ we again have the trivial result $Z(L,N) = 2^{LN}$ for the same reason as in the loop model case.

Furthermore, based on all $Z_{\rm Potts}(L,N)$ with $L,N \le 10$ we have made the observations
\begin{subequations}
\begin{eqnarray}
 Z_{\rm Potts}(L,N) &=& 0 \,, \quad \quad \quad \quad \quad \quad \ \ \mbox{for } \mathfrak{m} = -1 \,, \label{m-minus1-Potts}  \\
 Z_{\rm Potts}(L,N) &=& 2^{L N/2 + L + N - 2} \,, \quad \ \ \mbox{for } \mathfrak{m} = -2 \,. \label{m-minus2-Potts}
\end{eqnarray}
\end{subequations}
We can in fact derive (\ref{m-minus1-Potts}) by using results of section~\ref{sec:Potts-as-loop}. Since we have
\begin{equation}
 v = {\rm e}^K - 1 = \m = -1 \,,
\end{equation}
and $Q = \m^2 = 1$, it follows that we have a one-state Potts model in which a pair of equal neighbouring spins have the Boltzmann weight ${\rm e}^K = 0$.
Therefore, $Z_{\rm Potts}(L,N) = 0$ at $\m = -1$. Moreover, since the difference with the loop-model partition function $Z(L,N)$ is always
proportional to $Q-1$, as we have seen, this argument also proves (\ref{m-minus1-loop}).

On the other hand, the result (\ref{m-minus2-Potts}) remais elusive.

%%%%%%%%%%%%%%%%%%%%%%%%%%%%%%%%%%%%%%%%%%%%%%%%%%%%%%%%%%%%%%
\section{Rational $Q$-system of twisted spin chains}
\label{sec:Qsystem}
%%%%%%%%%%%%%%%%%%%%%%%%%%%%%%%%%%%%%%%%%%%%%%%%%%%%%%%%%%%%%%
In section~\ref{sec:XXZ}, we have shown that the XXZ spin chain (\ref{Pauliham}) can be written in terms of aTL generators.
Since both the loop model and the Potts model have been related to the aTL algebra, this implies that both models are intimately related to the XXZ spin chain.
The XXZ spin chain is integrable and can be solved by the Bethe ansatz. This opens the interesting possibility of studying the standard modules,
relevant for producing the torus partition function of the loop and Potts models, from the Bethe ansatz perspective.

In this section, we first review the twisted XXZ spin chain. Then we turn to the discussion of the physical solutions of the Bethe equations.
It is known that the Bethe equations contain unphysical solutions that do not solve the eigenvalue problem of the transfer matrix.
A better formulation that eschews unphysical solutions is the rational $Q$-system. This approach was first introduced as an efficient method of solving
the Bethe equations of the XXX spin chain in \cite{Marboe:2016yyn}. The properties of the $Q$-system and its relation to the physicality of solutions
were further clarified in \cite{Granet:2019knz}, and subsequently generalized to the XXZ case \cite{Bajnok:2019zub}.
We here go one step further and generalize the rational $Q$-system to both the XXX and XXZ spin chains with diagonal twists.

\subsection{Twisted XXZ spin chain and Bethe ansatz}
Up to an irrelevant global factor and a constant shift, we can define the XXZ spin chain (\ref{Pauliham}) by the following Hamiltonian
\begin{align}
\mathcal{H}_{\text{XXZ}}
=\sum_{m=1}^L\left(\sigma_m^+\sigma_{m+1}^-+\sigma_m^-\sigma_{m+1}^y+\Delta\,\sigma_m^z\sigma_{m+1}^z \right) \,,
\end{align}
where $L$ is the length of the spin chain, which is identified with the number of non-trivial generators of $\ATL{L}(\mm)$.
%We impose the twisted boundary condition
%\begin{align}
%\sigma_{m+L}^{\pm}=\kappa^{\pm}\sigma_m^{\pm}
%\end{align}
%that preserves integrability. Here $\kappa^{\pm}$ are two constant parameters. They are related to the phase factor $e^{i\phi}$ (\ref{def_rho}) by $\kappa^{\pm}=e^{\mp i\phi}$. We will work with generic $\kappa^{\pm}$ and specify to these values in a later stage. The twisted XXZ spin chain can be solved by Bethe ansatz.
We consider the algebraic Bethe ansatz. The $R$-matrix is given by
\begin{align}
R_{an}(u)\propto\left(
            \begin{array}{cccc}
              \sinh(u+\tfrac{i\gamma}{2}) & 0 & 0 & 0 \\
              0 & \sinh(u-\tfrac{\eta}{2}) & \sinh(i\gamma) & 0 \\
              0 & \sinh(i\gamma) & \sinh(u-\tfrac{i\gamma}{2}) & 0 \\
              0 & 0 & 0 & \sinh(u+\tfrac{i\gamma}{2}) \\
            \end{array}
          \right).
\end{align}
The precise normalization will be fixed by comparing results with the loop model given in section~\ref{sec:loopExplicit}. We consider diagonal twisted boundary condition, which can be implemented in the algebraic Bethe ansatz by a constant matrix
\begin{align}
K_a=\left(
      \begin{array}{cc}
        \kappa^+ & 0 \\
        0 & \kappa^- \\
      \end{array}
    \right)
\end{align}
in the auxiliary space. The monodromy matrix and the transfer matrix are then defined as
\begin{align}
\widetilde{\mathbb{M}}_a(u)=K_a\,R_{a1}(u)R_{a2}(u)\ldots R_{aL}(u) \,, \qquad \widetilde{\mathbb{T}}(u)=\text{tr}_a\widetilde{\mathbb{M}}_a(u) \,,
\end{align}
and seeing the former as a $2 \times 2$ matrix in auxiliary space of operators acting on the $L$ quantum spaces, we may write as usual
\begin{align}
 \widetilde{\mathbb{M}}_a(u) = \left[ \begin{array}{cc} \tilde{A}(u) & \tilde{B}(u) \\ \tilde{C}(u) & \tilde{D}(u) \end{array} \right] \,, \qquad \widetilde{\mathbb{T}}(u) = \tilde{A}(u) + \tilde{D}(u) \,.
\end{align}
The transfer matrix can be diagonalized by the Bethe ansatz. We start with the pseudovacuum $|\Omega\rangle=|\! \uparrow^L\rangle$, which is an eigenvector of $\tilde{A}$ and $\tilde{D}$, and is annihilated by $\tilde{C}$. We have
\begin{align}
\tilde{A}(u)|\Omega\rangle=a(u)|\Omega\rangle \,, \qquad \tilde{D}(u)|\Omega\rangle=d(u)|\Omega\rangle \,,
\end{align}
where
\begin{align}
a(u)=\kappa^+\,\left(\sinh(u+\tfrac{i\gamma}{2})\right)^L,\qquad d(u)=\kappa^-\,\left(\sinh(u-\tfrac{i\gamma}{2})\right)^L.
\end{align}
Bethe states $|\mathbf{u}\rangle$ are constructed by acting on $|\Omega\rangle$ by the $\tilde{B}(u)$ operators,
\begin{align}
|\mathbf{u}\rangle=\tilde{B}(u_1)\ldots \tilde{B}(u_M)|\Omega\rangle \,.
\end{align}
If the Bethe roots $\mathbf{u}=\{u_1,\ldots,u_M\}$ satisfy the Bethe ansatz equations
\begin{align}
\label{eq:BAEadditive}
\left(\frac{\sinh(u_j+\tfrac{i\gamma}{2})}{\sinh(u_j-\tfrac{i\gamma}{2})} \right)^L=-\frac{\kappa^-}{\kappa^+}\prod_{k=1}^M\frac{\sinh(u_j-u_k+i\gamma)}{\sinh(u_j-u_k-i\gamma)} \,,
\end{align}
the state $|\mathbf{u}\rangle$ diagonalizes the transfer matrix
\begin{align}
\widetilde{\mathbb{T}}(u)|\mathbf{u}\rangle=\tilde{\tau}(u)|\mathbf{u}\rangle \,.
\end{align}
The corresponding eigenvalue,
\begin{align}
\tilde{\tau}(u)\propto a(u)\frac{\tilde{Q}(u-i\gamma)}{\tilde{Q}(u)}
+d(u)\frac{\tilde{Q}(u+i\gamma)}{\tilde{Q}(u)} \,,
\end{align}
can be expressed in terms of the $Q$-function defined by
\begin{align}
\tilde{Q}(u)=\prod_{j=1}^M\sinh(u-u_j) \,.
\end{align}
To precisely relate the results to the loop model, we fix the normalization and take the eigenvalue of the transfer matrix to be
\begin{align}
\tilde{\tau}(u)=\kappa^+\left(\frac{\sinh(u+\tfrac{i\gamma}{2})}{\sinh(u-\tfrac{i\gamma}{2})}\right)^L\frac{\tilde{Q}(u-i\gamma)}{\tilde{Q}(u)}
+\kappa^-\frac{\tilde{Q}(u+i\gamma)}{\tilde{Q}(u)} \,.
\end{align}
To apply the computational algebraic-geometry method, it is more convenient to work with the multiplicative variables
\begin{align}
\label{mult-vars}
q=e^{i\gamma}\,,\qquad t=e^u\,,\qquad t_k=e^{u_k}\,,
\end{align}
in terms of which the Bethe equations (\ref{eq:BAEadditive}) become
\begin{align}
\label{eq:twistttBAE}
\left(\frac{t_j^2\, q-1}{t_j^2-q}\right)^L=-\frac{\kappa^-}{\kappa^+}\prod_{k=1}^M\frac{t_j^2\,q^2-t_k^2}{t_j^2-t_k^2\,q^2} \,.
\end{align}
The eigenvalue of the transfer matrix then takes the form
\begin{align}
\label{eq:transfertaut}
\tau(t)
%=&\,\kappa^+\left(\frac{t^2 q-1}{t^2-q}\right)^L\prod_{k=1}^M\frac{t^2-t_k^2q^2}{qt^2-qt_k^2}+\kappa^-\,\prod_{k=1}^M\frac{t^2q^2-t_k^2}{qt^2-qt_k^2}\\\nonumber
=\kappa^+\left(\frac{t^2 q-1}{t^2-q}\right)^L\frac{Q(t\,q^{-1})}{Q(t)}+\kappa^-\,\frac{Q(t\,q)}{Q(t)} \,,
\end{align}
where we have defined
\begin{align}
\label{eq:Qtprod}
Q(t)=\prod_{k=1}^M(t-t^{-1} t_k^2) \,.
\end{align}
$Q(t)$ is a Laurent polynomial of degree $M$ which can be expanded as
\begin{align}
\label{eq:sumQt}
Q(t)=t^M+\sum_{k=1}^{M-1}c_k\,t^{2k-M}+\frac{1}{t^Mc_0} \,.
\end{align}
To reproduce the result in the loop model, we take the twist parameters $\kappa^{\pm}=\rho^{\pm 1}$, with $\rho=e^{i\phi/2}$.

\subsection{Twisted $Q$-system of the XXX spin chain}
As mentioned before, to find the physical solutions of the Bethe quation, we apply the rational $Q$-system. To this end, we need to generalize the method of \cite{Granet:2019knz,Bajnok:2019zub} to include twisted boundary conditions. We first consider the XXX model in this subsection. The XXZ case will be discussed in the next subsection.

\begin{figure}[h!]
\centering
\includegraphics[scale=0.5]{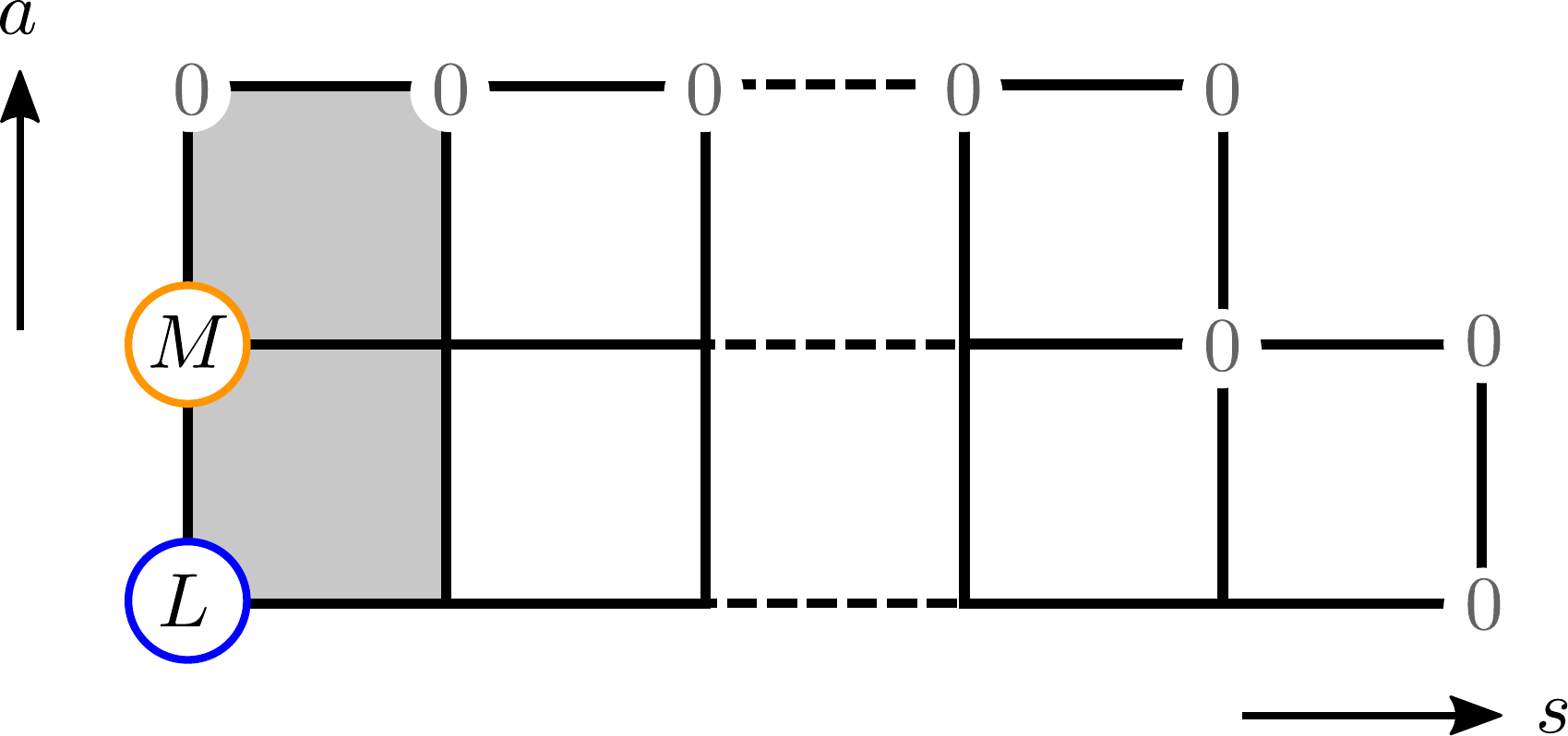}
\caption{The Young tableaux associated to the rational $Q$-system. At each node of the Young diagram we have one $Q$-function denoted by $Q_{a,s}$. The number in each circle denotes the power of the polynomial.}
\label{fig:YD}
\end{figure}

Let us recall briefly the twist-less rational $Q$-system of the XXX spin chain. Each $Q$-system is related to a Young tableaux as is shown in figure~\ref{fig:YD}.
For the $\mathfrak{su}(2)$ spin chain with length $L$ and magnon number $M$, the Young tableaux has two rows with the number of boxes given by $(M,L-M)$, where $L-M \ge M$. At each node of the Young tableaux, we define a $Q$-function. The $Q$-functions satisfy the following $QQ$-relation
\begin{align}
\label{eq:QQrelation0}
Q_{a+1,s}(u)Q_{a,s+1}(u)=Q_{a+1,s+1}^-(u)Q_{a,s}^+(u)-Q_{a+1,s+1}^+(u)Q_{a,s}^-(u) \,,
\end{align}
where we have introduced the notation $f^{\pm}(u)\equiv f(u\pm\frac{i}{2})$.%
\footnote{This notation only applies to $Q$-functions and has no relation to the superscripts used for the twist parameters $\kappa^\pm$ above.}
We impose the boundary conditions $Q_{a,s}=1$ at the upper-right boundary, together with $Q_{0,0}(u)=u^L$. One can then parameterize the $Q_{1,0}$ by
\begin{align}
\label{eq:defQ10}
Q_{1,0}(u)=\prod_{j=1}^M(u-u_j)=u^M+\sum_{k=0}^{M-1}c_k u^k.
\end{align}
Requiring all the $Q_{a,s}$ functions on the Young tableaux to be \emph{polynomials}, we obtain a set of equations for the coefficients $\{c_k\}$ in (\ref{eq:defQ10}), which can be solved analytically or numerically. The zeros of $Q_{1,0}(u)$ then give the corresponding Bethe roots.\par

A comment is in order here. The $QQ$-relations (\ref{eq:QQrelation0}) are in fact defined by proportionality. This means the $QQ$-relation
\begin{align}
Q_{a+1,s}(u)Q_{a,s+1}(u)={\kappa_{a}}\left(Q_{a+1,s+1}^+(u)Q_{a,s}^-(u)-Q_{a+1,s+1}^-(u)Q_{a,s}^+(u)\right) \,,
\end{align}
is equivalent to (\ref{eq:QQrelation0}) in the sense that they lead to the same Bethe roots.

Now we turn to the twisted spin chain. To incorporate the twist in the rational $Q$-system, we modify the $QQ$-relation \cite{Kazakov:2015efa} as follows
\begin{align}
\label{eq:QQrelation}
Q_{a+1,s}(u)Q_{a,s+1}(u)={\kappa_a}\,Q_{a+1,s+1}^-(u)Q_{a,s}^+(u)-Q_{a+1,s+1}^+(u)Q_{a,s}^-(u)\,,
\end{align}
where $\kappa_a$ are $u$ independent constants.
%A comment is in order here: The equation
%\begin{align}
%Q_{a+1,s}(u)Q_{a,s+1}(u)={\kappa_{a}}\left(Q_{a+1,s+1}^+(u)Q_{a,s}^-(u)-Q_{a+1,s+1}^-(u)Q_{a,s}^+(u)\right)
%\end{align}
%would give \emph{the same} $QQ$ relation. \textcolor{red}{[JJ: You mean the same as untwisted XXX? Surely the relative factor between the first and second term is related to the twist, right? Rewrite a bit: it is more logical to recall first the untwisted case, the discuss proportionality, and arrive at (\ref{eq:QQrelation}) in the end.]} The reason is that $QQ$-relation is actually defined by proportionality
%\begin{align}
%Q_{a+1,s}(u)Q_{a,s+1}(u)\propto Q_{a+1,s+1}^+(u)Q_{a,s}^-(u)-Q_{a+1,s+1}^-(u)Q_{a,s}^+(u)
%\end{align}
Notice that we multiply different factors onto the two terms on the right-hand side. We will call this deformed $QQ$-relation \eqref{eq:QQrelation} the $\kappa QQ$-relation.\par

The $\kappa QQ$-relation can be defined for the $\mathfrak{su}(N)$ spin chain and provides a method to solve the corresponding twisted Bethe equations. For our purpose, we consider only the $\mathfrak{su}(2)$ case where the Young tableaux have two rows. We impose the same boundary conditions $Q_{2,s}=1$ at the upper-right boundary, and $Q_{0,0}(u)=u^L$ in the lower-left corner. For this simple situation, the $\kappa QQ$ relations can be solved explicitly as follows.

First, consider the $\kappa QQ$ relation for $a=1$. By taking into account the boundary condition we find
\begin{align}
Q_{1,s+1}(u)=\kappa_1\,Q_{1,s}^+(u) - Q_{1,s}^-(u) \,.
\end{align}
This is a finite-difference equation whose solution is given by \cite{Granet:2019knz,Bajnok:2019zub}
\begin{align}
Q_{1,n}(u)=Q^{(n)}_{\kappa_1}(u)\equiv\widetilde{D}^{n}_{\kappa_1}Q(u) \,,
\end{align}
where the twisted discrete derivative is defined by
\begin{align}
\widetilde{D}_{\kappa}f(u)={\kappa} f(u+\tfrac{i}{2})-f(u-\tfrac{i}{2}) \,.
\end{align}
Taking then $a=0$ in the $\kappa QQ$-relation, we obtain
\begin{align}
Q_{0,n}=\frac{{\kappa_0}\,Q^{(n)-}_{\kappa_1} Q_{0,s}^+ -Q^{(n)+}_{\kappa_1} Q_{0,s}^- }{Q^{(n-1)}_{\kappa_1}} \,.
\end{align}

\paragraph{Bethe equations.} Let us derive the Bethe equations from the $\kappa QQ$-relations. Consider the following $\kappa QQ$-relations
\begin{subequations}
\label{eq:XXXBAE1}
\begin{align}
\label{eq:XXXBAE1-1}
Q_{1,1}(u)=&\,\kappa_1\,Q_{1,0}^+(u) -Q_{1,0}^-(u) \,, \\
\label{eq:XXXBAE1-2}
Q_{1,0}(u)Q_{0,1}(u)=&\,\kappa_0\,Q_{1,1}^-(u)Q_{0,0}^+(u) -Q_{1,1}^+(u)Q_{0,0}^-(u) \,.
\end{align}
\end{subequations}
The Bethe equations can be derived by requiring that all the $Q_{a,s}$ be polynomials in $u$. As for the non-twisted case, the zeros of $Q_{1,0}(u)$ are the Bethe roots, which we will denote by $\{u_k\}$. We evaluate the equations (\ref{eq:XXXBAE1}) at one of the Bethe roots $u_k$. From (\ref{eq:XXXBAE1-1}) we obtain
\begin{align}
Q_{1,1}^+(u_k)=\kappa_1\,Q_{1,0}^{++}(u_k)\,,\qquad Q_{1,1}^-(u_k)=-Q_{1,0}^{--}(u_k) \,,
\end{align}
where the double signs mean double shifts, $f^{\pm \pm}(u) \equiv f(u \pm i)$.
From (\ref{eq:XXXBAE1-2}) we have
\begin{align}
\kappa_0\,Q_{1,1}^-(u_k)Q_{0,0}^+(u_k)-Q_{1,1}^+(u_k)Q_{0,0}^-(u_k)=0 \,.
\end{align}
Combining these two equations, we find that
\begin{align}
\label{eq:QQBAE1}
\kappa_0 Q_{1,0}^{--}(u_k)Q_{0,0}^+(u_k)+\kappa_1 Q_{1,0}^{++}(u_k)Q_{0,0}^-(u_k)=0 \,.
\end{align}
Using the fact that
\begin{align}
Q_{0,0}(u)=u^L,\qquad Q_{1,0}(u)=\prod_{j=1}^M(u-u_j) \,,
\end{align}
we see that (\ref{eq:QQBAE1}) is precisely the twisted Bethe equation
\begin{align}
-\frac{\kappa_1}{\kappa_0}=\left(\frac{u_k+\tfrac{i}{2}}{u_k-\tfrac{i}{2}}\right)^L \prod_{j=1}^M\frac{u_k-u_j-i}{u_k-u_j+i} \,.
\end{align}
%Let us choose $\kappa_1=\kappa=e^{i\phi/2}$, $\kappa_0=1$, we find that
%\begin{align}
%\left(\frac{u_k+\tfrac{i}{2}}{u_k-\tfrac{i}{2}}\right)^L=-\kappa\prod_{j=1}^M\frac{u_k-u_j+i}{u_k-u_j-i}
%\end{align}
Similar to the untwisted case, the requirement that all $Q_{a,s}$ be polynomials in $u$ leads to a set of algebraic equations---the \emph{zero-remainder conditions} (ZRC)---for the coefficients of the $Q$-functions. In our case, the ZRC now depend on the twists. The ZRC for fixed values of $\kappa_0$ and $\kappa_1$ can be solved numerically. We find that for the system with length $L$ and magnon number $M$, the number of solutions for non-trivial twist is $L\choose M$. We have checked numerically that the solutions of the $\kappa QQ$-relation indeed solve the twisted Bethe equation.

\subsection{Twisted $Q$-system of the XXZ spin chain}
We are now ready to generalize the $\kappa QQ$-relation to the $q$-deformed case, for which the XXX chain just considered is the special case $q=1$. We consider the multiplicative variable $t = e^u$ as in (\ref{mult-vars}). The $\kappa QQ$-relation takes the same form
\begin{align}
Q_{a+1,s}(t)Q_{a,s+1}(t)={\kappa_a}\,Q_{a+1,s+1}^-(t)Q_{a,s}^+(t)-Q_{a+1,s+1}^+(u)Q_{a,s}^-(t) \,,
\end{align}
where now $f^{\pm}(t)\equiv f(t q^{\pm1/2})$, and we have introduced twist parameters $\kappa_0$ and $\kappa_1$. We focus on the $U_q(sl(2))$ case where the corresponding Young tableaux has two rows. The boundary conditions are\footnote{One might notice that the normalization of $Q_{1,0}(t)$ is different from the normalization of $Q(t)$ in (\ref{eq:sumQt}). Since the $\kappa QQ$-relation is defined by proportionality, this difference is harmless. }
\begin{align}
Q_{0,0}(t)=(t-t^{-1})^L\,, \qquad Q_{1,0}(t)=Q(t)=\prod_{j=1}^M(t t_j^{-1}-t^{-1}t_j) \,.
\end{align}
Both $Q$-functions are Laurent polynomials in $t$. The twisted discrete derivative in the $q$-deformed case is defined by
\begin{align}
\widetilde{D}_{q,\kappa}f(t)={\kappa}\,f(t q^{+1/2})-f(t q^{-1/2}) \,.
\end{align}
Again we can solve the $\kappa QQ$-relation explicitly, and find that $Q_{1,n}(t)=Q^{(n)}(t)=\widetilde{D}^n_{q,\kappa_1} Q(t)$. Similarly, $Q_{0,n}(t)$ is given by
\begin{align}
Q_{0,n}(t)=\frac{{\kappa_0}\,Q^{(n)+}Q_{0,n-1}^- - Q^{(n)-}Q_{0,n-1}^+}{Q^{(n-1)}} \,.
\end{align}
By requiring all $Q_{0,n}(t)$ to be polynomials in $t$ and $t^{-1}$, we obtain a set of zero-remainder conditions. Going through the same steps as the XXX case, we find
\begin{align}
\left(\frac{t_j^2\, q-1}{t_j^2-q}\right)^L =-\frac{\kappa_1}{\kappa_0}\prod_{k=1}^M\frac{t_j^2\,q^2-t_k^2}{t_j^2-t_k^2\,q^2} \,.
\end{align}
By identifying $\kappa_1=\kappa^-$ and $\kappa_0=\kappa^+$, we reproduce the Bethe equation of the twisted XXZ spin chain (\ref{eq:twistttBAE}). For generic $q$ and $\kappa^{\pm}$, the ZRCs can be solved numerically. We have checked that for length $L$ and magnon number $M$, the number of solutions of the ZRC is ${L\choose M}$ and they give the correct Bethe roots. The situation where $q$ is a root of unity is also interesting, but more subtle, and it will be analyzed elsewhere.

\subsection{An explicit example}
In order to illustrate that we can reproduce the result for the loop model from the twisted XXZ spin chain, we consider the example with $L=2$ in this subsection. In aTL language, there are two standard modules with $j=0,1$, which correspond to the two sectors of the XXZ spin chain with magnon numbers $M=1,0$ respectively. From now on, we take the twists to be $\kappa^{\pm}=\rho^{\pm 1}$.
\begin{itemize}
\item $j=1$. This corresponds to the sector $M=0$. We have $Q(t)=1$ and the transfer matrix (\ref{eq:transfertaut}) reads
\begin{align}
\label{eq:tau1p}
t_{1,\rho}(t)=\rho\,\left(\frac{t^2 q-1}{t^2-q}\right)^2+\rho^{-1} \,.
\end{align}
The transfer matrix of the loop model is given by fixing the spectral parameter to a special value $t^2=-1$. Plugging this into (\ref{eq:tau1p}), we obtain
\begin{align}
t_{1,\rho}\equiv t_{1,\rho}(i)=\rho+\rho^{-1} \,,
\end{align}
in agreement with (\ref{eq:t1L2}).
\item $j=0$. This corresponds to the sector $M=1$. We evaluate the transfer matrix at $t=i$,
\begin{align}
\label{eq:transferi}
t_0(i)=\rho\frac{Q(iq^{-1})}{Q(i)}+\rho^{-1}\frac{Q(iq)}{Q(i)} \,.
\end{align}
The $Q$-function is given by
\begin{align}
Q(t)=t+\frac{1}{c_0t} \,,
\end{align}
where $c_0$ satisfies the ZRC
\begin{align}
\label{eq:zrcsimple}
(q^2-\rho^2)c_0^2+2q(1-\rho^2)c_0+(1-q^2\rho^2)=0 \,.
\end{align}
Assuming for the moment $q^2-\rho^2\ne 0$, we have two solutions for $c_0$:
\begin{align}
c_0^{(1)}=-\frac{q\rho+1}{\rho+q}\,, \qquad c_0^{(2)} =-\frac{q\rho-1}{\rho-q} \,.
\end{align}
Plugging into the transfer matrix (\ref{eq:transferi}), the corresponding two eigenvalues are
\begin{align}
t_0^{(1)}=-2+(q+q^{-1})+(\rho+\rho^{-1})\,,\qquad t_0^{(2)}=2-(q+q^{-1})+(\rho+\rho^{-1})\,.
\end{align}
Now taking $\rho=-q$, we obtain
\begin{align}
\label{eq:eigentau2}
t_0^{(1)}=-2\,,\qquad t_0^{(2)}=2+2\mathfrak{m} \,,
\end{align}
in accord with (\ref{eq:evt0L2}), where $\mathfrak{m}$ is defined in (\ref{m_param}). We see that (\ref{eq:eigentau2}) gives precisely the two eigenvalues of the transfer matrix of the aTL loop model.
\end{itemize}
Before ending the section, let us make a comment here. We find that in the ZRCs (\ref{eq:zrcsimple}), the case $\rho=\pm q$ is special. In this case, the quadratic equation becomes a linear equation and the number of solutions is reduced from 2 to 1. As we will explain in more detail in the next section, this is a general phenomenon. When $\rho$ and $q$ satisfy certain relations, the number of solutions of the ZRC changes. These relations, as we shall see, are precisely the resonance conditions of the standard module.

%%%%%%%%%%%%%%%%%%%%%%%%%%%%%%%%%%%%%%%%%%%%%%%%%%%%%
\section{Resonance condition from algebraic geometry}
\label{sec:resonance}
%%%%%%%%%%%%%%%%%%%%%%%%%%%%%%%%%%%%%%%%%%%%%%%%%%%%%
As we have seen in section~\ref{sec:aTLalgebra}, for generic values of $q$ and $\rho$, the standard module $\mathcal{W}_{j,\rho^2}$ is irreducible with dimension $\hat{d}_j$. However, when $\rho$ satisfies the resonance condition, the representation becomes reducible and we can define the quotient module. From the Bethe ansatz perspective, the dimension of the standard module corresponds to the number of physical solutions of the Bethe equations. We therefore expect the structure of solutions to change at these special values of $\rho$. One simple example has been given in the previous section.\par

How can we see such a change in the number of solutions? One approach
is by using the Gr\"obner basis. For a given set of algebraic
equations, which are the ZRCs in our context, we can compute the
Gr\"obner basis by following the standard algorithm. The number of
solutions is given by the dimension of the quotient ring, whose basis
is characterized by the leading terms of the Gr\"obner
basis. Therefore, we expect that at special values, some coefficients
of the leading terms of the Gr\"obner basis vanish.\par

To test this expectation, we consider three explicit examples, $(L,M)=(4,1),(4,2)$ and $(L,M)=(6,3)$, in what follows. We will take a careful look at the leading terms of the Gr\"obner basis. At special value of the twists, some leading terms of the Gr\"obner basis vanish and the dimension of the quotient ring changes.

\subsection{Example 1: $L=4$, $M=1$}
For $L=4$, $M=1$, the $Q_{1,0}(t)$ is given by
\begin{align}
Q_{1,0}(t)=t+\frac{1}{c_0t} \,.
\end{align}
The rational $Q$-system leads to the following ZRC
\begin{align}
0=&\,(\rho^2-q^4)(q^2-1)\textcolor{blue}{c_0^4}+4q(\rho^2-q^2)(q^2-1)\textcolor{blue}{c_0^3}+6q^2(q^2-1)(\rho^2-1)\textcolor{blue}{c_0^2}\\\nonumber
&\,+4q(\rho^2q^2-1)(q^2-1)\textcolor{blue}{c_0}+(\rho^2 q^4-1)(q^2-1) \,.
\end{align}
This is a single-variable polynomial equation. Note that for $q=\pm1$, the right-hand side is zero identically, which implies that there are infinitely many solutions to the ZRC in this case. This is nothing but a manifestation of the subtleties at $q$ being a root of unity. Here we focus on generic $q$, therefore we can strip off the global factor $q^2-1$. The Gr\"obner basis can be computed straightforwardly and is given by
\begin{align}
g_1=(q^4-\rho^2)\textcolor{blue}{c_0^4}+4q(q^2-\rho^2)\textcolor{blue}{c_0^3}-6q^2(\rho^2-1)\textcolor{blue}{c_0^2}
+4q(1-q^2\rho^2)\textcolor{blue}{c_0}+(1-q^4\rho^2)\,.
\end{align}
The leading term of the Gr\"obner basis is
\begin{align}
\label{eq:LT41}
\text{LT}[g_1]=(q^4-\rho^2)\textcolor{blue}{c_0^4} \,,
\end{align}
from which we can read off the quotient-ring basis
\begin{align}
\{c_0^3,c_0^2,c_0,1\}.
\end{align}
This implies that there are 4 physical solutions to the twisted Bethe equation with $L=4$, $M=1$.\par

\paragraph{Special twist.} It is easy to see that the leading term's coefficient (\ref{eq:LT41}) vanishes at the specific values of the twist $\rho^2=q^4$. This hints that if we set $\rho=\pm q^2$, the number of solutions may be different from the generic case. Indeed, taking $\rho=q^2$, we find the following ZRC
\begin{align}
0=&\,4q^3(q^2-1)^2\,\textcolor{blue}{c_0^3}+6q^2(q^2+1)(q^2-1)^2\,\textcolor{blue}{c_0^2}\\\nonumber
&\,+4q(1-q^2-q^6+q^8)\,\textcolor{blue}{c_0}+(1-q^2-q^8+q^{10})\,.
\end{align}
The corresponding Gr\"obner basis is given by
\begin{align}
g_1=4q^3\,\textcolor{blue}{c_0^3}+6q^2(1+q^2)\,\textcolor{blue}{c_0^2}+4q(1+q^2+q^4)\,\textcolor{blue}{c_0}+(1+q^2)(1+q^4)\,.
\end{align}
The leading term is
\begin{align}
\text{LT}[g_1]=4q^3\,\textcolor{blue}{c_0^3}\,,
\end{align}
which leads to the quotient ring basis
\begin{align}
\label{eq:qrspecial}
\{c_0^2,c_0,1\}\,.
\end{align}
Therefore, we see that at $\rho=q^2$, the number of solution is 3.

\paragraph{Resonance condition.} Let us now compare what we have found with the resonance condition of the aTL representation discussed in section~\ref{sec:resonance}. At the special value $\rho=q^2$, the corresponding standard module $\mathcal{W}_{1,q^2}$ satisfies the resonance condition (\ref{eq:emb-cond}) with $(j,z)=(1,q^2)$ and $(j',z')=(2,q)$. The quotient module is
\begin{align}
\overline{\mathcal{W}}_{1,q^2}=\mathcal{W}_{1,q^2}/\mathcal{W}_{2,q}\,,
\end{align}
with the corresponding dimension
\begin{align}
\bar{d}_1=\text{dim}\,\overline{\mathcal{W}}_{1,q^2}={4\choose 2+1}-{4\choose 2+2}=3\,.
\end{align}
This matches nicely with the dimension of the quotient ring (\ref{eq:qrspecial}).

\subsection{Example 2: $L=4$, $M=2$}
Now we consider a more complicated case with $L=4$, $M=2$. In this case, the $Q_{1,0}$ function is given by
\begin{align}
Q_{1,0}(t)=t^2+c_1+\frac{1}{c_0\,t^2}\,.
\end{align}
The ZRC gives a set of algebraic equations for $\{c_0,c_1\}$. These equations are too long to be presented here.
The Gr\"obner basis of the ZRC in this case, with respect to
degree-reverse lexicographic ordering, consists of four polynomials $\{g_1,g_2,g_3,g_4\}$. The leading terms are given by
\begin{align}
\label{eq:Gr4}
&\text{LT}[g_1]=q^2(q^2\rho^2-1)^3\,\textcolor{blue}{c_1^3}\,,\\\nonumber
&\text{LT}[g_2]=q^2(\rho^2-1)(q^2\rho^2-1)\,\textcolor{blue}{c_0c_1^2}\,,\\\nonumber
&\text{LT}[g_3]=(q^2-\rho^2)^2\,\textcolor{blue}{c_0^2c_1}\,,\\\nonumber
&\text{LT}[g_4]=(q^4-\rho^2)(q^2-\rho^2)^3\,\textcolor{blue}{c_0^3} \,,
\end{align}
where the blue-colored parts are the monomials in $c_0$ and $c_1$, and the rest are coefficients. At generic values of $\rho$ and $q$, the basis of the quotient ring is given by
\begin{align}
\{c_0^2,\,c_1^2,\,c_0c_1,\,c_0,\,c_1,\,1\} \,,
\end{align}
which implies that the corresponding twisted Bethe equations have 6 physical solutions.

\paragraph{Special twists and resonance conditions.} From
(\ref{eq:Gr4}), we see that one or more of the leading terms' coefficients vanish when
\begin{align}
\rho^2=1\,,\qquad \rho^2=q^{\pm2}\,,\qquad \rho^2=q^4\,.
\end{align}
The case $\rho^2=1$ corresponds to the twist-less case. The other two cases, $\rho=q^{\pm 1},q^{\pm2}$, correspond exactly to the two resonance conditions $(j,j')=(0,1),(0,2)$ given in (\ref{eq:emb-cond}).

Since the coefficient vanish at the special points, they need to be treated separately. We can set $\rho$ to these special values in the $\kappa QQ$-relation, and then calculate the Gr\"obner basis. For $\rho=q$, the leading terms for the Gr\"obner basis now become
\begin{align}
&\text{LT}[g_1]=(q^2+1)\textcolor{blue}{c_1}\,,\\\nonumber
&\text{LT}[g_2]=q^2(q^2+4q+1)(q^2-4q+1)\textcolor{blue}{c_0^2}\,,
\end{align}
from which it is easy to see that the dimension of the quotient ring is 2. This matches the dimension of the quotient module given in (\ref{eq:dbarj}) with $L=4$ and $(j,j')=(0,1)$. \par

Taking instead $\rho=q^2$, the leading terms of the Gr\"obner basis become
\begin{align}
&\text{LT}[g_1]=6q^6\,\textcolor{blue}{c_0^2}\,,\\\nonumber
&\text{LT}[g_2]=(q^2-q+1)^3(q^2+q+1)^3\,\textcolor{blue}{c_1^3}\,,\\\nonumber
&\text{LT}[g_3]=(q^2-q+1)(q^2+q+1)\,\textcolor{blue}{c_0c_1^2}\,,
\end{align}
from which we find that the dimension of the quotient ring is now 5. This matches the dimension of the quotient module given in (\ref{eq:dbarj}) with $L=4$ and $(j,j')=(0,2)$.

\subsection{Example 3: $L=6$, $M=3$}
In this case, the $Q_{1,0}$ function reads
\begin{align}
Q_{1,0}(t)=t^3+c_2\,t+\frac{c_1}{t}+\frac{1}{c_0t^3}\,.
\end{align}
The ZRC leads to a set of algebraic equations in $\{c_0,c_1,c_2\}$. The corresponding Gr\"obner basis consists of 12 polynomials $\{g_1,\ldots,g_{12}\}$. The leading terms are given by
\begin{align}
&\text{LT}[g_1]=(q^2-\rho^2)(q^4-\rho^2)\,\textcolor{blue}{c_0^2c_1}\,,\\\nonumber
&\text{LT}[g_2]=q^4(q^2-\rho^2)(q^2\rho^2-1)^2(q^4\rho^2-1)\,\textcolor{blue}{c_2^4}\,,\\\nonumber
&\text{LT}[g_3]=q^4(q^2\rho^2-1)^3 G_1(q,\rho)\,\textcolor{blue}{c_1c_2^3}\,,\\\nonumber
&\text{LT}[g_4]=q^4(q^2-\rho^2)(\rho^2-1)(q^2\rho^2-1)\,\textcolor{blue}{c_0c_2^3}\,,\\\nonumber
&\text{LT}[g_5]=q^6(\rho^2-1)^2(q^2\rho^2-1)^3G_2(q,\rho)\,\textcolor{blue}{c_1^2c_2^2}\,,\\\nonumber
&\text{LT}[g_6]=q^4(\rho^2-1)^2(q^2\rho^2-1)\,\textcolor{blue}{c_0c_1c_2^2}\,,\\\nonumber
&\text{LT}[g_7]=q^6(\rho^2-1)^2(q^4-\rho^2)(q^2\rho^2-1)^4(q^4\rho^2-1)^2G_1(q,\rho)G_2(q,\rho)\,\textcolor{blue}{c_1^3c_2}\,,\\\nonumber
&\text{LT}[g_8]=q^2(\rho^2-1)(q^2-\rho^2)(q^4-\rho^2)(q^2\rho^2-1)^2\,\textcolor{blue}{c_0c_1^2c_2}\,,\\\nonumber
&\text{LT}[g_9]=(q^2-\rho^2)^4(q^4-\rho^2)^3(q^4\rho^2-1)\,\textcolor{blue}{c_0^3c_2}\,,\\\nonumber
&\text{LT}[g_{10}]=q^4(\rho^2-1)(q^2-\rho^2)(q^4-\rho^2)(q^2\rho^2-1)^4(q^4\rho^2-1)^4G_1(q,\rho)G_2(q,\rho)\,\textcolor{blue}{c_1^4}\,,\\\nonumber
&\text{LT}[g_{11}]=q^4(\rho^2-1)(q^2-\rho^2)^2(q^4-\rho^2)(q^2\rho^2-1)^2(q^4\rho^2-1)\,\textcolor{blue}{c_0c_1^3}\,,\\\nonumber
&\text{LT}[g_{12}]=(q^2-\rho^2)^4(q^4-\rho^2)^4(q^6-\rho^2)(q^4\rho^2-1)G_1(q,\rho)G_2(q,\rho)\,\textcolor{blue}{c_0^4}\,,
\end{align}
where the polynomials $G_1(q,\rho)$ and $G_2(q,\rho)$ are given by
\begin{align}
\label{eq:G12}
G_1(q,\rho)=&\,-2q^6\,\rho^4+(q^8+2q^6-2q^4+2q^2+1)\rho^2-2q^2\,,\\\nonumber
G_2(q,\rho)=&\,q^6\rho^6+(q^{10}-3q^8-2q^6+3q^4-q^2-1)\rho^4\\\nonumber
&+(q^{10}+q^8-3q^6+2q^4+3q^2-1)-q^4\,.
\end{align}
\paragraph{Special twists and resonance conditions.} We observe that
there are several cases where the leading terms' coefficient vanish. Firstly, $\rho^2=1$ corresponds to the twist-less case. Secondly we have the following conditions
\begin{align}
\label{eq:resonanceRho}
\rho^2=q^{\pm2}\,,\qquad \rho^2=q^{\pm4}\,,\qquad \rho^2=q^{\pm6}\,,
\end{align}
which correspond to the resonance conditions $(j,j')=(0,1),(0,2),(0,3)$ in (\ref{eq:emb-cond}).

In this more complicated example, we also have the possibility that at certain values of $\rho$, the rather complicated polynomials $G_1(q,\rho)$ and $G_2(q,\rho)$ vanish. It is important to notice that the nature of these two types of special values of $\rho$ are different. By tuning $\rho$ to the special values $q^{\pm1},q^{\pm2},q^{\pm3}$ which satisfy the resonance conditions (\ref{eq:resonanceRho}) in the $\kappa QQ$-relation, we can verify that the dimensions of the resulting quotient rings are $5,14,19$ respectively. These are indeed the dimensions of the quotient modules.\par

On the other hand, we find numerically that if we take $\rho$ to be one of the values where $G_1(q,\rho)$ or $G_2(q,\rho)$ vanish, the dimension of the quotient ring does not change, and remains equal to 20. Let us take a more careful look at this point.

For generic value of $\rho$ and $q$,
%the leading term monomials are
%\begin{align}
%\{c_0^2c_1,c_2^4,c_1c_2^3,c_0c_2^3,c_1^2c_2^2,c_0c_1c_2^2,c_1^3c_2,c_0c_1^2c_2,c_0^3c_2,c_1^4,c_0c_1^3,c_0^4\}.
%\end{align}
the corresponding quotient-ring basis is given by
\begin{align}
\label{eq:quotientA}
\left\{c_{0}^2 c_{2}^2,c_0^3,c_0c_1^2,c_1^3,c_0^2 c_2,c_0c_1c_2,\textcolor{red}{c_1^2 c_2},c_0c_2^2,c_1c_2^2,c_2^3,c_0^2,c_0c_1,c_1^2,c_0c_2,c_1 c_2,c_2^2,c_0,c_1,c_2,1\right\}\,.
\end{align}
Now we consider the special value of $\rho$ such that $G_1(q,\rho)$ is vanishing. The leading-term monomials are then given by
\begin{align}
\left\{c_1^2 c_2,c_0^2 c_1,c_2^4,c_0c_2^3,c_0c_1c_2^2,c_0^3 c_2,c_1^4,c_0c_1^3,c_0^4\right\} \,.
\end{align}
We see that there are now only 9 polynomials in the Gr\"obner basis. Nevertheless, the quotient ring is still of dimension 20, with the following basis
\begin{align}
\label{eq:quotientC}
\left\{c_0^2 c_2^2,\textcolor{red}{c_1c_2^3},c_0^3,c_0c_1^2,c_1^3,c_0^2 c_2,c_0c_1c_2,c_0c_2^2,{c_1c_2^2},c_2^3,c_0^2,c_0c_1,c_1^2,c_0c_2,c_1 c_2,c_2^2,c_0,c_1,c_2,1\right\} \,.
\end{align}
We find that the two quotient-ring bases are not exactly the same in
(\ref{eq:quotientA}) and (\ref{eq:quotientC}). The difference stems
from the red-colored monomials.\par

An alternative way to see the difference between the two types of
special values of $\rho$ is by considering different orderings. For a
different ordering, the Gr\"obner basis in general takes a different
form. Nevertheless, at the special values which satisfy resonance
conditions, the coefficients of the leading terms still vanish, such
that the dimension of the quotient ring basis is different. On the
other hand, for a differential ordering, the zeros of $G_1(q,\rho)$
and $G_2(q,\rho)$ in (\ref{eq:G12}) in general no longer lead to vanishing leading terms of the new Gr\"obner basis. Therefore, these values are simply artifacts of the specific ordering that we choose.

\subsection{General comments}
From the above three examples, and by exploring more examples, we can make the following observations:
\begin{enumerate}
\item The resonance conditions of the aTL algebra can be obtained by requiring the coefficients of the leading terms of the Gr\"obner basis to vanish.
\item There are additional values of $\rho$ at which some of the
  leading terms' coefficients vanish. Such additional values are
  artifacts of the specific basis we are working with. By choosing a
  different monominal ordering, they do not lead to vanishing
  coefficients anymore. Thus these values do not reflect the intrinsic properties of the equations themselves.
\item We can set the values of $\rho$ to satisfy the resonance condition in the $\kappa QQ$-relation. The number of solutions for such $\kappa QQ$-relation is precisely the dimension of the quotient module. We can compute the transfer matrices of the quotient module using these $\kappa QQ$-relations. There is then agreement between the dimension of the quotient ring and the dimension of the quotient module \eqref{Wbar-def}.
\end{enumerate}

We emphasize that these are observations based on explicit examples and guided by physical intuitions. It would be very interesting to provide a more rigorous and general proof for these observations.

%%%%%%%%%%%%%%%%%%%%%%%%%%%%%%%%%%%%%%%%%%%%%%%%%%%%%%%%%%%%%%%%%%
\section{Algebraic-geometry approach to the torus partition function}
\label{sec:TorusG}
%%%%%%%%%%%%%%%%%%%%%%%%%%%%%%%%%%%%%%%%%%%%%%%%%%%%%%%%%%%%%%%%%%
In this section, we present the algebraic-geometry approach to computing the torus partition function of the loop model and the Potts model. Using this approach, we obtain closed-form expressions for the partition function $Z(L,N)$ for $L=4,6$ and arbitrary integer $N$, which will be presented in section~\ref{sec:Closedform}. For larger $L$, we can compute the partition function for fixed $L$ up to 10 and for large $N$. The latter results are too long to be presented in the paper. Instead, for such cases we give the zeros of the partition functions. The zeros condense on certain curves in the $N\to\infty$ limit. Such condensation curves can be generated by numerical methods and perfectly match our results. This will be discussed in section~\ref{sec:zeros}.

\subsection{General strategy}
The main quantities to be computed are the companion matrices of the eigenvalues of the transfer matrices. For the chain on $L$ sites, the possible values of $j$ are $j=0,1,\ldots,\tfrac{L}{2}$. For the  loop model, the value $j$ is the number of through-lines. It is related to the XXZ spin chain magnon numbers by $j=\tfrac{L}{2}-M$.

For each fixed $L$ and $j\ne0$, the values of the twists are given by $\rho^2=\omega_k$ (\ref{eq:quantizerho2}).%
\footnote{Notice that the quantization condition is for $\rho^2$. For the value of $\rho$, we can choose two possible signs, according to \eqref{eq:emb-cond}. This raises the question which branch we should choose.
For the partition functions with even $L$ and $M$, the choice of the branch does not matter and we can choose either sign.}
%\begin{align}
%\rho^2=\zeta_k\equiv e^{\frac{2\pi i k}{j}},\qquad k=1,2,\ldots,j.
%\end{align}
We denote the companion matrix of the sector $j$ with the twist $\rho^2=\omega_k$ by $\mathbf{T}_{j,\omega_k}$.

The case $j=0$ is special. For $j=0$, the twist is given by $\rho^2=q^2$, which satisfies the resonance condition. This implies that the corresponding standard module $\mathcal{W}_{0,q^2}$ is reducible and we have $\overline{\mathcal{W}}_{0,q^2}=\mathcal{W}_{0,q^2}/\mathcal{W}_{1,1}$. Therefore, to compute the companion matrices of the transfer matrix, we can perform the computation over $\overline{\mathcal{W}}_{0,q^2}$ and $\mathcal{W}_{1,1}$ separately and then take the direct sum. More precisely, we denote the companion matrix of the transfer matrix over $\overline{\mathcal{W}}_{0,q^2}$ by $\overline{\mathbf{T}}_{0,q^2}$. We then have
\begin{align}
\mathbf{T}_{0,q^2}=\overline{\mathbf{T}}_{0,q^2}\oplus \mathbf{T}_{1,1}.
\end{align}
As discussed in the previous section, the quotient ring of the ZRC which corresponds to $\overline{\mathcal{W}}_{0,q^2}$ is constructed by taking $\rho^2=q^2$ in the ZRC and then performing the standard algebraic-geometry computation.

\paragraph{Momentum decomposition.} The computation of the companion matrix can be simplified further by decomposing the quotient ring. In the current context, one natural way proceeds by performing a decomposition by the \emph{total lattice momentum} of the spin chain. This method has been applied previously to the XXX spin chain \cite{Jacobsen:2018pjt}. The momentum decomposition for the twisted spin chain can be obtained as follows. The Bethe equations (\ref{eq:BAEadditive}) can be written as
\begin{align}
\label{eq:BAE2add}
e^{ip(u_l)L}=-\rho^{-2}\prod_{k=1}^M\frac{\sinh(u_l-u_k+\eta)}{\sinh(u_l-u_k-\eta)}\,,\qquad l=1,2,\ldots,M\,,
\end{align}
where $p(u_l)$ is the momentum of each magnon. Taking the product of (\ref{eq:BAE2add}) for all $l$, we obtain
\begin{align}
e^{iP_ML}=\rho^{-2M} \,,
\end{align}
where $P_M$ is the total momentum
\begin{align}
P_M=p(u_1)+p(u_2)+\cdots+ p(u_M) \,.
\end{align}
Therefore we find that
\begin{align}
e^{iP_M}=e^{\frac{2\pi i n}{L}}\rho^{-\frac{2M}{L}}\,,\qquad n=1,2,\ldots,L\,.
\end{align}
On the other hand, we have
\begin{align}
e^{iP_M}=\prod_{j=1}^M\frac{\sinh(u_l+\tfrac{\eta}{2})}{\sinh(u_l-\tfrac{\eta}{2})}=\prod_{j=1}^M\frac{t_l^2q-1}{t_l^2-q}
=\frac{Q(q^{-1/2})}{Q(q^{+1/2})} \,.
\end{align}
%For the sector with $j$ through lines and $\rho^2=\zeta_k$, we can decompose the corresponding quotient ring by the conditions
Therefore, the momentum-decomposition condition is given by
\begin{align}
\frac{Q(q^{-1/2})}{Q(q^{+1/2})}=e^{\frac{2\pi i n}{L}}\rho^{-\frac{2M}{L}}\,,\qquad n=1,2,\ldots\,.
\end{align}
%where $Q(t)$ is given by (\ref{eq:Qtprod}).
%The function $Q(t)$ is related to the eigenvalue of the transfer matrix
%\begin{align}
%\label{eq:tauQ}
%\tau_M=\rho\,\frac{Q(iq^{-1})}{Q(i)}+\rho^{-1}\frac{Q(iq^{+1})}{Q(i)}.
%\end{align}
%As our choice of the twist are
%\begin{align}
%\rho^2=e^{\frac{2\pi i k}{j}}.
%\end{align}
%The decomposition condition can be written as
%\begin{align}
%\frac{\rQ_M(q^{-1/2})}{\rQ_M(q^{+1/2})}=q^{-M}e^{\frac{2\pi i p}{L}}e^{-\frac{M}{L}\frac{2\pi i k}{j}},\qquad p=1,\cdots,L.
%\end{align}
Thus, for a given through-line number $j\ne0$, we have two independent decompositions. Each sector in the double decomposition is labeled by two integers $(n,k)$ where $n=1,\ldots,L$ and $k=1,\ldots,j$.
%Let us denote the corresponding companion matrix in the sector $(p,k)$ by $\mathbf{T}_{L,j}^{(p,k)}$. The task of AG computation is to construct all the companion matrices $\mathbf{T}_{L,j}^{(p,k)}$.
%The $j=0$ case is special and need to be considered separately. For $j=0$, there is no through line and no quantization condition. We choose the twist to be $\rho^2=q^2$. The transfer matrix in the quotient module is denoted by $\overline{\mathbf{T}}_{L,0}^{(p)}$ where the decomposition according to total momentum can still be performed. The other eigenvalues of the transfer matrix is obtained by taking $j=1$ and $k=1$. We denote this companion matrix by $\widetilde{\mathbf{T}}_{L,1}^{(p,1)}$.

%\paragraph{Singular solutions} (\textcolor{red}{YF: Maybe explain better.}) One important comment associated to the momentum decomposition is that we need to consider the special case
%\begin{align}
%Q(q^{-1/2})=0,\qquad Q(q^{+1/2})=0.
%\end{align}
%If such solution exists, the momentum decomposition condition
%%\begin{align}
%%\rQ_M(q^{-1/2})-q^{-M}e^{\frac{2\pi i(p-M)}{L}}\rQ_M(q^{+1/2})=0
%%\end{align}
%is satisfied for any $M$. This means it will appear in each momentum sector and we will over count such solutions. To avoid this problem, we first select out the solutions that satisfy this condition. Then we perform the momentum decomposition by requiring
%\begin{align}
%Q(q^{-1/2})\ne 0.
%\end{align}
%Technically this can be achieved by adding another equation $w\,Q(q^{-1/2})+1$ to the ideal at the price of introducing an auxiliary variable $w$.

\subsection{$TQ$-relation of the twisted spin chain}
If we apply (\ref{eq:transfertaut}) to compute the companion matrix of the transfer matrix, we need to take the inverse of the companion matrix of $Q(i)$; see for example (\ref{eq:transferi}). There are two drawbacks for this strategy: (1) It is time consuming to take the inverse of the companion matrix analytically, especially when the dimension of the matrix is large; (2) Sometimes the determinant of the companion matrix of $Q(i)$ vanishes, so the matrix does not have an inverse, even though the eigenvalues of the transfer matrix are perfectly well-defined.\par

To avoid these problems, we can use the $TQ$-relation to express the eigenvalues of the transfer matrix directly in terms $\{c_k\}$, which are the coefficients of $Q(t)$. The procedure is as follows. Consider the transfer matrix (with a different normalization) for generic spectral parameter $t$. We have the following $TQ$-relation
\begin{align}
\label{eq:TQrelation}
\tilde{\tau}(t)Q(t)=\rho(t^2q-1)^L\,Q(tq^{-1})+\rho^{-1}(t^2-q)^LQ(t q) \,.
\end{align}
The eigenvalue of the transfer matrix $\tilde{\tau}(t)$ is a polynomial in $t$,
\begin{align}
\label{eq:expandT}
\tilde{\tau}(t)=\sum_{k=0}^L a_k\,t^{2k} \,.
\end{align}
Plugging this and (\ref{eq:Qtprod}) into the $TQ$-relation (\ref{eq:TQrelation}), and comparing the coefficients of different powers in $t$, we obtain a set of algebraic equations in $\{a_k\}$ and $\{c_k\}$. This set of equations are linear both in $\{a_k\}$ and $\{c_k\}$ and can be solved straightforwardly for $\{a_k\}$ in terms of $\{c_k\}$. Plugging back into (\ref{eq:expandT}), we can write $\tilde{\tau}(t)$ in terms of $\{c_k\}$. Once this is done, the transfer matrices that we need for the computation of the torus partition functions are given by
\begin{align}
\tau_M=\frac{\tilde{\tau}(i)}{(-q-1)^L} \,.
\end{align}
The algebraic-geometry computation gives the companion matrix of $\{c_k\}$. Using the $TQ$-relation as described above, we obtain the companion matrix of $\tau_M$, which is the $\mathbf{T}_{j,\omega_k}$ we are after.\par

Compared to the transfer matrices appearing in (\ref{Z-loop}), which are constructed using the link-pattern basis for the standard module, our companion matrix $\mathbf{T}_{j,\omega_k}$ takes a different form, but it share the same eigenvalues, viz., the matrices are related by a similarity transformation. Therefore we have
\begin{align}
\text{tr}_{\mathcal{W}_{j,\rho^2=\omega_k}}\left[t\left(\frac{\gamma}{2}\right)^N\right]=\text{tr}\,\mathbf{T}_{j,\omega_k}^N
\end{align}
for any $N$.
\subsection{The full partition function}
In order to compute the torus partition function, we need to take the Markov trace of the transfer matrices. This is done just as before. Therefore the torus partition function can be computed as in (\ref{Z-loop}),
\begin{align}
Z_{\text{loop}}(L,N)=\text{tr}\,\mathbf{T}_{0,q^2}^N+\sum_{j=1}^{L/2}\sum_{m|j}\Lambda_{j,m}\sum_{k\in\kappa_j(m)}\text{tr}\,\mathbf{T}_{j,\omega_k}^N \,.
\end{align}
The same ingredients can be used to compute the torus partition function of Potts model,
\begin{align}
Z_{\text{Potts}}(L,N)=\mathfrak{m}^{LN/2}\left\{\text{tr}\,\mathbf{T}_{0,q^2}^N+\frac{Q-1}{2}\text{tr}\,{\mathbf{T}_{0,-1}^N}+
\sum_{j=1}^{L/2}\sum_{m|j}\widetilde{\Lambda}_{j,m}\sum_{k\in\kappa_j(m)}\text{tr}\,\mathbf{T}_{j,\omega_k}^N\right\} \,,
\end{align}
where $\widetilde{\Lambda}_{j,m}$ has been defined in (\ref{eq:shiftedLambda}).\par

Before presenting the explicit results, let us make a comment on the computational procedure. In principle, we can follow the standard algorithm and compute the Gr\"obner basis and companion matrices with generic parameters $\rho$ and $q$. In practice, however, keeping $\rho$ and $q$ generic slows down the computation considerably because bulky analytic expressions are generated in the intermediate steps. Therefore, we employ a slightly different strategy in the actual computation. The basic idea is to compute the results for different sets of values of $(\rho,q)$ and then interpolate the final result using this data. We refer to Appendix~\ref{app:AGbasic} for details.

%%%%%%%%%%%%%%%%%%%%%%%%%%%%%%%%%%%%%%%%%%%%%%%%%%%%%%%%%%%%%%
\section{Analytical results in closed form}
\label{sec:Closedform}
%%%%%%%%%%%%%%%%%%%%%%%%%%%%%%%%%%%%%%%%%%%%%%%%%%%%%%%%%%%%%%
In this section, we present the exact closed-form results for $L=4,6$ and general $N$.

%\subsection{Analytic result for $L=4$}
%\label{sec:L4}
%The partition function for $L=4$ is given by
%\begin{align}
%Z(4,M)=\text{tr}\,(t_0)^M+\Lambda_{1,1}\,\text{tr}\,(t_{1,1})^M+\Lambda_{2,1}\,\text{tr}\,(t_{2,1})^M+\Lambda_{2,2}\,\text{tr}\,(t_{2,i})^M
%\end{align}
%In fact, this case is rather simple and we do not need to perform the decomposition to find the eigenvalues of the transfer matrix, although it makes the computation more efficient. Using the AG computation, we reproduced the closed form partition function given in Jesper's note.

\subsection{Analytic result for the loop model}
\label{sec:L6}
The analytic result for $L=4$ has been given in (\ref{Z4M}). This case is rather simple and we do not need to perform the decomposition to find the eigenvalues of the transfer matrix, although it makes the computation slightly more efficient. Using the algebraic-geometry approach, we reproduced the closed-form partition function (\ref{Z4M}), which we will not repeat here.\par

Let us now consider the $L=6$ case. The partition function is given by
\begin{align}
\label{eq:Z6Nanalytic}
Z_{\text{Loop}}(6,N)=&\,\text{tr}\,\mathbf{T}_{0,q^2}^N+\Lambda_{1,1}\,\text{tr}\,\mathbf{T}_{1,1}^N+\Lambda_{2,1}\,\text{tr}\,\mathbf{T}_{2,1}^N
+\Lambda_{2,2}\,\text{tr}\,\mathbf{T}_{2,-1}^N\\\nonumber
&\,+\Lambda_{3,1}\,\text{tr}\,\mathbf{T}_{3,1}^N+\Lambda_{3,3}\left[\mathbf{T}_{3,e^{2i\pi/3}}^N+\mathbf{T}_{3,e^{4i\pi/3}}^N \right] \,.
\end{align}
The multiplicities $\Lambda_{i,j}$ are given by
\begin{align}
\Lambda_{1,1}=&\,\mathfrak{m}^2-2,\\\nonumber
\Lambda_{2,1}=&\,\frac{1}{2}\mathfrak{m}^2(\mathfrak{m}^2-3)\,,\\\nonumber
\Lambda_{2,2}=&\,\frac{1}{2}(\mathfrak{m}^4-5\mathfrak{m}^2+4)\,,\\\nonumber
\Lambda_{3,1}=&\,\frac{1}{3}(\mathfrak{m}^6-6\mathfrak{m}^4+11\mathfrak{m}^2-6)\,,\\\nonumber
\Lambda_{3,3}=&\,\frac{1}{3}\mathfrak{m}^2(\mathfrak{m}^4-6\mathfrak{m}^2+8)\,.
\end{align}
To obtain an analytic result, we need to find the eigenvalues of the companion matrices in the four sectors with $j=0,1,2,3$.
The dimension of these companion matrices are $20,15,6,1$ respectively.

\paragraph{The $j=0$ case.} For $j=0$, there are no through-lines. We take the value of the twist to be $\rho=-q$. Such choice of the twist satisfies the resonance condition and we can compute the eigenvalues for the quotient module $\overline{\mathcal{W}}_{0,q^2}$ and the standard module $\mathcal{W}_{1,1}$.\par

The dimension of the quotient module $\overline{\mathcal{W}}_{0,q^2}$ is 5. The 5 eigenvalues of $\mathbf{T}_{0,q^2}$ are given by
\begin{align}
\lambda_{0,-q}^{\text{quo}}(1)=&\,-\mathfrak{m}^2+2 \,,\\\nonumber
\lambda_{0,-q}^{\text{quo}}(2)=&\,-\mathfrak{m}^2+2 \,,\\\nonumber
\lambda_{0,-q}^{\text{quo}}(3)=&\,-\mathfrak{m}^3-6\mathfrak{m}^2-8\mathfrak{m}-2 \,,\\\nonumber
\lambda_{0,-q}^{\text{quo}}(4)=&\,\frac{1}{2}(\mathfrak{m}^3+8\mathfrak{m}^2+28\mathfrak{m}+28)-\frac{1}{2}(\mathfrak{m}+2)^2\sqrt{\mathfrak{m}^2+48} \,,\\\nonumber
\lambda_{0,-q}^{\text{quo}}(5)=&\,\frac{1}{2}(\mathfrak{m}^3+8\mathfrak{m}^2+28\mathfrak{m}+28)+\frac{1}{2}(\mathfrak{m}+2)^2\sqrt{\mathfrak{m}^2+48} \,.
\end{align}
The dimension of the standard module $\mathcal{W}_{1,1}$ is 15. To find the eigenvalues $\mathbf{T}_{1,1}$ analytically, we perform the momentum decomposition. The value of the twist is taken to be $\rho=-1$. We can take the momentum quantum number to be $p=1,2,\ldots,6$ for non-singular solutions $\rQ_M(q^{1/2})\ne 0$, together with the singular case $\rQ_M(q^{1/2})=\rQ_M(q^{-1/2})=0$. We denote the eigenvalues in each sector by $\lambda_{0,-q}^{\text{std},p}(j)$ and $\lambda_{0,-q}^{\text{std},\text{s}}(j)$. We list the eigenvalues in each sector in what follows.
\begin{itemize}
\item $p=1$. There are 2 eigenvalues:
\begin{align}
\lambda_{0,-q}^{\text{std},1}(1)=&\,-\frac{1}{2}(\mathfrak{m}^2+4\mathfrak{m}+8)+\frac{1}{2}(\mathfrak{m}+2)\sqrt{\mathfrak{m}^2+12}\,,\\\nonumber
\lambda_{0,-q}^{\text{std},1}(2)=&\,-\frac{1}{2}(\mathfrak{m}^2+4\mathfrak{m}+8)-\frac{1}{2}(\mathfrak{m}+2)\sqrt{\mathfrak{m}^2+12}\,.
\end{align}
\item $p=2$. There are 3 eigenvalues. The explicit forms of the eigenvalues are somewhat bulky and we first give the characteristic equation
\begin{align}
&x^3+(4\mathfrak{m}^2+20\mathfrak{m}+30)x^2\\\nonumber
&+2(\mathfrak{m}^4+10\mathfrak{m}^3+30\mathfrak{m}^2+40\mathfrak{m}+30)x
+4(\mathfrak{m}^5+5\mathfrak{m}^4+10\mathfrak{m}^3+10\mathfrak{m}^2+4\mathfrak{m}+2)=0 \,.
\end{align}
We can use the standard formula to find the solution of the cubic equation. The solutions are given by
\begin{align}
\lambda_{0,-q}^{\text{std},2}(k)=-\frac{1}{3}\left(b+\xi^k C^{\tfrac{1}{3}}+\frac{\Delta}{\xi^k C^{\tfrac{1}{3}}} \right),\qquad k=1,2,3
\end{align}
where $\xi=(-1+i\sqrt{3})/2$, and
\begin{align}
b=&\,4\mathfrak{m}^2+20\mathfrak{m}+30\,,\\\nonumber
\Delta=&\,10(\mathfrak{m}+2)^2(\mathfrak{m}^2+6\mathfrak{m}+18)\,,\\\nonumber
C=&\,2(\mathfrak{m}+2)^3(14\mathfrak{m}^3+153\mathfrak{m}^2+594\mathfrak{m}+1188)\\\nonumber
&+6i\sqrt{3}(\mathfrak{m}+2)^4\sqrt{2\mathfrak{m}^4+9\mathfrak{m}^2+432}\,.
\end{align}
\item $p=3$. There are 2 eigenvalues, which are the same as in the $p=1$ case.
\item $p=4$. There are 3 eigenvalues. The characteristic equation is
\begin{align}
&x^3-2(\mathfrak{m}^2+5\mathfrak{m}+3)x^2\\\nonumber
&+(-\mathfrak{m}^4-\mathfrak{m}^3+6\mathfrak{m}^2-4\mathfrak{m}-12)x
+\mathfrak{m}^5+2\mathfrak{m}^4-2\mathfrak{m}^3+4\mathfrak{m}^2+16\mathfrak{m}+8=0\,.
\end{align}
The solutions are again given by
\begin{align}
\lambda_{0,-q}^{\text{std},4}(k)=-\frac{1}{3}\left(b+\xi^k C^{\tfrac{1}{3}}+\frac{\Delta}{\xi^k C^{\tfrac{1}{3}}} \right),\qquad k=1,2,3
\end{align}
where now
\begin{align}
b=&\,-2(\mathfrak{m}^2+5\mathfrak{m}+3)\,,\\\nonumber
\Delta=&\,(\mathfrak{m}+2)^2(7\mathfrak{m}^2+15\mathfrak{m}+18)\,,\\\nonumber
C=&\,-\frac{1}{2}(\mathfrak{m}+2)^3(34\mathfrak{m}^3+117\mathfrak{m}^2+216\mathfrak{m}+108)\\\nonumber
&+\frac{3i\sqrt{3}}{2}(\mathfrak{m}+2)^4\sqrt{8\mathfrak{m}^4+9\mathfrak{m}^2+108} \,.
\end{align}
\item $p=5$. There is 1 eigenvalue,
\begin{align}
\lambda_{0,-q}^{\text{std},5}(1)=2\mathfrak{m}^2+6\mathfrak{m}+2 \,.
\end{align}
\item $p=6$. There are 3 eigenvalues, which are the same as in the $p=4$ case.
\item Singular solution. There is 1 eigenvalue which is given by
\begin{align}
\lambda_{0,-q}^{\text{std},\text{s}}=2\mathfrak{m}+2\,.
\end{align}
\end{itemize}

\paragraph{The $j=1$ case.} For $j=1$, we have 2 through-lines and we take the value of the twist to be $\rho=-1$. This gives precisely the module $\mathcal{W}_{1,1}$. The eigenvalues of $\mathbf{T}_{1,1}$ has already been discussed in the $j=0$ case.

\paragraph{The $j=2$ case.} For $j=2$, we have 4 through-lines and we need to consider two values of the twist: $\rho=1$ and $\rho=i$. The companion matrices $\mathbf{T}_{2,1}$ and $\mathbf{T}_{2,-1}$ are 6-dimensional. The corresponding eigenvalues are given by:
\begin{itemize}
\item Eigenvalues of $\mathbf{T}_{2,1}$:
\begin{align}
\{-\mm,-\mm,-\mm,-\mm,-\mm,12+5\mm\}\,.
\end{align}
\item Eigenvalues of $\mathbf{T}_{2,-1}$:
\begin{align}
\{\pm(\mm+2),\pm(2+\sqrt{3})(\mm+2),\pm(2-\sqrt{3})(\mm+2)\}\,.
\end{align}
\end{itemize}

\paragraph{The $j=3$ case.} For $j=3$, we have 6 through-lines. This case is trivial and for generic value of twist $\rho$, we have
\begin{align}
\mathbf{T}_{3,\rho^2}=\rho+\rho^{-1}\,.
\end{align}
We need to take $\rho=e^{i\pi/3},e^{2i\pi/3},1$ and thus
\begin{align}
\mathbf{T}_{3,1}=2\,,\qquad \mathbf{T}_{3,e^{2i\pi/3}}=1\,,\qquad \mathbf{T}_{3,e^{4i\pi/3}}=-1\,.
\end{align}
Plugging all the eigenvalues of the transfer matrices into (\ref{eq:Z6Nanalytic}) gives the analytic expression of the partition function $Z_{\text{loop}}(6,N)$ of the loop model for any $N$.

\subsection{Analytic result for the Potts model}
To construct the partition function for the Potts model we need an additional piece, which is $\mathbf{T}_{0,-1}$ in the $j=0$ sector. For $L=4,6$, the companion matrix $\mathbf{T}_{0,-1}$ has dimension 6 and 20, respectively. We can find the analytic eigenvalues of the transfer matrix for $L=4$, but not for $L=6$. Therefore, we will write down the analytic expression for $L=4$
\begin{align}
\frac{Z_{\text{Potts}}(4,N)}{\mm^{2N}}=\text{tr}\,\mathbf{T}_{0,q^2}^N+\frac{\mm^2-1}{2}\text{tr}\,\mathbf{T}_{0,-1}^N
+\tilde{\Lambda}_{1,1}\,\text{tr}\,\mathbf{T}_{1,1}^N
+\tilde{\Lambda}_{2,1}\,\text{tr}\,\mathbf{T}_{2,1}^N+\tilde{\Lambda}_{2,2}\,\text{tr}\,\mathbf{T}_{2,-1}^N\,,
\end{align}
where we have
\begin{align}
\tilde{\Lambda}_{1,1}=-1\,,\qquad \tilde{\Lambda}_{2,1}=\frac{1}{2}\mm^2(\mm^2-3)\,,\qquad \tilde{\Lambda}_{2,2}=\frac{1}{2}(\mm^2-2)(\mm^2-1)\,.
\end{align}
There are 6 eigenvalues of $\mathbf{T}_{0,-1}$, given by
\begin{align}
\left\{0,0,\frac{1}{2}(\mm+2)\big(\mm+4\pm\sqrt{\mm^2+16}\big),-\frac{1}{2}(\mm+2)\big(\mm+4\pm\sqrt{\mm^2+16}\big)\right\}.
\end{align}
Assuming $N$ is even, we find that the contribution of $\mathbf{T}_{1,1}^N$ cancels the same contribution in $\mathbf{T}_{0,q^2}^N$ and we find finally
\begin{align}
Z_{\text{Potts}}(4,N)=&\,\mm^{2N}\left\{(\mm^2+7\mm+8)^N+(\mm^2+\mm)^N+2^{N-1}\mm^2(\mm^2-3)\right.\\\nonumber
&\,\left.+(\mm^2-1)(\mm+2)^N\left[(\mm+4+\sqrt{\mm^2+16})^N+(\mm+4-\sqrt{\mm^2+16})^N \right]\right\}.
\end{align}

%%%%%%%%%%%%%%%%%%%%%%%%%%%%%%%%%%%%%%%%%%%%%%%%%%%%%%%%%%%%%%
\section{Zeros of partition functions}
\label{sec:zeros}
%%%%%%%%%%%%%%%%%%%%%%%%%%%%%%%%%%%%%%%%%%%%%%%%%%%%%%%%%%%%%%

In this section we investigate the zeros of the partition function $Z(L,N)$ in the plane of complex $\m$.
For simplicity, we shall focus on the loop model, where we recall that $\m$ is the weight given to each loop;
we leave the Potts model aside for future work.
We are particularly interested in the partial thermodynamic limit, where $L$ is fixed and finite, but $N \gg L$.

\subsection{Condensation curves}

A convenient tool for describing the partial thermodynamic limit is provided by the Beraha-Kahane-Weiss (BKW) theorem \cite{BKW75}.
It states that in the limit $N/L \to \infty$ the partition function zeros condense on certain curves in the complex $\m$-plane,
henceforth called {\em condensation curves}.

Let $\{\Lambda_i(\m)\}$ with $i=1,2,\ldots$ denote the union of eigenvalues of all the transfer matrices that build up the torus partition function
\eqref{Z-loop}. As we have seen, they can be obtained by diagonalizing $t(\gamma/2)$ in any one of the standard modules
entering \eqref{Z-loop}. For a given $\m$, we order these eigenvalues by norm,
$|\Lambda_1(\m)| \ge |\Lambda_2(\m)| \ge \cdots$, and we say that $\Lambda_i(\m)$ is {\em dominant}
at $\m$ if no other eigenvalue has a strictly greater norm.
The BKW theorem \cite{BKW75} then states that the condensation curves
are given by the loci where there are (at least) two dominant eigenvalues, $|\Lambda_1(u)| = |\Lambda_2(u)|$.

We shall use a numerical technique for tracing out the condensation curves that has already been detailed in our previous paper \cite{Jacobsen:2018pjt}.
It combines an efficient method for the numerically exact diagonalization of the sparse transfer matrices with
a direct-search method that allows us to trace out the condensation curves. The sparse matrix decomposition follows directly from
the writing \eqref{transfer} of $t(u)$ as a product of $R$-matrices. The geometrical action can then be inferred from \eqref{splittings},
where we take a link-pattern basis for each standard module. To take into account the possibility that the dominant eigenvalues may
come from sectors with different through-line numbers we actually diagonalize the direct sum of $t(u)$ over all standard modules entering \eqref{Z-loop}.

\subsection{Results}

\begin{figure}[H]
%\begin{center}
\hspace{-1.5cm}\includegraphics[scale=0.35]{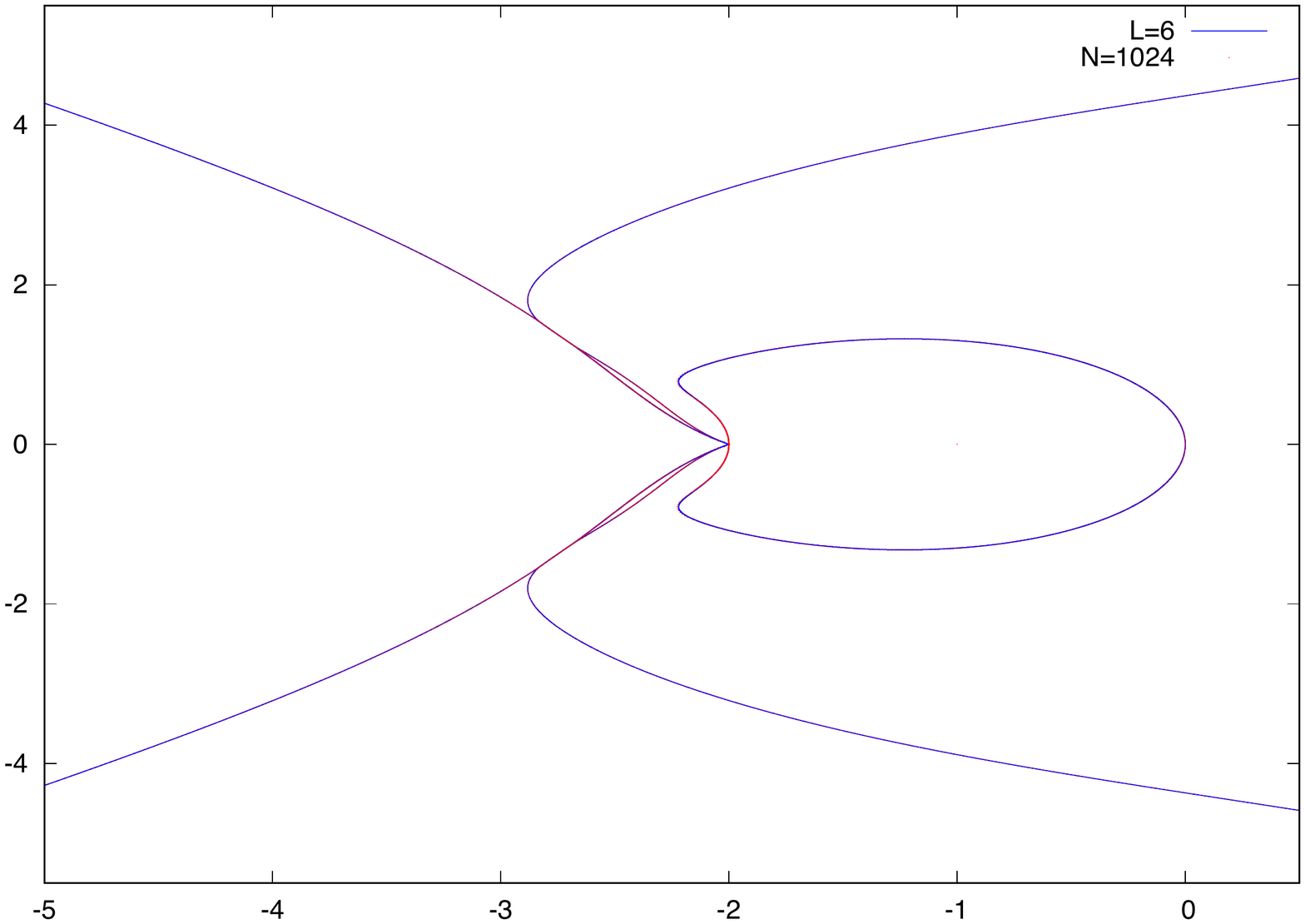} \hspace{-1.5cm}\includegraphics[scale=0.35]{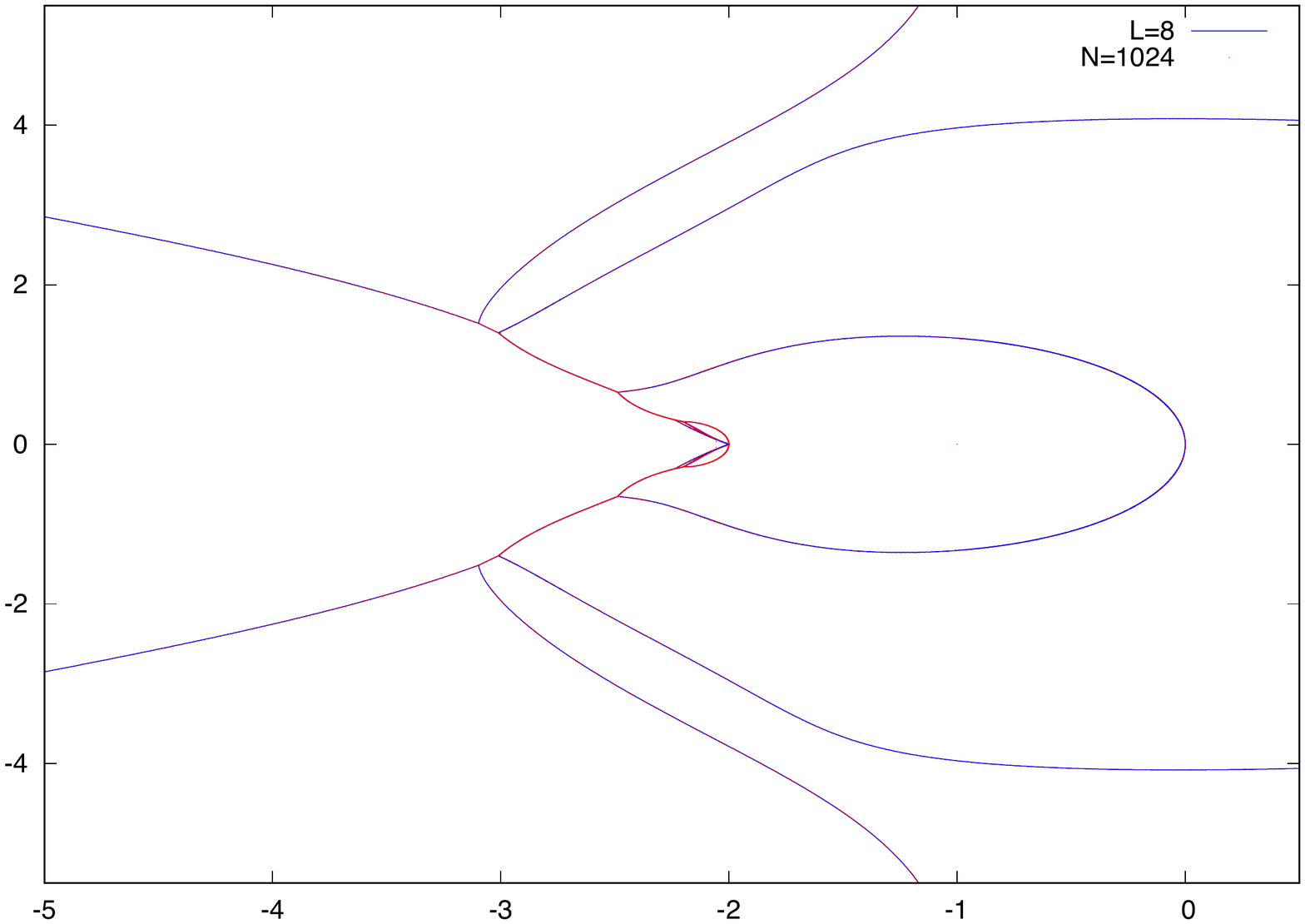} \\

\vspace{-1.5cm}
\hspace{-1.5cm}\includegraphics[scale=0.35]{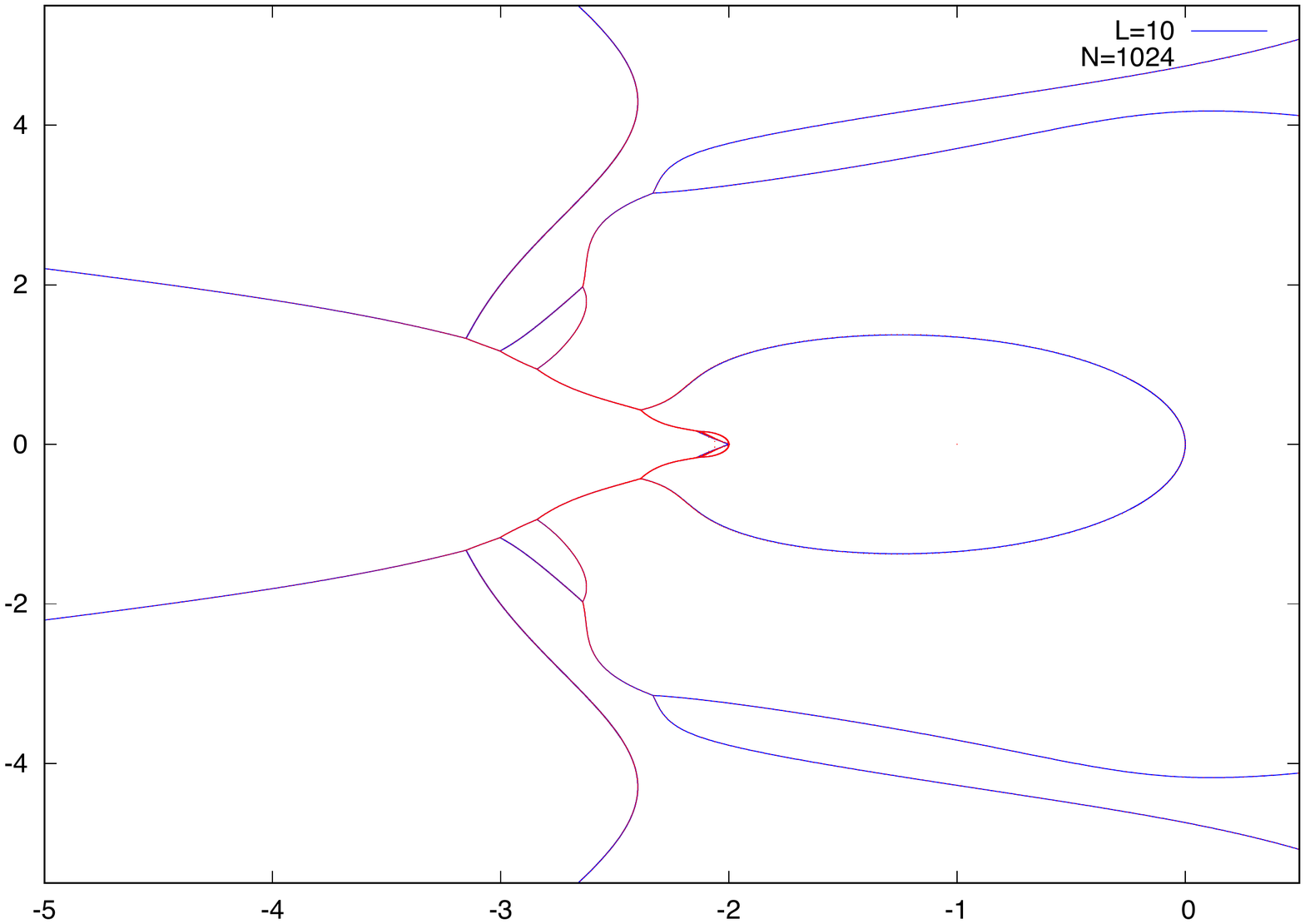} \hspace{-1.5cm}\includegraphics[scale=0.35]{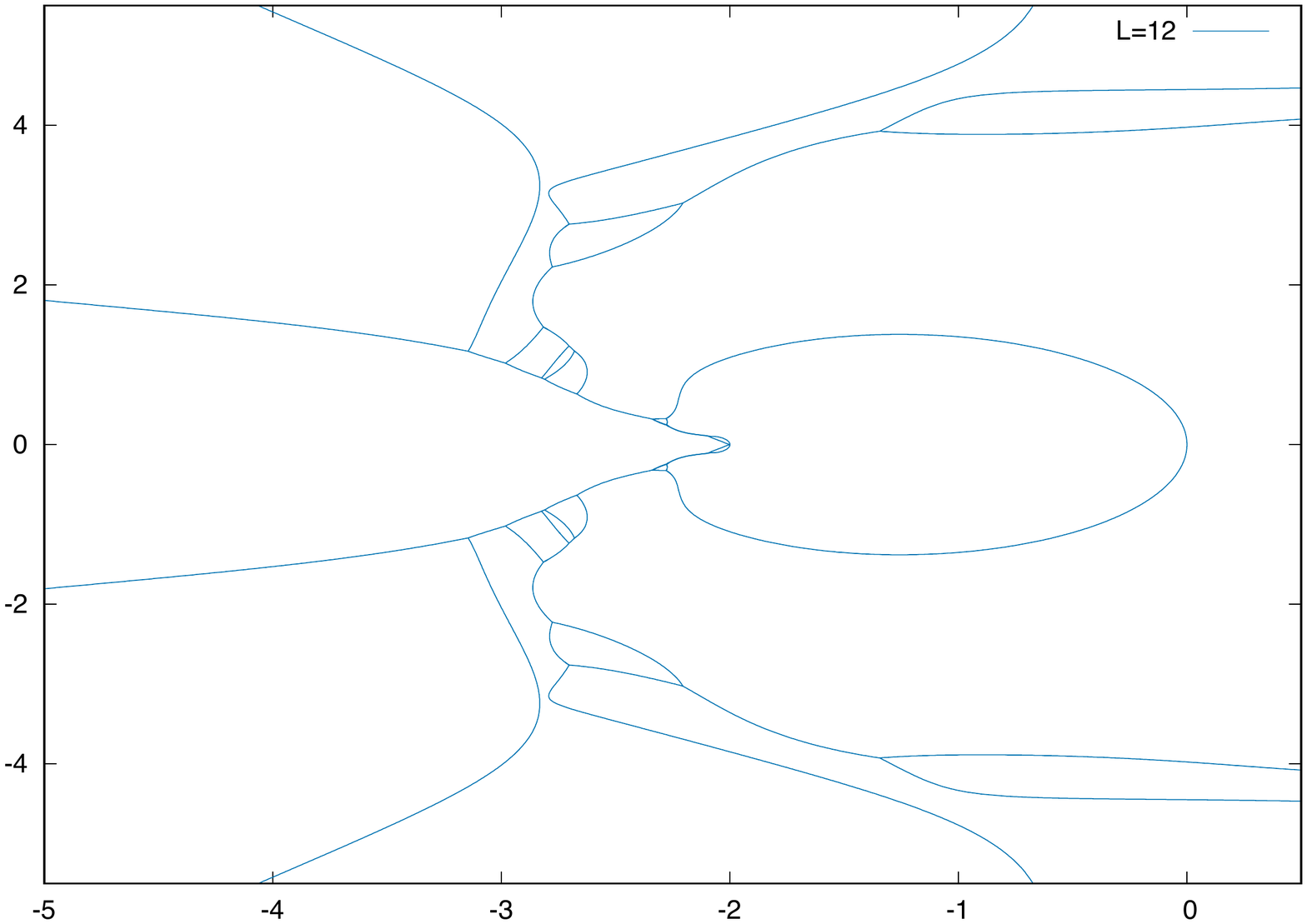} \\

\vspace{-1.5cm}
\hspace{-1.5cm}\includegraphics[scale=0.35]{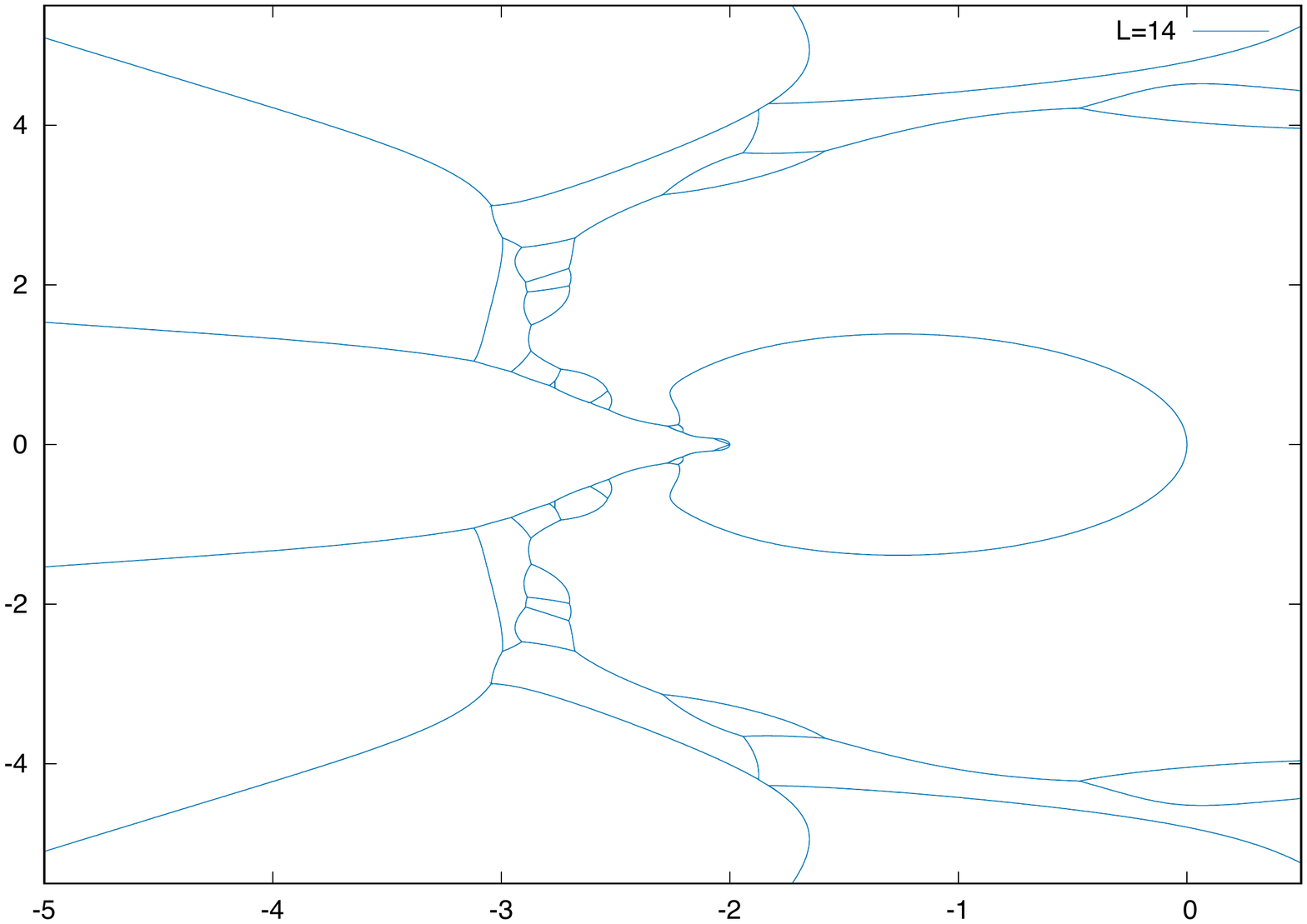} \hspace{-1.5cm}\includegraphics[scale=0.35]{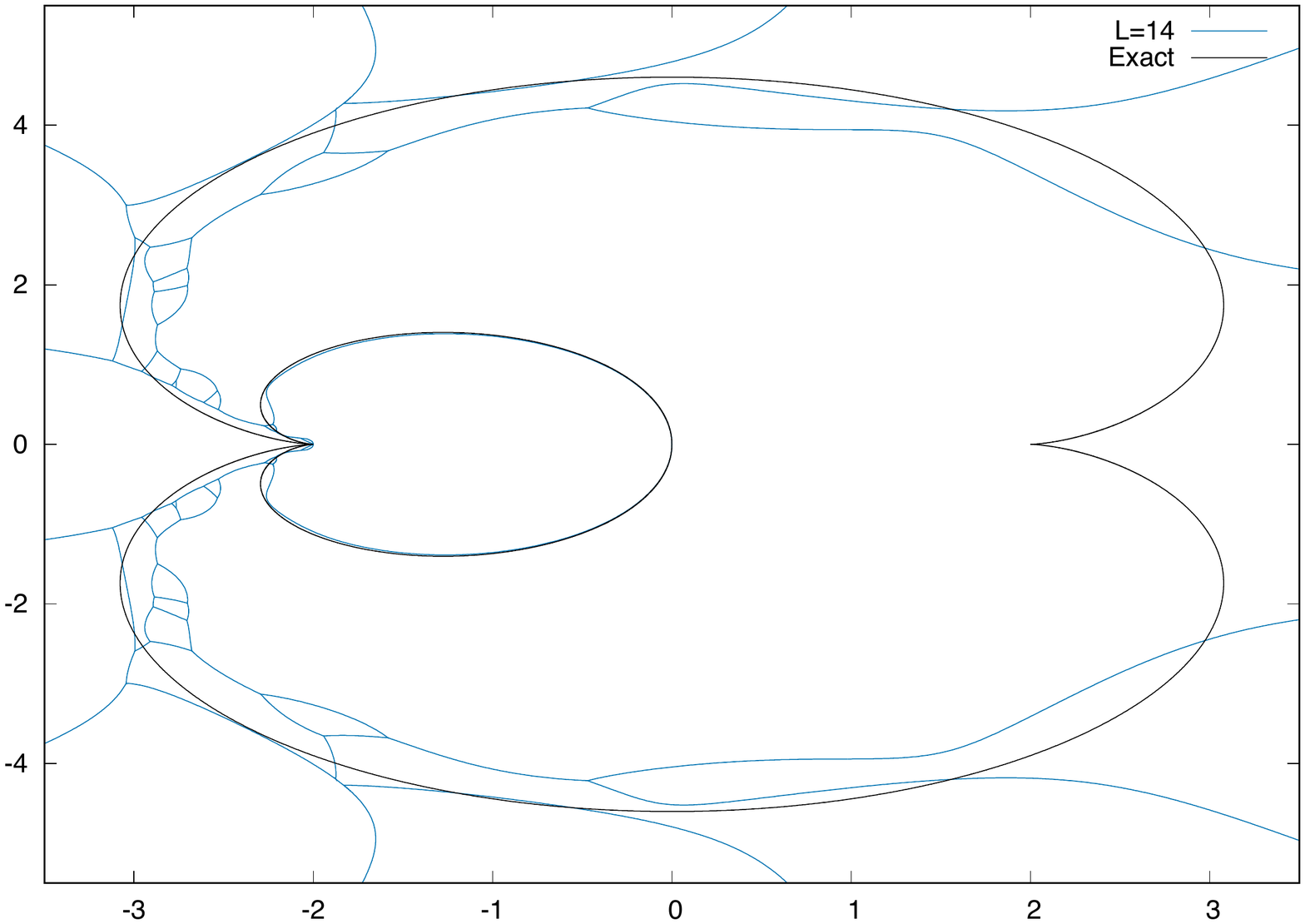}
\vspace{-1.2cm}
\caption{Condensation curves in the complex $\m$-plane for partition-function zeros of the loop model on $L \times N$ tori, in the limit $N \to \infty$.
The panels show, in reading direction, the cases $L=6,8,10,12,14$. The first three panels also superpose the partition-function zeros
for a finite system with $N=1024$ (in red color). The last panel compares the $L=14$ curves with an exact prediction for the thermodynamic
limit ($L\to\infty$). % \textcolor{red}{[JJ: finish last panel, clean up legends.]}
}
\label{fig:cond}
%\end{center}
\end{figure}

The results for the condensation curves with $L=6,8,10,12,14$ are shown in Figure~\ref{fig:cond}.
Notice that the curves are invariant under changing the sign of $\Im\, \m$.
The first three cases, $L=6,8,10$, are compared to the partition function zeros for a system of large
aspect ratio ($N=1024$), obtained from the algebraic-geometry computations. The agreement is seen
to be perfect, but it should be noticed that the density of zeros on the curves exhibits considerable variation.

Apart from this, the condensation curves display several noteworthy features. For any $L$ they go through the points $\m=0$ and $\m=-2$ exactly.
There is a central bubble ${\cal B}_1$ containing those two points, a number of branches extending out towards infinity, and a ``necklace'' region with many bifurcations
and small enclosed regions that appears to produce, as $L$ increases, a larger bubble ${\cal B}_2$ that surrounds ${\cal B}_1$.

We can provide the exact expressions of ${\cal B}_1$ and ${\cal B}_2$ in the thermodynamic limit ($L \to \infty$) by the following argument.
The loop model is known to be critical along a segment of the real line, ${\cal S} = \{ \m | -2 < \m \le 2 \}$, and its continuum limit is described by the Coulomb Gas (CG)
approach to conformal field theory \cite{CGreview}. It is natural to assume \cite{JRS22} that the model remains critical in a larger region ${\cal R}$ of the complex plane,
with ${\cal S} \subset {\cal R}$, and that within ${\cal R}$ all the CG results remain valid by analytic continuation.

It is customary to parametrize the loop weight as
\begin{equation}
 \m = - 2 \cos(\pi g) \,,
\end{equation}
where $g$ is called the CG coupling constant. Inside ${\cal S}$ we have $0 < g \le 1$. The critical exponents of the loop model are conveniently
written in the Kac-table parametrization of conformal weights
\begin{equation}
\label{Kac}
 h_{r,s} = \frac{(r g - s)^2 - (g-1)^2}{4 g} \,.
\end{equation}
The model contains \cite{CGreview} in particular the identity operator $I$ of weight $h_{1,1} = 0$, the energy operator $\varepsilon$ of weight $h_{3,1}$,
and for any $j \in \mathbb{N}^*$ the $2j$-leg operators ${\cal O}_j$ of weight $h_{j,0}$. Notice that ${\cal O}_j$ is the dominant operator inside the
standard module with $2j$ through-lines.

For $-2 < \m < 0$, or $0 < g < \frac12$, we have that $h_{1,0} < 0$, so the 2-leg operator is {\em more} relevant than the identity. Moreover, it is the operator of most
negative conformal weight among the ${\cal O}_j$. The power-law behavior of correlation functions is governed by the real part of the conformal weights, so we can expect a phase
transition to take place when $\Re\, h_{1,0} = 0$. We hypothesize that the locus of this transition is precisely ${\cal B}_1$, in the thermodynamic limit.
Since $h_{1,0} = \frac12 - \frac{1}{4g}$ by \eqref{Kac}, we arrive at the prediction
\begin{equation}
\label{B1-exact}
 {\cal B}_1 = -2 \cos \left( \frac{\pi}{2 + i \mathbb{R}} \right) \,.
\end{equation}
This is in excellent agreement with the condensation curves, as witnessed by the comparison with the $L=14$ curve in the last panel of Figure~\ref{fig:cond}.

The termination of critical behavior in the CG is related \cite{CGreview} to the fact that the energy operator $\varepsilon$, of weight $h_{3,1} = 2 g - 1$, becomes marginal when $g \uparrow 1$ (\emph{i.e.}, $\m \uparrow 2$).
We therefore expect termination of critical behavior when $\Re\, h_{3,1} = 1$. Hypothesizing that the locus of this transition is precisely ${\cal B}_2$ we arrive at our second prediction,
\begin{equation}
 \label{B2-exact}
 {\cal B}_2 = -2 \cos \left( \frac{\pi}{1 + i \mathbb{R}} \right) \,.
\end{equation}
As seen from the last panel of Figure~\ref{fig:cond}, this cuts right through the ``necklace'' region, and although finite-$L$ effects are definitely larger than in the case of ${\cal B}_1$,
we expect this part of the condensation curves to converge to \eqref{B2-exact} when $L \to \infty$. We finally identify the region of criticality ${\cal R}$ with the interior of ${\cal B}_2$.

%%%%%%%%%%%%%%%%%%%%%%%%%%%%%%%%%%%%%%%%%%%%%%%%%%%%%%%%%%%%%%
\section{Conclusions and discussion}
\label{sec:conclude}
%%%%%%%%%%%%%%%%%%%%%%%%%%%%%%%%%%%%%%%%%%%%%%%%%%%%%%%%%%%%%%
In this paper, we discovered a precise relation between representation theory of the aTL algebra and the algebraic structure of the solutions to the Bethe equations of the XXZ spin chain with a diagonal twist. While such a connection might seem surprising at the first glance, it turns out to be quite natural. One of the reasons that this connection appears to have been left unnoticed hitherto is presumably that it would be rather difficult to see it by solving the Bethe equation numerically. This demonstrates the power and usefulness of computational algebraic geometry in the study of integrable models.

Using this relation, we were able to compute the torus partition function of the O($\mathfrak{m}$) loop model and the $Q$-state Potts model in two very different ways, which we called the \emph{geometric algebra} and \emph{algebraic geometry} approaches, respectively. The final results of these two methods match nicely. We studied the distribution of partition-function zeros for $L=6,8,10$ in the partial thermodynamic limit and found that the zeros condense on certain curves. The condensation curves can be constructed directly in the partial thermodynamic limit and match beautifully with our numerical results of partition function zeros. We would like to highlight the fact that part of the condensation curves can be understood from conformal field theory in the full thermodynamic limit, $L \to \infty$. This is quite remarkable because usually condensation curves can only be determined numerically, while here we can write down the analytic formulae (\ref{B1-exact})--(\ref{B2-exact}) for these curves.

As a byproduct of the algebraic-geometry approach, we constructed the rational $Q$-system for the \emph{twisted} Bethe equations, both for the XXX and the XXZ spin chains. The main advantage of the rational $Q$-system, compared to the original Bethe equations, is that it leads to physical solutions only, and hence we do not need to impose further conditions to exclude unphysical ones. We expect that our construction generalizes naturally to
higher-rank cases, some of which have recently been shown to harbor random-geometry applications generalizing those of the loop model \cite{Lafay1,Lafay2}.

From the perspective of the CAG (Computational Algebraic Geometry)-BAE (Bethe Ansatz Equations) program, this work is the first step in applying CGA techniques to XXZ-type BAE, which are not polynomial equations in their original form, but which can nevertheless be rationalized by a proper change of variables. These equations contain free parameters, which are the anisotropic parameter $q$ and the twist $\rho^2$ in our case. On the one hand, these parameters bring richer structures to the equations because the solutions depend on the parameters. There can be special loci in the parameter space where the structure of the solutions change drastically and exhibit noteworthy features. The resonance condition which we analyzed in this paper is an example. On the other hand, it brings new challenges in the CAG computations. To start with, the zero-remainder conditions associated with the rational $Q$-system are more involved. At the same time, although the computation of Gr\"obner bases and companion matrices are purely algebraic and straightforward, the standard algorithm leads to rather bulky analytic expressions in intermediate steps. This fact slows down the computation considerably. This prohibits us from obtaining explicit analytic expression for $L>10$ by the algebraic-geometry approach.

A number of natural future directions can be pursued following the
current work. One important immediate question is to investigate the
case where $q$ is a root of unity. It has been known for a while that
the representation theory of Temperley-Lieb algebra has an intricate structure \cite{Pasquier:1989kd} in this case. Already in the non-periodic
case, this leads to indecomposable structures in the lattice model \cite{RidoutSaintaubin,Dubail-b} which entail
strong connections to the corresponding structures in the logarithmic CFT that govern the continuum limit \cite{Vasseur}. The case of periodic boundary conditions
is even richer and, as yet, the resulting incomposable structures are only partially understood (see the review \cite{LCFTreview} and references therein).
Given the relation established in the current paper for generic $q$,
we expect that these structures should also be encoded in the
solutions of the corresponding Bethe equations. Indeed, we can already
see that all the Gr\"obner bases that we constructed in this paper
exhibit vanishing leading coefficients when $q$ is an appropriate root of unity. Can we see more fine structures by using proper algebro-geometric tools? We leave this fascinating question for future studies.

Another natural generalization is to consider the XYZ spin chain. This case is even less investigated, as it is significantly more complicated. Following the strategy of the current paper, we need to seek for a proper change of variables to rationalize the Bethe equations which involve elliptic function. This seems to be possible following the treatment in \cite{baxter2002completeness}, but with new features.\par

Apart from developing algebro-geometric methods for more general types of Bethe equations, one can also find applications of our method in broader contexts. So far we have mainly applied our methods to compute partition functions of important statistical-mechanical models, such as the six-vertex model \cite{Jacobsen:2018pjt} and the Potts model. In recent years, it has been discovered that integrable spin chains have deep connections with supersymmetric gauge theories in various setups. We give one potential application as an example. By the Bethe/gauge duality \cite{Nekrasov:2009uh,Nekrasov:2009rc,Nekrasov:2014xaa}, Bethe-ansatz techniques can be applied to compute twisted superconformal indices \cite{Closset:2016arn}. The Bethe equations arising in this context are precisely those of the twisted XXZ spin chain which we studied in this paper. The algebraic-geometry approach developed here can be applied straightforwardly to compute twisted superconformal indices analytically.

\section*{Acknowledgements}
We would like to thank S.\ Ribault and H.\ Saleur for inspiring discussions.
The research of JLJ was supported by the European Research Council through the grant NuQFT
and by the Agence Nationale de la Recherche through the grant CONFICA (grant No.\ ANR-21-CE40-0003).
YZ received support from the NSF
of China through Grants No.\ 11947301, 12047502, 12075234
and the Key Research Program of the Chinese Academy of Sciences through Grant No.\ XDPB15. YZ is also
grateful to the Institute of Theoretical Physics, Chinese Academy of Sciences, for its hospitality
through the ``Peng Huanwu visiting professor program''.
Gef\"ordert durch die Deutsche Forschungsgemeinschaft (DFG) - Projektnummer 286237555 - TRR 195 [Funded by the Deutsche Forschungsgemeinschaft (DFG, German Research Foundation) - Project- ID 286237555 - TRR 195]. The work of JB was supported by Project B5 of SFB-TRR 195.

\appendix
%%%%%%%%%%%%%%%%%%%%%%%%%%%%%%%%%%%%%%%%%%%%%%%%%%%%%%%%%%%%%%
\section{More on computational algebraic geometry}
\label{app:AGbasic}

In this appendix, we provide an overview of the methods of computational algebraic geometry used in the main text. Our discussion contains two parts. In the first part, we introduce the basic notions of computational algebraic geometry with examples. Most of the contents in this part is standard. For more details we refer to the textbooks \cite{GP,CA}. When the algebraic equations under consideration contain extra free parameters, like the ZRC of the twisted XXZ spin chain, direct computations lead to rather complicated analytic expressions in the intermediate steps. This slows down the computations considerably. To improve the efficiency, we need to employ a slightly different strategy, which will be discussed in the second part.

%\section{Algebraic sets}

\subsection{Basic notions of computational algebraic geometry}
\begin{definition}
An \textbf{affine algebraic set} is the common zero set%
\[
V(f_{1},\ldots,f_{r})=\left\{  p\in K^{n}\mid f_{1}(p)=0,\ldots,f_{r}%
(p)=0\right\}
\]
of polynomials $f_{1},\ldots,f_{r}\in R=K[x_{1},\ldots,x_{n}]$.
\end{definition}

Is there a structure in algebra which corresponds to the geometry? If
$f_{1}\left(  p\right)  =0,\ldots,f_{s}\left(  p\right)  =0$ for $p\in K^{n}$,
then also every $R$-linear combination of the $f_{i}$ vanishes on $p$, that is%
\[
\left(  \sum_{i=1}^{s}r_{i}\cdot f_{i}\right)  \left(  p\right)  =\sum
_{i=1}^{s}r_{i}\left(  p\right)  f_{i}\left(  p\right)  =0
\]
for all $r_{i}\in R$. This leads to a key concept in algebra, the notion of an ideal:

\begin{definition}
\label{Def Ideal}Let $R$ be a commutative ring with $1$. An \textbf{ideal} is
a non-empty subset $I\subset R$ with $a+b\in I$ and $ra\in I$ for all $a,b\in
I$ and $r\in R$.

For $S\subset R$ we write $\left\langle S\right\rangle $ for the \textbf{ideal
generated} by $S$, which consists of all finite $R$-linear combinations of
elements of $S$.
\end{definition}

So it makes sense to generalize our definition of a vanishing loci to
arbitrary subsets $S\subset R$, and, as seen above, we have $V(f_{1}%
,\ldots,f_{s})=V(\left\langle f_{1},\ldots,f_{s}\right\rangle )$. On the other hand,
if $I\subset R$ is an ideal, then $V(I)$ is an algebraic set: Every ideal
$I\subset R$ is finitely generated, that is, of the form $I=\left\langle
f_{1},\ldots,f_{s}\right\rangle $ with $f_{i}\in R$. Rings with this property are
called Noetherian.

\begin{definition}
A commutative ring $S$ with $1$ is called \textbf{Noetherian} if every ideal
$I\subset S$ is finitely generated, equivalently, if every ascending chain of
ideals $I_{1}\subset I_{2}\subset\ldots$ becomes stationary.
\end{definition}

\begin{theorem}
[Hilbert Basis Theorem]\label{thm Hilbert basis theorem}If $S$ is a Noetherian
ring, then also $S\left[  x\right]  $ is Noetherian. In particular, if $K$ is
a field then, by induction, $K\left[  x_{1},\ldots,x_{n}\right]  $ is Noetherian.
\end{theorem}

%\section{Division with remainder and Gr\"{o}bner bases}

\begin{definition}
\label{def L}Let $K$ be a field. An element
\[
x^{\alpha}:=x_{1}^{\alpha_{1}}\cdot \ldots \cdot x_{n}^{\alpha_{n}}\in R
\]
is called a \textbf{monomial}, and an element $c\cdot x^{\alpha}$ with $c\in K$ is
called a \textbf{term}. Every polynomial $f\in R$ is a finite sum $f=%
%TCIMACRO{\tsum \nolimits_{\alpha}}%
%BeginExpansion
{\textstyle\sum\nolimits_{\alpha}}
%EndExpansion
c_{\alpha}x^{\alpha}$ of terms. The \textbf{degree} $\deg(f)$ of $f$ is the
maximal $\left\vert \alpha\right\vert :=\alpha_{1}+\ldots+\alpha_{n}$ with
$c_{\alpha}\neq0$.

If $>$ is a total ordering on the set of monomials of $R$ and $0\neq f\in R$,
then the largest term $\operatorname{LT}(f)=c\cdot x^{\alpha}$ is called the
\textbf{lead term} of $f$, $\operatorname{LC}(f)=c$ the \textbf{lead
coefficient} of $f$, and $L(f)=x^{\alpha}$ the \textbf{lead monomial} of $f$.
For $f=0$ we set $L(0)=LT(0)=LC(0)=0$.
\end{definition}

\begin{definition}
A \textbf{monomial ordering} on the semigroup of monomials $x^{\alpha}$ is a
total ordering $>$ which respects multiplication, that is, $x^{\alpha
}>x^{\beta}\Rightarrow x^{\alpha}x^{\gamma}>x^{\beta}x^{\gamma}$ for all
$\alpha,\beta,\gamma$. A monomial ordering is \textbf{global }if it is a well
ordering,\footnote{That is, any non-empty set of monomials has a smallest
element.} equivalently if $x_{i}>1$ for all $i$.
\end{definition}

\begin{definition}
The most important global monomial orderings are the \textbf{lexicographical
ordering} (lp), where $x^{\alpha}>x^{\beta}$ if the leftmost nonzero entry of
$\alpha-\beta$ is positive, and the \textbf{degree reverse ordering} (dp),
where we prefer the monomial of larger degree, and break a tie via $x^{\alpha
}>x^{\beta}$ if there is an $1\leq i\leq n$ such that $\alpha_{n}=\beta
_{n},\ldots,\alpha_{i+1}=\beta_{i+1},\alpha_{i}<\beta_{i}$. While the former
is the key to solve multivariate polynomial systems, the latter usually leads
to the best performance in computations.
\end{definition}

\begin{example}
\label{ex divrem}Using $lp$ we divide $x^{2}y+x$ by $y-1$ and $x^{2}-1$:%
\[
\boldsymbol{x}^{2}\boldsymbol{y}+x=x^{2}\left(  \boldsymbol{y}-1\right)
+1\cdot\left(  \boldsymbol{x}^{2}-1\right)  +\underline{x+1}\text{.}%
\]
Here the lead terms are shown in bold and the remainder is underlined.
\end{example}

\begin{definition}
Given a monomial ordering $>$ and a subset $G\subset R$, we define the\textbf{
leading ideal} of $G$ as%
\[
L(G)=\left\langle L(f)\mid f\in G\backslash\{0\}\right\rangle \subset
R\text{.}%
\]

\end{definition}

\begin{definition}
[Gr\"{o}bner bases]Given an ideal $I$ and a global monomial ordering $>$, then
for a finite subset $G\subset I$ with $0\notin G$ obviously $L(G)\subset
L(I)$, and we call $G$ a \textbf{Gr\"{o}bner basis} of $I$ with respect to
$>$, if $L(G)=L(I)$.
\end{definition}

\begin{theorem}
Every ideal $I\subset K[x_{1},\ldots,x_{n}]$ has a Gr\"{o}bner basis.
\end{theorem}

\begin{definition}
\label{def NF}Given a set $G=\{g_{1},\ldots,g_{s}\}\subset R$, a
\textbf{normal form} is a map $\operatorname*{NF}(-,G):R\rightarrow R$ with

\begin{enumerate}
\item $\operatorname*{NF}(0,G)=0$.

\item If $\operatorname*{NF}(f,G)\neq0$ then $L(\operatorname*{NF}(f,G))\notin
L(G)$.

\item For every $0\neq f\in R$ there are $a_{i}\in R$ with%
\[
f-\operatorname*{NF}(f,G)=%
%TCIMACRO{\tsum \nolimits_{i=1}^{s}}%
%BeginExpansion
{\textstyle\sum\nolimits_{i=1}^{s}}
%EndExpansion
a_{i}g_{i}%
\]
and $L(f)\geq L(a_{i}g_{i})$ for all $i$ with $a_{i}g_{i}\neq0$. Such an
expression we call a
\index{standard represenation}%
\textbf{standard represenation} of $f$.
\end{enumerate}
\end{definition}

\begin{lemma}
Given a set $G=\{g_{1},\ldots,g_{s}\}\subset R$ and any order of preference of
the $g_{i}$ in the division, the \textbf{Buchberger division} yields a normal
form $\operatorname*{NF}(-,G)$, the \textbf{Buchberger normal form}.
\end{lemma}

\begin{example}
\label{ex divrem2}If in Example \ref{ex divrem}, we prefer the divisor
$x^{2}-1$ over $y-1$, we obtain:%
\[
\boldsymbol{x^{2}y}+x=y\cdot\left(  \boldsymbol{x^{2}}-1\right)
+\underline{x+y}\ \
\]
So depending on the choice made, the remainder will be $x+1$ or $x+y$.
\end{example}

\begin{theorem}
[Ideal membership]\label{thm ideal membership}Let $I\subset R$ be an ideal and
$f\in R$. If $G=\left\{  g_{1},\ldots,g_{s}\right\}  $ is a Gr\"{o}bner basis
of $I$ and $\operatorname*{NF}$ is a normal form, then%
\[
f\in I\Longleftrightarrow\operatorname*{NF}(f,G)=0\text{.}%
\]

\end{theorem}

\begin{corollary}
\label{cor gb generates}If $G\ $is a Gr\"{o}bner basis of $I$, then
$I=\left\langle G\right\rangle $.
\end{corollary}

\begin{definition}
An element $f\in R$ is called \textbf{reduced} with respect to a set $G\subset
R$, if no term of $f$ is contained in $L(G)$. A normal form
$\operatorname*{NF}(-,G)$ is called a \textbf{reduced normal form} if
$\operatorname*{NF}(f,G)$ is reduced with respect to $G$ for all $f$.
\end{definition}

\begin{example}
If we in the Buchberger normal form put non-divisible terms into the remainder
also in intermediate steps, we obtain a reduced normal form, the
\textbf{reduced Buchberger normal form}.
\end{example}

\begin{example}
\label{ex divrem3}We can continue the Example \ref{ex divrem2}:%
\[
\boldsymbol{x}^{2}\boldsymbol{y}+x=y\cdot\left(  \boldsymbol{x}^{2}-1\right)
+\underline{x}+1\cdot(\boldsymbol{y}-1)+\underline{1}%
\]

\end{example}

Indeed, the remainder is now unique provided we divide by a Gr\"{o}bner basis:

\begin{theorem}
\label{thm reduced unique}Let $>$ be a global ordering, $I\subset R$ an ideal,
$f\in R$ and $G$ a Gr\"{o}bner basis of~$I$. If $\operatorname*{NF}(-,G)$ is a
reduced normal form, then the map $\operatorname*{NF}(-,G)$ is uniquely
determined by $>$ and~$I$. We then also write $\operatorname*{NF}(-,I)$.
\end{theorem}

\begin{remark}
Even when using a reduced normal form and a Gr\"{o}bner basis, although the
remainder is unique, the generated expression in the generators may not be. In
the Examples \ref{ex divrem} and \ref{ex divrem3} we obtain\vspace{-0.1cm}%
\[
\boldsymbol{x}^{2}\boldsymbol{y}+x=y\cdot\left(  \boldsymbol{x}^{2}-1\right)
+1\cdot(\boldsymbol{y}-1)+\underline{x+1}\vspace{-0.1cm}%
\]
and%
\[
\boldsymbol{x}^{2}\boldsymbol{y}+x=x^{2}\left(  \boldsymbol{y}-1\right)
+1\cdot\left(  \boldsymbol{x}^{2}-1\right)  +\underline{x+1}%
\]
respectively.
\end{remark}

Gr\"{o}bner bases can be computed via Buchberger's algorithm, which is derived from

\begin{theorem}
[Buchberger's criterion]\label{thm buchberger criterion}%
\index{Buchberger's criterion}%
If $I\subset R$ is an ideal, $\operatorname*{NF}$ a normal form and $0\notin
G=\{g_{1},\ldots,g_{s}\}\subset I$, then $G$ is a Gr\"{o}bner basis of $I$, if
and only if $I=\left\langle G\right\rangle $ and $\operatorname*{NF}%
(\operatorname*{spoly}(g_{i},g_{j}),G)=0$ for all $i\neq j$, where%
\[
\operatorname*{spoly}(f,g)=\frac{\operatorname{lcm}(L(f),L(g))}{LT(f)}%
f-\frac{\operatorname{lcm}(L(f),L(g))}{LT(g)}g
\]
is constructed to cancel the lead terms.
\end{theorem}

\begin{corollary}
By computing $r=\operatorname*{NF}(\operatorname*{spoly}(g_{i},g_{j}),G)$ for
all $i,j$, if $r\neq0$ adding $r$ to the generating set and iterating until
all $r$ vanish, one obtains a Gr\"{o}bner basis of $\left\langle
G\right\rangle $. This process terminates in finitely many steps since $R\ $is
Noetherian, and is called \textbf{Buchberger's algorithm}.
\end{corollary}

Buchberger's algorithm generalizes both Gaussian elimination (for reducing
linear systems of equations) and the Euclidean algorithm (for computing
greatest common divisors of univariate polynomials) to higher-degree
multivariate polynomial equations. Alternative algorithms to obtain
Gr\"{o}bner bases include Faug\`{e}re's F4/F5 algorithms \cite{faugere}. For
our computations, we make use of the implemenation of Buchberger's algorithm
in the open-source computer algebra system \textsc{Singular} \cite{DGPS}.

\begin{example}
\label{ex bb}Buchberger's algorithm using the ordering $lp$ transforms the
following given system%
\[%
\begin{tabular}
[c]{r}%
$2x^{2}-xy+2y^{2}-2=0$\\
$2x^{2}-3xy+3y^{2}-2=0$%
\end{tabular}
\ \ \ \ \longmapsto\ \ \ \
\begin{tabular}
[c]{r}%
$3y+8x^{3}-8x=0$\\
$4x^{4}-5x^{2}+1=0$%
\end{tabular}
\ \ \ \
\]
such that the resulting system allows us to read off the four solutions
$(x,y)=\left(  \pm1,0\right)  ,(\pm\frac{1}{2},\pm1)$, see Figure
\ref{fig two ell}. This behaviour is typical for systems with zero-dimensional
solution sets when using $lp$.%
%TCIMACRO{\FRAME{fhFU}{4.2258in}{1.8672in}{0pt}{\Qcb{Buchberger's algorithm for
%the intersection of two ellipses}}{\Qlb{fig two ell}}{twoell2.eps}%
%width 4.2258in;  height 1.8672in;  depth 0pt;  original-width 6.6002in;
%original-height 2.8966in;  cropleft "0";  croptop "1";  cropright "1";
%cropbottom "0";  filename 'twoell2.eps';file-properties "XNPEU";}}}%
%BeginExpansion
\begin{figure}[h]
\begin{center}
\includegraphics[
height=1.8672in,
width=4.2258in
]%
{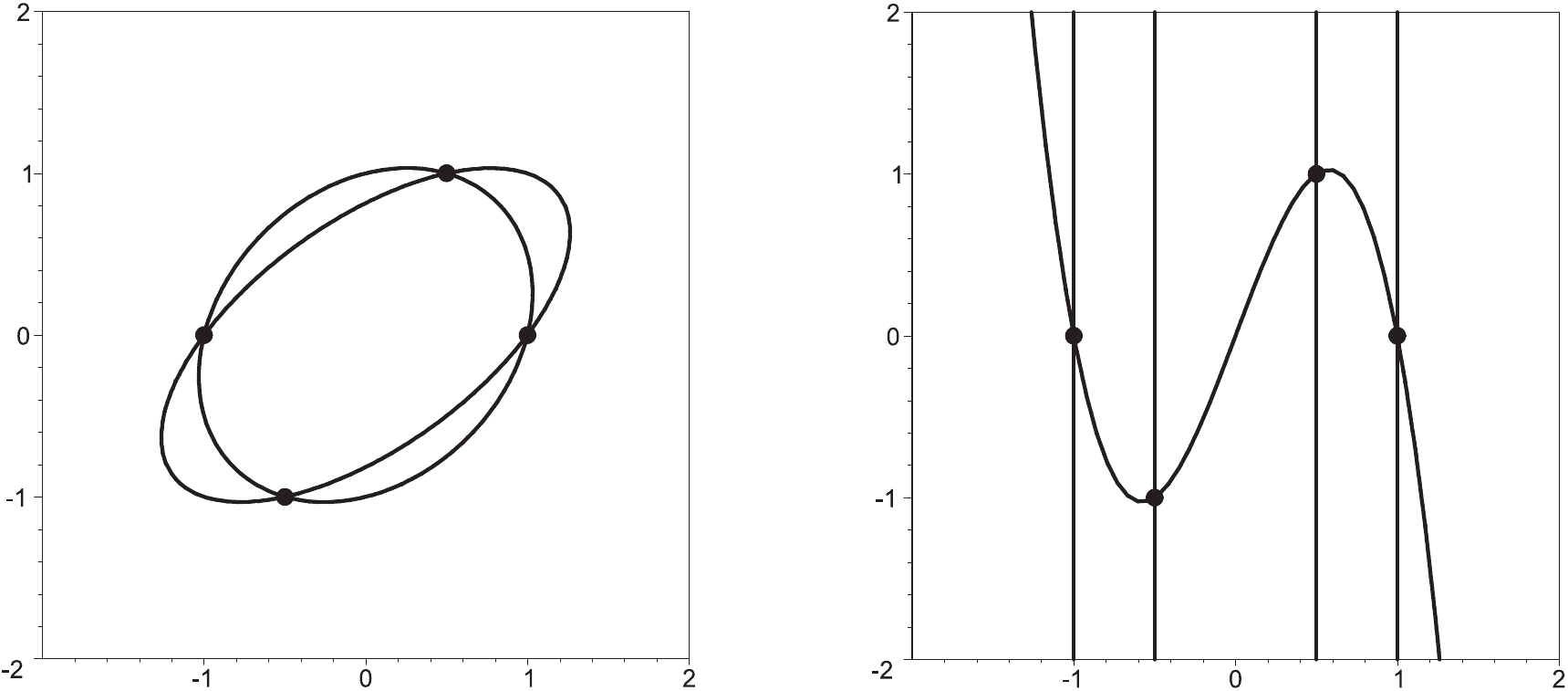}%
\caption{Buchberger's algorithm for the intersection of two ellipses.}%
\label{fig two ell}%
\end{center}
\end{figure}
%EndExpansion

\end{example}

%\section{Coordinate ring}

\begin{definition}
Let $S$ be a commutative ring with $1$ and $I\subset S$ an ideal. The set
$S/I=\{\overline{f}=f+I\mid f\in S\}$ is a commutative ring with
one\footnote{$\overline{f}=f+I=\{f+g\mid g\in I\}$. Note that $\overline{f}$ is
the class of $f$ with respect to the equivalence relation $f\sim
g\Leftrightarrow$ $f-g\in I$.} with representative-wise addition $\overline
{f}+\overline{g}:=\overline{f+g}$ and multiplication $\overline{f}%
\cdot\overline{g}:=\overline{f\cdot g}$.
\end{definition}

\begin{definition}
For our usual notation, and $I\subset R$ an ideal, the quotient ring $R/I$ is
called the \textbf{coordinate ring} of $V(I)$.
\end{definition}

In particular, if, as in Example \ref{ex bb}, the vanishing locus of a given
ideal $I$ is finite, the coordinate ring $R/I$ allows us to directly analyize
$V(I)$. The coordinate ring becomes computationally accessible via Gr\"{o}bner bases.

\begin{lemma}
If $I\subset R$ is an ideal and $\operatorname*{NF}(-,I)$ a reduced normal
form, then%
\[
\phi:R/I\longrightarrow V=_{K}\left\langle x^{\alpha}\mid x^{\alpha}\notin
L(I)\right\rangle \text{, \quad}\overline{f}\longmapsto\operatorname*{NF}%
(f,I)
\]
is an isomorphism of $K$-vector spaces. With the multiplication
on $V$ induced by applying $\operatorname*{NF}(-,I)$ after the respective
operation in $R$, it becomes an isomorphism of $K$-algebras.
\end{lemma}

\begin{theorem}
\label{thm zero dim solve}Let $K=\overline{K}$, $I\subset R$. Then the
following are equivalent:

\begin{enumerate}
\item $\left\vert V\left(  I\right)  \right\vert <\infty$

\item If $G$ is a Gr\"{o}bner basis of $I$, then for every $i$ there is a
$g\in G$ with $L(g)=x_{i}^{\alpha_{i}}$ and $\alpha_{i}\geq0$.

\item $\dim_{K}(R/I)<\infty$.
\end{enumerate}
\end{theorem}

\begin{theorem}
In the setting of the previous theorem,
\[
\left\vert V\left(  I\right)  \right\vert \leq\dim_{K}\left(  R/I\right)
\]
and equality holds if $I$ is a radical ideal.\footnote{The radical of $I$ is
$\sqrt{I}=\left\{  f\in R\mid\exists a\in\mathbb{N}\text{ with }f^{a}\in
I\right\}  $, and $I$ is racial if $I=\sqrt{I}$. If $K=\overline{K}$ and
$V(I)$ is finite, this amounts to $I$ not having multiple zeros.}
\end{theorem}

%\section{Companion matrix}

\begin{definition}
With respect to the basis $\mathcal{B}=\left\{  x^{\alpha}\mid x^{\alpha
}\notin L(I)\right\}  =:\{m_{1},\ldots,m_{r}\}$ of $V$, multiplication
$\overline{f}\cdot$ by $\overline{f}$ in $R/I$ is given by matrix
multiplication with the representing matrix%
\[
M_{f}:=M_{\mathcal{B}}^{\mathcal{B}}(\overline{f}\cdot)=(a_{ij})_{i,j=1,\ldots
,r}%
\]
defined by%
\begin{equation}
\overline{f}\cdot\bar{m}_{j}=\sum_{i=1}^{r}a_{ij}\bar{m}_{i}%
\label{companion_matrix}
\end{equation}
The matrix $M_{f}$ is called the \textbf{companion matrix} of $f$.
\end{definition}

\begin{remark}
Passing from $\overline{f}\cdot$ to $M_{f}$ is a well-defined isomorphism of
$K$-algebras from $R/I$ to a subalgebra of $\operatorname*{Mat}_{n\times
n}(K)$: We have $M_{f}=M_{g}$ if an only if $\overline{f}=\overline{g}$, and
$M_{f+g}=M_{f}+M_{g}$ and $M_{f\cdot g}=M_{f}\cdot M_{g}=M_{g}\cdot M_{f}$.
\end{remark}

\begin{definition}
If $M_{g}$ is invertible, then the companion matrix of the rational
function $f/g$ $\in\operatorname{quot}(R)$ is $M_{f/g}=M_{f}\cdot
M_{g}^{-1}$.
\end{definition}

\begin{remark}
Assume that $K=\overline{K}$ and $I$ is radical, and write $V(I)=\left\{
p_{1},\ldots,p_{r}\right\}  $. Then the sum of the values of $f\in R$ over the
$p_{i}$ can be computed as the trace%
\begin{equation}
\sum_{i=1}^{r}f(p_{i})=\operatorname{Tr}M_{f}\text{.}%
\label{eq:sum_companion}
\end{equation}
\end{remark}

\subsection{Implementation notes}
From the general introduction above, we see that if a physical quantity is
expressed as a rational function summed over all Bethe roots, it can
be calculated as the trace of the companion matrix
\eqref{eq:sum_companion}. In practice, the efficient implementation
of this method is a non-trivial task, especially for our case with two
parameters (anisotropy parameter $q$ and the twist $\rho^2$). This is mainly due to two reasons. To start with, the ZRC from the rational $Q$-system, which are obtained by taking polynomial remainders, involve complicated coefficients which are rational functions of $q$ and $\rho$. Second, the algebro-geometric computation makes such complications proliferate and lead to bulky expressions in the intermediate steps. On the other hand, we know the final results that we are after have some simple structures, \emph{e.g.} they are a polynomial in $\m$ with integer coefficients. Therefore our strategy is to perform the computation for different sets of fixed numbers $(q,\rho)$ and then use proper interpolation formulae to find the final result.

In more details, when computing the Gr\"obner basis, we consider {\it
  $\mathbb Q$-valued} $\m$ and a constant $\rho$, and perform the Gr\"obner-basis
computation. Here $\m= -q-q^{-1}$ in the usual parametrization. Note
that the choice of $\m$ should avoid the singular values of
$q$. The Gr\"obner-basis computation with differnt
  $\mathbb Q$-valued $\m$ and a constant $\rho$ can be parallelized on different
cores or computer nodes. We observe that the computation time for
different values is roughly the same, so the parallelization is very
efficient. The Gr\"obner-basis computation is powered by the computer
algebra system
{\sc Singular} \cite{DGPS}.

Secondly, for each value of $\m$ and $\rho$, once the Gr\"obner basis is
obtained, we use \eqref{companion_matrix} to get the companion
matrix. Then the formula \eqref{eq:sum_companion} is used to
compute the corresponding value of the partition function. In this
step, the linear-algebra related computations were done using {\sc Singular}. Here we emphasize  one subtlety in this step:
the companion matrix of a polynomial's power, $M_{f^n}$, is
needed for our computation. In practice, $f^n$ may be a huge
polynomial with a high
degree and to handle such an object requires a lot of RAM. One might seek the help from the property of a companion matrix,
\begin{equation}
  \label{eq:1}
  M_{f^n} =(M_f)^n,
\end{equation}
and carry out the power computation on matrices  instead of on
polynomials. However, in practice, we find that $M_f$ may be a
dense matrix and such a matrix-power computation is not particularly
efficient. Therefore, we consider the remainder $\bar f$ of $f$
divided by the Gr\"obner basis and the following relation,
\begin{equation}
  \label{eq:2}
  M_{f^n}= M_{\bar f^n}\,.
\end{equation}
For the computation of $\bar f$'s power, after each
multiplication, we divide the intermediate result towards the
Gr\"obner basis and take the remainder. In the way, the power computation is both efficient and RAM
economical.

Finally, from the previous two steps, we get the partition function's
values for a list of $\mathbb Q$-values of $\m$ and the constant $\rho$. Note that the
partion function is {\it not} a polynomial in $q$ but a polynomial in
$\m$. Then we do a standard Newton-polynomial interpolation
in $\m$ to get the fully analytic partition function. This step is
also powered by {\sc Singular}.

We observe that these computations over {\it finite fields} are much
faster than over rational numbers. The results over finite fields can
be further lifted to rational numbers. Although in this paper, our
computation is done with rational numbers, the finite-field method is
promising for future computations.

%%%%%%%%%%%%%%%%%%%%%%%%%%%%%%%%%%%%%%%%%%%%%%%%%%%%%%%%%%%%%%

\section{Partition function results}

In the appendix, we give some results for the partition functions on $L \times L$ tori.

\subsection{Loop model}
\label{app:resloop}

\begin{subequations}
\begin{scriptsize}
\begin{eqnarray}
 Z(2,2) &=& 8 \m + 8 \m^2 \,, \\
 Z(4,4) &=& 8192 \m + 21480 \m^2 + 21632 \m^3 + 10870 \m^4 + 2912 \m^5 + 416 \m^6 + 32 \m^7 + 2 \m^8 \,, \\
 Z(6,6) &=& 933120000 \m + 4805838000 \m^2 + 11273241408 \m^3 + 16045483644 \m^4 + 15559888944 \m^5 + 10937711238 \m^6 + \nonumber \\
 & & 5789102208 \m^7 + 2366852316 \m^8 + 761946512 \m^9 + 196445178 \m^{10} + 41330040 \m^{11} + 7264344 \m^{12} + \nonumber \\
 & & 1093680 \m^{13} + 142350 \m^{14} + 15504 \m^{15} + 1296 \m^{16} + 72 \m^{17} + 2 \m^{18} \,, \\
 Z(8,8) &=& 11399736556781568 \m + 98753367284702880 \m^2 + 408215404199487488 \m^3 + 1073458486777028536 \m^4 + \nonumber \\
 & & 2019645508709576192 \m^5 + 2898901544581297904 \m^6 + 3305315996372032512 \m^7 + 3077975536301413966 \m^8 + \nonumber \\
 & & 2388746188868407168 \m^9 + 1568981821963717920 \m^{10} + 882883392305392000 \m^{11} + 429902632210623280 \m^{12} + \nonumber \\
 & & 182694107341546624 \m^{13} + 68277701699851360 \m^{14} + 22602553762014464 \m^{15} + 6675367972759608 \m^{16} + \nonumber \\
 & & 1772015524475520 \m^{17} + 426101046596224 \m^{18} + 93529351075584 \m^{19} + 18862409602432 \m^{20} + \nonumber \\
 & & 3507873891968 \m^{21} + 601033262944 \m^{22} + 94278773888 \m^{23} + 13385143248 \m^{24} + 1693508480 \m^{25} + \nonumber \\
 & & 187378704 \m^{26} + 17724928 \m^{27} + 1392704 \m^{28} + 87296 \m^{29} + 4096 \m^{30} + 128 \m^{31} + 2 \m^{32} \,,
\end{eqnarray}
\end{scriptsize}
\end{subequations}

\subsection{Potts model}
\label{app:respotts}

\begin{subequations}
\begin{scriptsize}
\begin{eqnarray}
 Z(2,2) &=& 4 \m^3 + 7 \m^4 + 4 \m^5 + \m^6 \,, \\
 Z(4,4) &=& 4096 \m^9 + 14056 \m^{10} + 20064 \m^{11} + 15866 \m^{12} + 7904 \m^{13} + 2716 \m^{14} + 688 \m^{15} + 129 \m^{16} + 16 \m^{17} + \m^{18} \,, \\
 Z(6,6) &=& 466560000 \m^{19} + 2787706800 \m^{20} + 7672413504 \m^{21} + 2973997052 \m^{22} + 15177608016 \m^{23} + 13124121918 \m^{24} + \nonumber \\
 & & 8762270208 \m^{25} + 4665716646 \m^{26} + 2034402892 \m^{27} + 742926501 \m^{28} + 231451420 \m^{29} + 62280555 \m^{30} + \nonumber \\
 & & 14527800 \m^{31} + 2918460 \m^{32} + 496704 \m^{33} + 69786 \m^{34} + 7788 \m^{35} + 649 \m^{36} + 36 \m^{37} + \m^{38} \,, \\
 Z(8,8) &=& 5699868278390784 \m^{33} + 54114294708625056 \m^{34} + 245768814667966464 \m^{35} + 712035958563348408 \m^{36} + \nonumber \\
 & &1480662138337422848 \m^{37} + 2357818409374775024 \m^{38} + 2995910569524062208 \m^{39} + 3125648118282379662 \m^{40} + \nonumber \\
 & & 2735085239079547520 \m^{41} + 2040940779811857696 \m^{42} + 1316441555189606912 \m^{43} + 742458511348946736 \m^{44} + \nonumber \\
 & & 369838333858415104 \m^{45} + 164188589066409952 \m^{46} + 65497210088055296 \m^{47} + 23650757646619036 \m^{48} + \nonumber \\
 & & 7779745847048704 \m^{49} + 2343002229704412 \m^{50} + 648253398180864 \m^{51} + 165020839221248 \m^{52} + \\
 & & 38632494033536 \m^{53} + 8297778619424 \m^{54} + 1628522160640 \m^{55} + 290394241736 \m^{56} + 46707916288 \m^{57} + \nonumber \\
 & & 6714987800 \m^{58} + 853018752 \m^{59} + 94339168 \m^{60} + 8906112 \m^{61} + 698400 \m^{62} + 43712 \m^{63} + 2049 \m^{64} + 64 \m^{65} + \m^{66} \,. \nonumber
\end{eqnarray}
\end{scriptsize}
\end{subequations}

%\bibliographystyle{utphys}
%\bibliography{refs}

\providecommand{\href}[2]{#2}\begingroup\raggedright\endgroup

\end{document}